\documentclass[format=acmsmall, review=false, screen=true]{acmart}
\usepackage{booktabs} 
\usepackage{graphicx}
\usepackage{caption}
\usepackage{subcaption}
\usepackage{makecell}
\usepackage{slashbox}
\usepackage{array,multirow}
\usepackage{color}
\usepackage{xcolor}
\usepackage{hyperref}
\usepackage{amssymb}
\usepackage{rotating}
\usepackage{dirtytalk}
\usepackage{csquotes}
\usepackage{amsmath, bm}
\usepackage[sans]{dsfont}
\usepackage{upquote}
\usepackage{media9}
\usepackage[rightcaption]{sidecap}

\usepackage[ruled,linesnumbered]{algorithm2e} 
\usepackage{parskip}

\SetAlFnt{\small}
\SetAlCapFnt{\small}
\SetAlCapNameFnt{\small}
\SetAlCapHSkip{0pt}
\IncMargin{-\parindent}

\setcopyright{acmlicensed}

\begin{document}
\title[Learning to Engage with Interactive Systems: A Field Study on DRL in a Public Museum]{Learning to Engage with Interactive Systems: A Field Study on Deep Reinforcement Learning in a Public Museum}

\author{Lingheng Meng}
\orcid{0000-0001-9419-5774}
\affiliation{%
  \institution{University of Waterloo}
  \streetaddress{200 University Avenue West}
  \city{Waterloo}
  \state{Ontario}
  \postcode{N2L 3G1}
  \country{Canada}}
\email{lingheng.meng@uwaterloo.ca}

\author{Daiwei Lin}
\orcid{ }
\affiliation{%
  \institution{University of Waterloo}}
\email{daiwei.lin@uwaterloo.ca}

\author{Adam Francey}
\orcid{ }
\affiliation{%
  \institution{University of Waterloo}}
\email{alzfrancey@uwaterloo.ca}

\author{Rob Gorbet}
\orcid{ }
\affiliation{%
  \institution{University of Waterloo}}
\email{rbgorbet@uwaterloo.ca}

\author{Philip Beesley}
\orcid{ }
\affiliation{%
  \institution{University of Waterloo}}
\email{pbeesley@uwaterloo.ca}

\author{Dana Kuli\'c}
\orcid{ }
\affiliation{%
  \institution{University of Waterloo}}
\affiliation{%
    \institution{Monash University}}
\email{dana.kulic@monash.edu}

\begin{abstract}
Physical agents that can autonomously generate engaging, life-like behaviour will lead to more responsive and interesting robots and other autonomous systems. Although many advances have been made for one-to-one interactions in well controlled settings, future physical agents should be capable of interacting with humans in natural settings, including group interaction. In order to generate engaging behaviours, the autonomous system must first be able to estimate its human partners' engagement level.  In this paper, we propose an approach for estimating engagement during group interaction by simultaneously taking into account active and passive interaction, i.e. occupancy, and use the measure as the reward signal within a reinforcement learning framework to learn engaging interactive behaviours. The proposed approach is implemented in an interactive sculptural system in a museum setting.  We compare the learning system to a baseline using pre-scripted interactive behaviours. Analysis based on sensory data and survey data shows that adaptable behaviours within an expert-designed action space can achieve higher engagement and likeability.
\end{abstract}

%
%


%
%


\maketitle

\renewcommand{\shortauthors}{L. Meng, D. Lin, A. Francey, R. Gorbet, P. Beesley and D. Kuli\'c}

\section{Introduction}
\label{sec:INTRODUCTION}

As robots enter human environments, engaging with human occupants in a suitable manner becomes increasingly important \cite{sidner2004look, rich2010recognizing, bohus2014managing, castellano2009detecting, sidner2003engagement, khosla2017human, anzalone2015evaluating, ivaldi2017towards}. To facilitate long-term interaction, interactive systems should be able to continuously adapt in order to generate engaging and appropriate behaviour. Although for simple interactive systems it may be possible to manually design engaging behaviours, this approach is time-consuming and sometimes unfeasible for complex non-anthropomorphic interactive systems with many degrees of freedom (DOF), e.g. hundreds of DOF. In addition, manually designed behaviours constrain the system to a limited set of reactions, whereas autonomous generation of actions provides the possibility of adaption and continuous behavioural evolution in a nonstationary environment. Therefore, understanding how to autonomously generate engaging behaviour is necessary for long-term Human Robot Interaction (HRI), and learning algorithms can be advantageous. 

When introducing learning into an interactive system, human input could be explicit or implicit.  An example of explicit input is a human taking a teacher role, giving feedback to guide the robot to achieve some goals. With implicit input, a learning signal is generated based on observations of human behaviours during interaction, without requiring active feedback. For successful long-term interaction in social and public settings, the implicit model is more appropriate than the explicit teacher model, because in the implicit model learning is not dependent on human expertise or willingness to train the system.  Finally, for successful interaction in social and public settings, the system should successfully engage with and learn from both indiviudal and group interactions.

\begin{figure}[htbp]
        \begin{subfigure}[t]{.34\linewidth}
            \centering
            \includegraphics[width=1\linewidth]{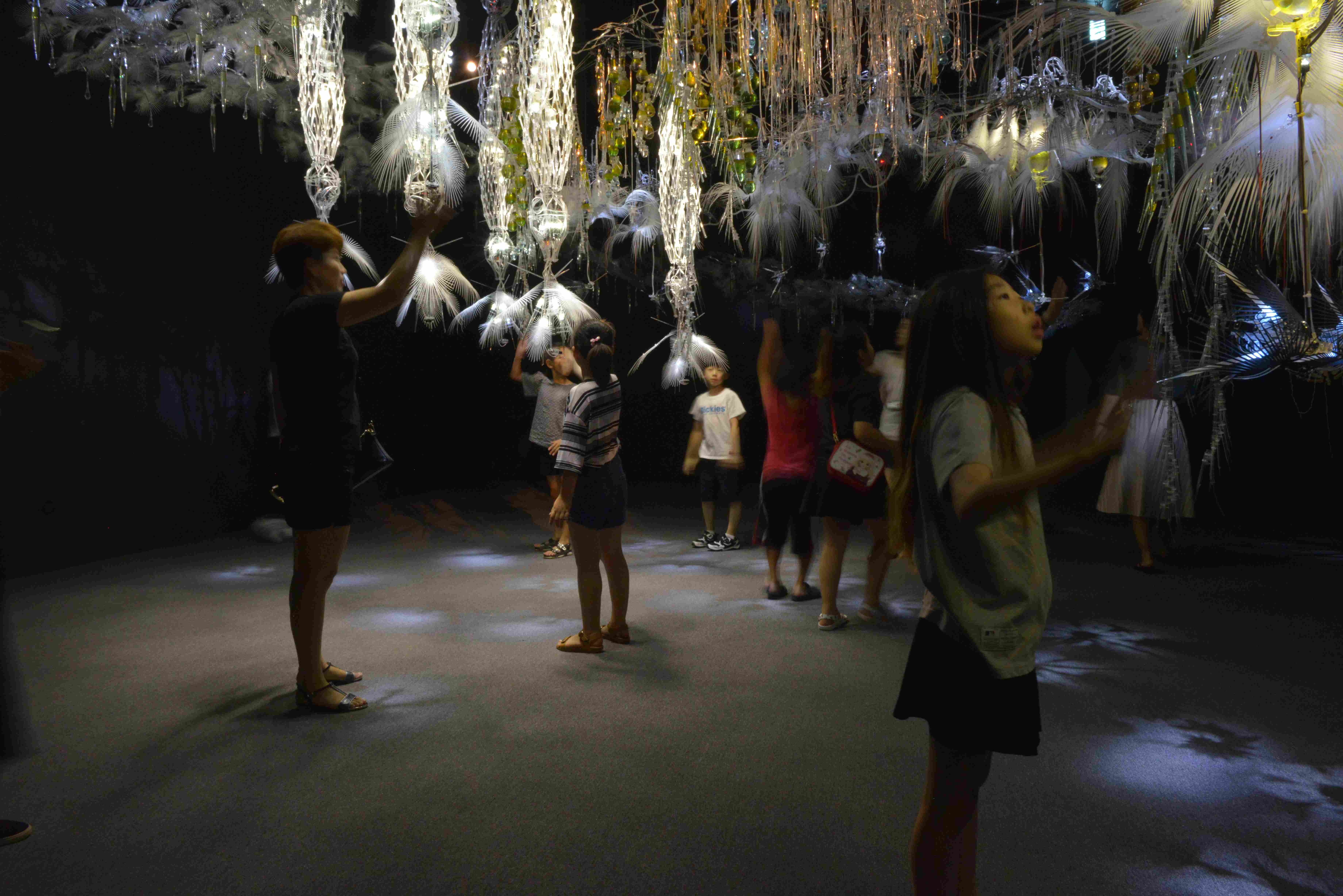}
            \caption{\emph{Radiant Soil}, installed in Daejeon Museum of Art in Daejeon, South Korea, 2018.}
            \label{fig:Radiant_Soil_2017}
        \end{subfigure}
        \hspace{5pt}%
        \begin{subfigure}[t]{.34\linewidth}
            \centering
            \includegraphics[width=1\linewidth]{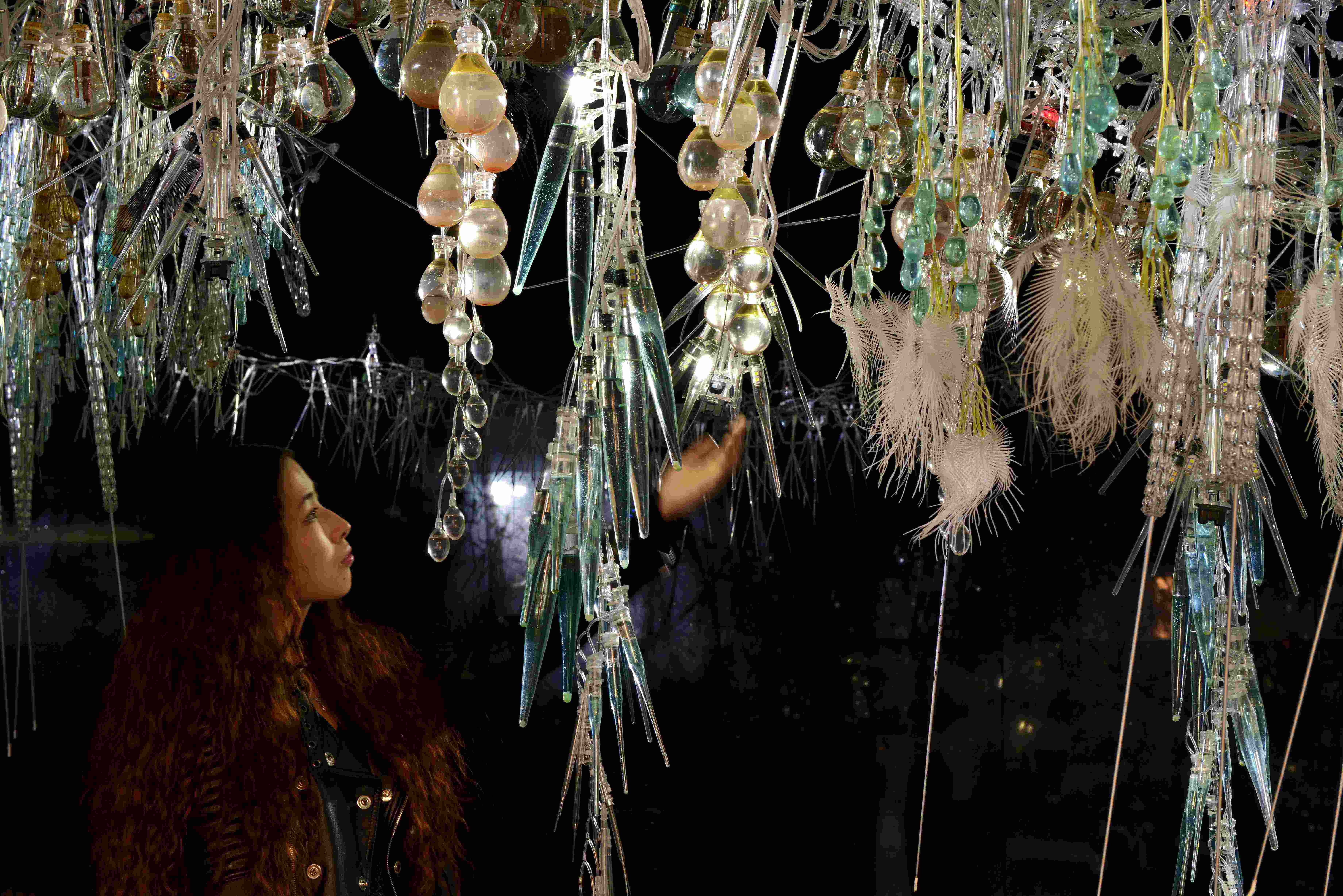}
            \caption{\emph{Epiphyte Spring}, installed in Design Institute of Landscape and Architecture in Hangzhou, China, 2015.}
            \label{fig:Epiphyte_Spring_2015}
        \end{subfigure}
        \hspace{5pt}%
        \begin{subfigure}[t]{.24\linewidth}
            \centering
            \includegraphics[width=1\linewidth]{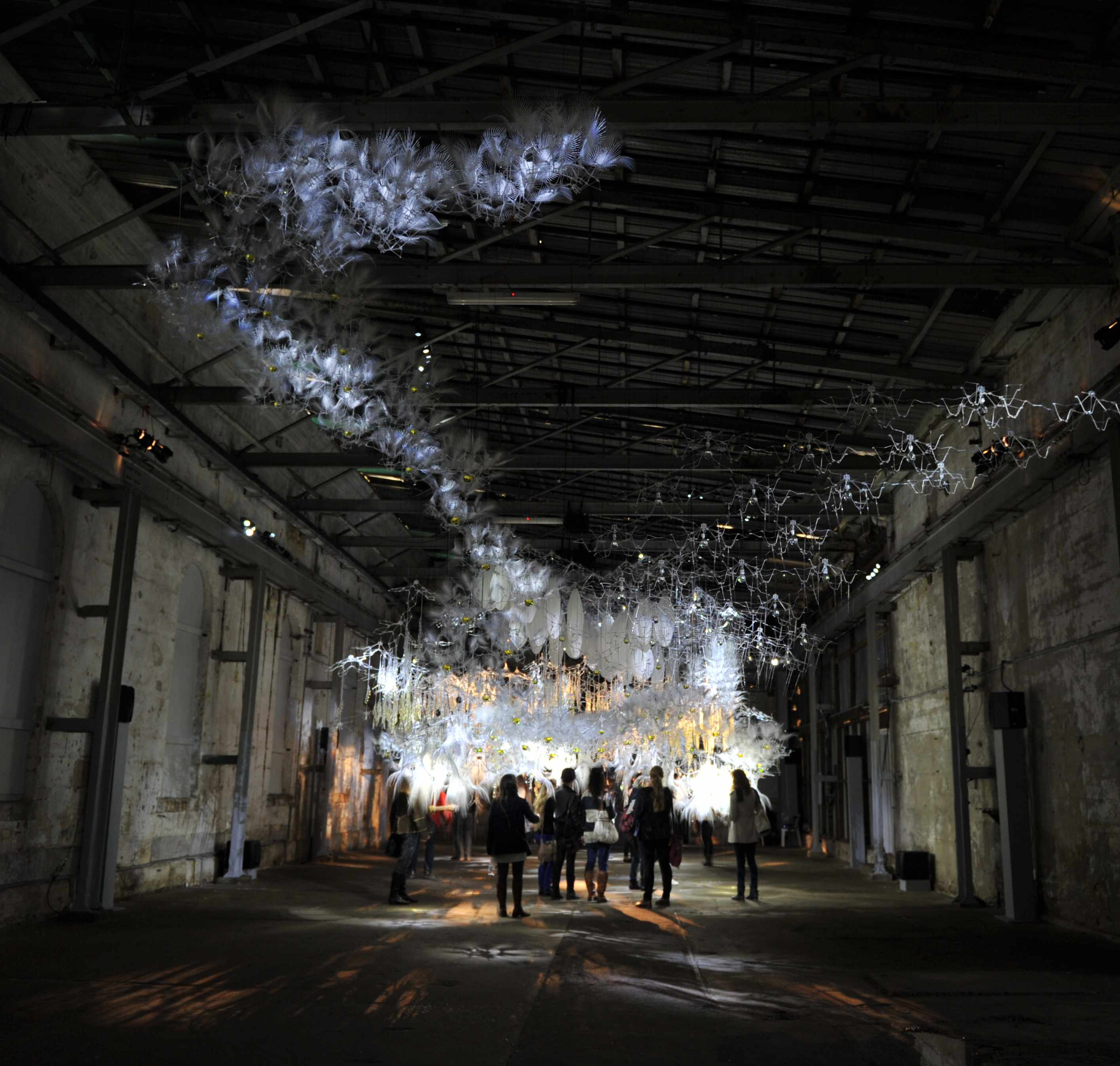}
            \caption{\emph{Hylozoic Series: Sibyl}, installed in the Industrial Precinct in Sydney, Australia, 2012.}
            \label{fig:Hylozoic_Seriesr_2012}
        \end{subfigure}%
        \caption{Living Architecture Systems. (Photos courtesy of Philip Beesley Architect Inc.)}
        \label{fig:Living_Architecture_Systems}
    \end{figure}

We study the challenge of long term autonomy with Living Architecture Systems (LAS).  LAS are interactive systems at architectural scale that emulate living environments aiming to engage occupants in long-term interaction (see Fig. \ref{fig:Living_Architecture_Systems} for sample installations by the Living Architecture Systems Group (LASG) and Philip Beesley Architect Inc. (PBAI)\footnote{More LAS environments by LASG/PBAI can be found at \url{http://www.philipbeesleyarchitect.com/sculptures/}}). An immersive LAS can have dozens of sensors and hundreds of actuators with various modalities to enable interaction with visitors, which poses a challenge when designing effective interaction control strategies. Historically LASG/PBAI environments have used pre-scripted, human-designed behaviours. In this work, we are interested in the potential for adaptive machine learning behavioural algorithms to generate interactive behaviours.  LAS aims for long-term continuous interaction, and this necessitates the capability of LAS to evolve with and to learn in a non-stationary environment, rather than using a single pre-scripted solution or assuming the existence of a single optimal solution.

\begin{sloppypar}
Although LAS aims for long-term engagement with occupants, the architectural scale and non-anthropomorphic, immersive nature of LAS makes it distinctive compared to robots more typically studied in HRI \cite{sheridan2016human, haddadin2016physical, kanda2016human, tsarouchi2016human, goodrich2008human, zlotowski2015anthropomorphism}. Since LAS is intended to create life-like behaviours at architectural scale, it facilitates accommodating multiple occupants and group interaction, rather than the one-to-one interaction most commonly studied in HRI. In addition, unlike HRI studies with humanoid robots that can be directly inspired by human-human interactions, LAS is a non-anthropomorphic robot, making the design of interactive behaviour less intuitive. While it is possible to manually design life-like interactive behaviours to some extent, it is very time-consuming, and the resulting behaviours are non-adaptive. Therefore, implementing interactive systems that can adapt and learn in dynamic crowd settings and can autonomously generate engaging behaviour is critically important, not only because it is useful in many applications, including public spaces, schools, workplaces, and family residences etc, but also because HRI in such interactive systems could extend our understanding of socially acceptable and engaging interactive behaviour.
\end{sloppypar}

Manually designed behaviour patterns can generate engaging life-like behaviours based on the designer's understanding \cite{beesley2010hylozoic, beesley2015evolving}, but this can be time-consuming and result in behaviours that become predictable during long term interaction. To overcome this limitation, machine learning may be used to automatically generate behaviour and study whether such behaviour is attractive or engaging for participants. For example, in \cite{chan2015curiosity,  chan2016interacting}, a Curiosity-Based Learning Algorithm (CBLA) was implemented in the LAS to automatically generate behaviour based on a computational notion of curiosity \cite{oudeyer2009intrinsic} of the LAS. However, the action produced by the CBLA is purely intrinsically motivated by the curiosity mechanism and does not consider any extrinsic motivation, e.g., maximizing measures of human response, which could play a more important role when an interactive system aims to engage participants. Another issue with \cite{chan2015curiosity,  chan2016interacting} and many works on social robots \cite{kanda2016human,breazeal2016social} is that the proposed approach is only designed and tested for one-to-one interaction in a controlled setting, rather than group interactions in natural settings, i.e., multiple people interacting with an interactive system simultaneously without instruction or guidance by the researchers.

Learning interactive behaviour for an extrinsically motivated LAS in natural settings with group interactions is a difficult challenge that is not typically considered in Reinforcement Learning (RL) tasks in controlled settings with only one-to-one interaction. Firstly, LAS is an interactive system at architectural scale, with very large state and action spaces, typically having dozens of sensors and hundreds of actuators. The learning challenge is exacerbated by a  complicated and non-stationary environment, where the LAS might be manipulated by occupants with different cultural backgrounds, interests and personalities. In addition, interactions happening in the real world happen less frequently than interactions in the simulators or video games typically studied in RL. This results in fewer interactions, which poses a huge challenge for data-driven learning algorithms. Last but not least, there is no standard measurement for estimating engagement, i.e., the extrinsic motivation of LAS. If a measurement of engagement is sparse and time-delayed, learning interactive behaviour becomes more difficult.

In this paper, we investigate extrinsically motivated learning in natural settings with group interaction. Specifically, we use sensors embedded in the interactive system to estimate the overall occupancy and engagement of the visitors in the LAS, and use this estimate as the extrinsic reward for learning. We investigate whether we can exploit the designer's pre-scripted behaviours to bootstrap learning. We use the Deep Deterministic Policy Gradient (DDPG) algorithm to train an Actor-Critic agent which acts in the designer-parameterized action space, which we call the Parameterized Learning Agent (PLA). Both fixed pre-scripted and learning behaviours are evaluated over three weeks at a public exhibition at the Royal Ontario Museum (ROM), and all data is collected from museum visitors. We evaluate how engaging the behaviours are based on a quantitative analysis of estimated engagement and the number of active interactions, and on qualitative analysis of human survey results.

\section{RELATED WORK}
\label{sec:RELATED_WORK}

\textbf{\textit{Adaptive Control}} \hspace{5pt} Traditional robotic systems operating in environments where the system dynamics are unknown or varying have used adaptive control \cite{aastrom2013adaptive}. In the domain of human-robot interaction, \cite{wang2016adaptive} applied adaptive control to robot navigation, where a social proxemics potential field is constructed and used to design a robot motion controller which is able to adapt its desired trajectory smoothly and at the same time comply with the proxemics contraints. Nakazawa et al. \cite{nakazawa2013unified} proposed a potential field imposing a repulsive fin to allow adaptive control of an accompanying robot, where the robot is able to adapt its relative position to the accompanied human in the presence of obstacles. Vitiello et al. \cite{vitiello2017neuro} proposed a Neuro-Fuzzy-Bayesian approach for adaptive control of a robot's proxemics behaviour, where recognized human activities and human personality acquired by questionnaires are input into an Adaptive Neuro Fuzzy Inference Engine (ANFIS) to determine a robot's stopping distance. Adaptive control has also been applied to rehabilitation robots \cite{zhang2015passivity}, robots driven by Series Elastic Actuators \cite{li2016adaptive}, and many other applications. However, adaptive control relies on a known structure of the dynamics model. In our case, it is difficult to apply adaptive control  because we do not have access to a good model of the environment dynamics, since the number of visitors is unpredictable and visitors' interaction style varies.

\textbf{\textit{Learning with Human as Explicit Teacher}} \hspace{5pt} Interactive Reinforcement Learning (IRL) studies RL algorithms in the context of HRI, where humans explicitly provide rewards or actions to guide an agent in RL.  Isbell et al. \cite{isbell2001social} applied RL in a complex human online social environment, where a human interacts with a learning agent by providing a reward signal, and highlighted that many of the standard assumptions, such as stationary rewards, Markovian behaviour, appropriateness of average reward, for RL are clearly violated. Thomaz \cite{thomaz2005real, thomaz2006reinforcement} proposed IRL for training human-centric assistive robots through natural interaction, where a human coach's feedback is used to shape the predefined reward. In addition, Thomaz \cite{thomaz2005real, thomaz2006reinforcement} introduced anticipatory cues to allow the human coach to predict the robot's action and provide timely feedback. In \cite{suay2011effect}, IRL was first studied in real-world robotic systems, and showed that the positive effects of human guidance increase with state space size. Other works converting human feedback to reward include \cite{knox2010combining, knox2012reinforcement}. Griffith et al. \cite{griffith2013policy} proposed an algorithm (Advise) to formalize the meaning of human feedback as policy feedback, which is more effective and robust than using human feedback as extra reward.  Krening et al. \cite{krening2018interaction} studied the effect of different teaching methods with the same underlying RL algorithm on human teachers' experience in terms of frustration. This work emphasizes that high transparency will decrease frustration and increase perceived intelligence. Different from most of these works where the human plays the direct role of teacher, i.e., directly providing either a reward or a policy advice in an interactive way, in this paper we aim to learn how to engage visitors without requiring them to consciously teach or train the interactive system. Therefore, in this paper the learning agent does not explicitly receive rewards from visitors, but uses a measure of engagement as reward, which is calculated based on observed visitor behaviours, without any constraints on the frequency and consistency of interaction on the visitor.

\textbf{\textit{Learning with Human as Implicit Teacher}} \hspace{5pt} RL algorithms have been deployed in social robotics to enable a robot to acquire socially acceptable skills, where a predefined reward function is employed to implicitly infer the reward signal from sensed human behaviour. In \cite{kumagai2018towards}, emotion recognition based on videos captured during participants' interaction with a robot is exploited as a reward within an RL framework to adapt the robot's behaviour towards participants' personal preferences. Leite et al. \cite{leite2011modelling} proposed to model the user's affective state by combining facial expression recognition and game state in a supervised way.  The predicted affective state is then used to calculate the reward in an RL framework to personalize response to users. Macharet et al. \cite{macharet2013learning} investigated the effectiveness of using RL to learn socially acceptable approaching behaviour for a mobile robot. Gordon et al. \cite{gordon2016affective} studied affective personalization in an integrated intelligent tutoring system, where in a one-to-one interaction setting, facial expression based engagement and valence estimation are used as reward in a standard SARSA algorithm. These studies all work on relatively small state and action spaces without exploiting deep neural networks. Recently, \textbf{Deep Reinforcement Learning (DRL)} has also been investigated in social robotics. Qureshi et al. \cite{qureshi2016robot} proposed Multimodal Deep Q-network (MDQN) for a handshake robot, where a deep neural network was employed to approximate the Q-value function based on both image and depth information. Chen et al. \cite{chen2017socially} studied Social Aware Collision Avoidance Deep Reinforcement Learning (SA-CADRL), where the reward function was designed to penalize actions leading to states where a robot is too close to nearby pedestrians and a deep neural network is used to approximate the Q-value function. Different from these works where either DRL is not exploited or the action space is discrete, our work applies continuous control DRL to an interactive social robot.

\textbf{\textit{One-to-one HRI In Natural Settings Without Learning}} \hspace{5pt} Interactions in natural settings, e.g., public spaces, are too complex to be simulated in controlled laboratory settings, so to understand natural HRI it is important to conduct field studies. Many HRI systems have been tested in public spaces such as hospitals \cite{mutlu2008robots}, train stations \cite{hayashi2007humanoid}, service points \cite{kaipainen2018nice,keizer2014machine}, airports \cite{joosse2017guide}, shopping malls \cite{shi2018robot}, hotels \cite{chung2018your} and museums \cite{shiomi2006interactive}, among others. Although robots investigated in these papers are sometimes surrounded by a group of people in a public place, these robots typically interact with only one person at a specific time, i.e., one-to-one interaction. Unlike these works, the larger scale of LAS enables LAS to simultaneously accommodate interactions with multiple people. In addition, the behaviours of robots studied in most of these works are pre-designed by researchers without continuous learning and adaptability.

\textbf{\textit{Interactive Artworks Without Learning}} \hspace{5pt} Interactive artworks are outcomes of combining arts and engineering, and have brought a new research direction for understanding HRI, especially with non-anthropomorphic robots, not only from the robotics  perspective but also from an artistic perspective. LAS in this paper and previous installations \cite{beesley2015evolving} are examples of such immersive interactive systems that promote roboticists' and architects' understanding of life-like interactive behaviour. Other examples include \cite{levillain2017interacting}, who investigated visitors' interaction and observation behaviours with mobile pianos in a museum. \cite{st2017control,st2010human} studied flying cubes in multiple publicly accessible spaces where flying cubes play the role of living creatures. In \cite{cuan2018time}, roboticists collaborated with a professional dancer to design interactive dancing robots. \cite{wakefield2013spatial} created spatial interactions with immersive virtual reality technology. \cite{augugliaro2013methods} studied dancing quadcopters, without an interactive component. Close collaboration between artists and roboticists has been fostering the creation of richer modes of interaction and extending the scope of studies in HRI. Most works mentioned here rely heavily on the design of choreographers, while in this paper we add adaptability on top of choreography by learning engaging behaviour based on the action space defined by choreographers.

\textbf{\textit{Engagement Estimation in HRI}} \hspace{5pt} As the intention of HRI is to engage human participants in interactions, detecting and measuring initial and ongoing engagement in HRI is critical for initiating and maintaining interaction \cite{doherty2018engagement,glas2015definitions,o2008user}. Many approaches for measuring engagement rely on rich sensors, such as cameras, lasers, sonar and audio sensors, and on sophisticated facial expression, gesture, gaze, body movements, physiological signals and speech recognition technologies \cite{kulic2007affective,ROB-049,anzalone2015evaluating}. Sanghvi et al.\cite{sanghvi2011automatic} proposed to estimate engagement by extracting features of body motion from video by training an engagement prediction model in a supervised way, but the proposed method requires a lateral camera and users to wear a specific coloured coat. Keizer et al. \cite{keizer2014machine} trained an engagement classifier based on multimodal corpus in a supervised way, which classifies engagement into three levels i.e. \textit{NotSeekingEngagement}, \textit{SeekingEngagement} and \textit{Engaged}.  Michalowski et al. \cite{michalowski2006spatial} studied a spatial model of engagement for a social robot and tested a robotic receptionist to engage visitors, and suggested that direction and speed of motion are more appropriate measures of engagement than location or distance alone. However, this work used distance to determine the engagement level and what pre-scripted behaviour to take, rather than using the distance as a learning signal. Bohus et al. \cite{bohus2009models,bohus2009open} studied the engagement model for a multiparty open-world dialog system, but the system can only actively engage in at most one interaction even though it could simultaneously keep track of additional, suspended interactions. Engagement in \cite{bohus2009models,bohus2009open} is viewed as several states. Behaviours corresponding to each engagement state are pre-defined. Even though the authors claim the system is a multiparty interaction, the number of participants is limited because of the field of view of the camera and the interaction screen. Khamassi et al. \cite{khamassi2018robot} and Velentzas et al. \cite{velentzas2018adaptive} proposed to estimate engagement from human gaze and use it as a reward function in an RL framework to allow fast adaption to environment change, where experiments are conducted in a well controlled setting. Most works studying engagement in HRI focus on humanoid social robots, one-to-one interactions, and discrete measures of engagement.  In this paper we study a non-anthropomorphic robotic environment, group interaction, and measure engagement as a continuous value. Most importantly, we use a measure of engagement as the extrinsic reward to drive learning, rather than using the measure of engagement as a contingent event to trigger a predefined behaviour. In addition, since our environment has highly varying lighting conditions and the potential for severe occlusions, camera based methods may be less reliable.

Interactive systems, whether humanoid robots or non-anthropomorphic robots like LAS, are human oriented and aim for continuous engagement. These characteristics intrinsically require agents to be able to autonomously generate behaviours and continuously adapt to their human partner(s). In this paper, we exploit the standard reinforcement learning (RL) framework and a novel reward function estimating engagement to realize autonomous agents, and study if this reward could generate behaviour that is more engaging than pre-scripted behaviour.  All experiments are conducted in a public museum exhibit with museum visitors.

\section{The Interactive Architecture Testbed}
\label{sec:The_Interactive_Architecture_Testbed}

In this section, we describe the physical system used as the testbed in this paper, and the design of \emph{Pre-scripted Behaviours (PBs)} that drive the interactive behaviour of the system.  The PBs, which are designed by expert architects and interactive system designers, are the baseline we use to compare to the learning system described in Section \ref{sec:Proposed_Approaches} below.

\begin{figure}[thpb]
        \begin{minipage}{0.48\textwidth}
            \centering
            \includegraphics[height=0.48\textwidth]{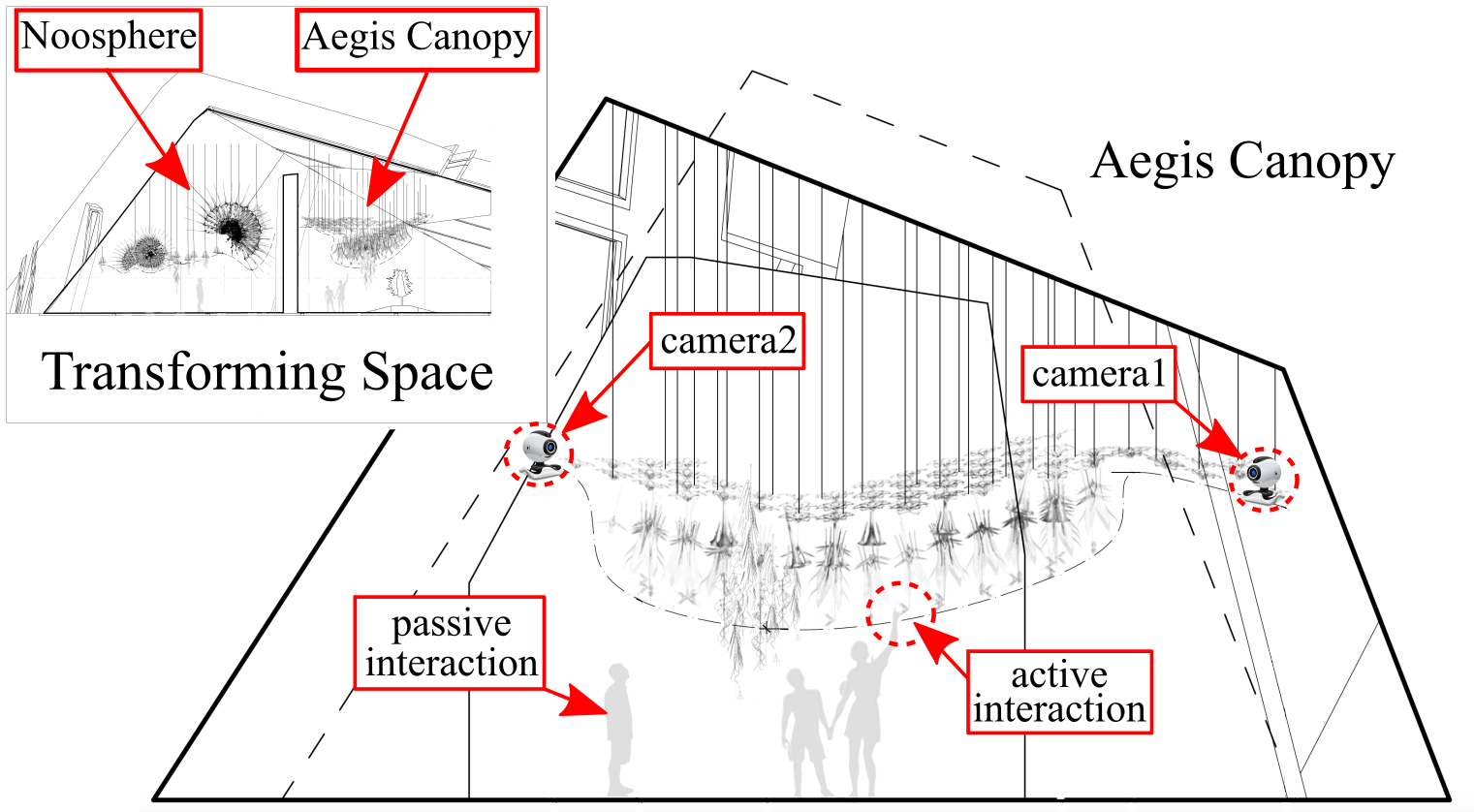}
            \caption{Installation Diagram and Interaction Types}
            \label{fig:Installation_Diagram_and_Interaction_Types}
        \end{minipage}
        \begin{minipage}{0.48\textwidth}
            \centering
            \begin{subfigure}[t]{.5\linewidth}
                 \centering
                 \includegraphics[height=1\textwidth]{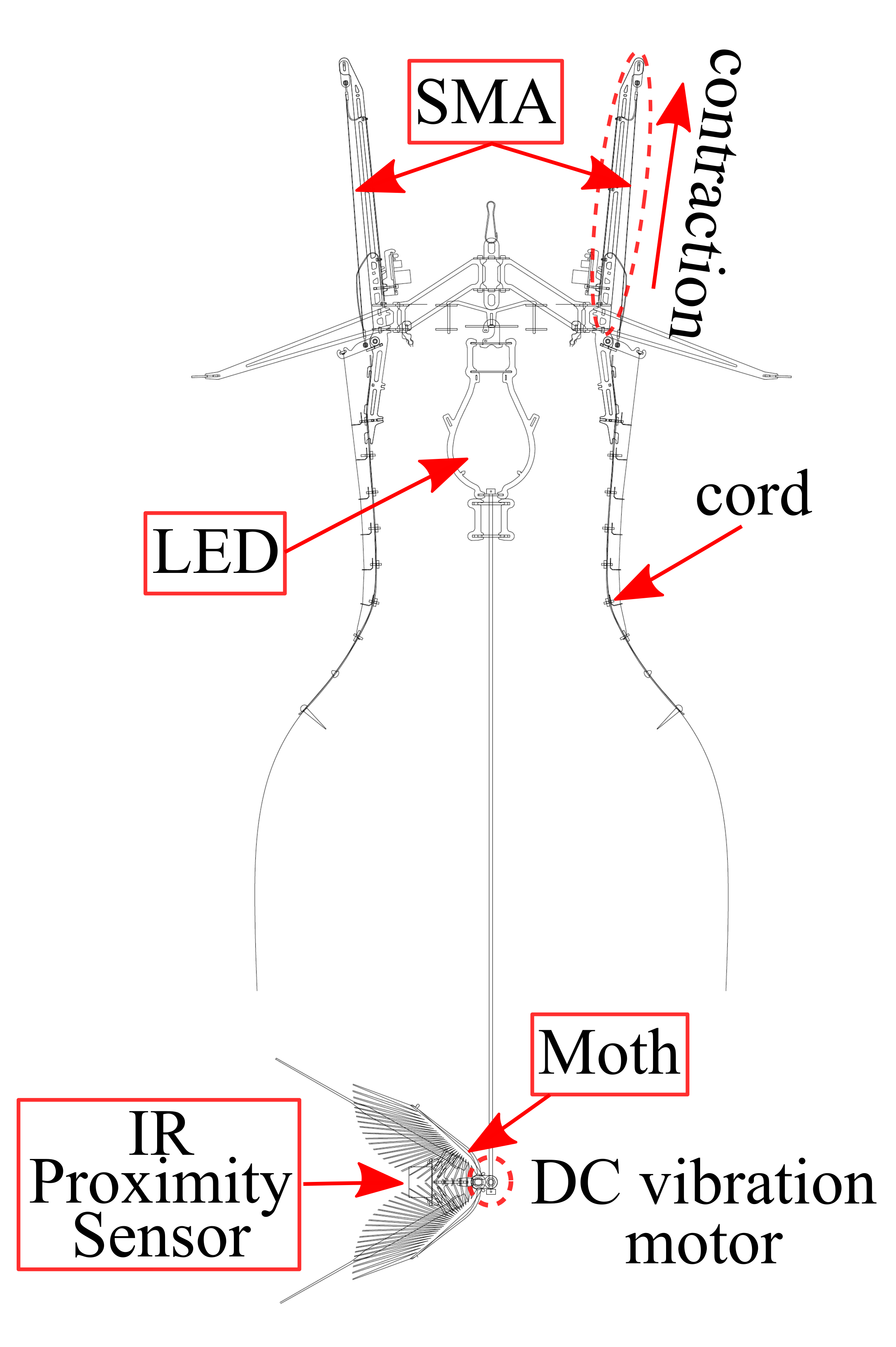}
                 \caption{Node Diagram}
                 \label{fig:Node_Diagram}
            \end{subfigure}%
            \begin{subfigure}[t]{.5\linewidth}
                \centering
                \includegraphics[height=1\textwidth]{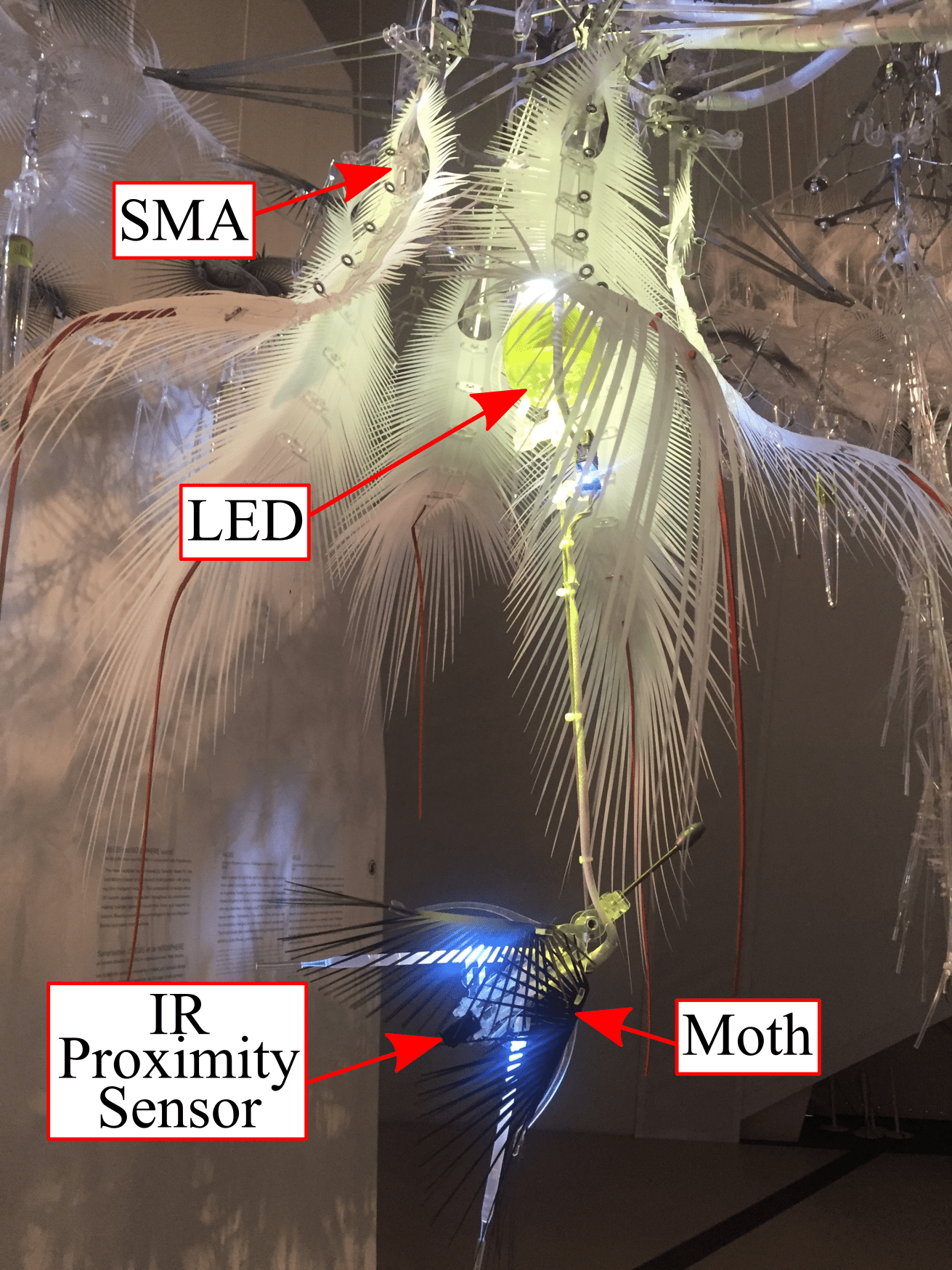}
                \caption{Fully Assembled Node}
                \label{fig:Fully_Assembled_Node}
            \end{subfigure}
            \caption{Diagram of Node in the LAS}
            \label{fig:Diagram_of_Node_in_the_LAS}
        \end{minipage}
    \end{figure}

Our testbed {\em Aegis Canopy} is part of the exhibition Transforming Space which consists of two sculptures, i.e. the Aegis Canopy and Noosphere, as shown in the top-left sub-figure in Fig. \ref{fig:Installation_Diagram_and_Interaction_Types}. The exhibition was exhibited at the Royal Ontario Museum (ROM) in Toronto, Canada from June 2 to October 8, 2018 (\url{https://www.rom.on.ca/en/philip}) and was publicly accessible to over 60,000 visitors who had toured the exhibit. 

Since this research mainly used the Aegis Canopy part of the installation, we will describe the design of the Aegis Canopy in detail, and subsequently refer to the Aegis Canopy part of the installation as the Living Architecture System (LAS).
    
\subsection{Physical Living Architecture System}
\label{subsec:Physical_Living_Architecture_System}

The LAS hangs overhead within the Aegis Canopy space, with an approximate height of 1.8 meters. Fig. \ref{fig:Installation_Diagram_and_Interaction_Types} shows the front view of the LAS. The active part of the system is composed of eight speakers and 24 nodes. 

Each node in the LAS consists of six Fronds, one Moth, and one high-power LED as its actuated systems and one infrared (IR) sensor, as shown in Fig. \ref{fig:Diagram_of_Node_in_the_LAS}. Each Frond includes a shape memory alloy (SMA) wire which contracts when voltage is applied, pulling a cord attached to a flexible co-polyester sheet, as illustrated in Fig. \ref{fig:Node_Diagram}. The contraction generates a smooth and gentle movement, and when the applied voltage is removed the SMA slowly relaxes to its original shape.\footnote{A video illustrating the SMA assembly contracting and relaxing can be found in \url{https://youtu.be/YcreirXrRF4}.}  The Moth consists of wing-like flexible flaps attached to a small DC vibration motor that vibrates when activated, making the moth appear to be flapping its wings.  The Moth also houses two small LED lights which illuminate during vibration. The single high-power LED located beneath the central flask can be faded to illuminate the coloured liquid in the flask. The IR sensor senses the proximity of visitors, and generates a continuous reading proportional to the distance between any part of the body of a visitor and the sensor location in the Aegis Canopy.

There are eight speakers distributed throughout the LAS. These speakers play two types of sound samples. The first sound is a background sound played on a continuous loop. The second sound is triggered by the IR sensors. These speakers are independently controlled by specialized software, so here we treat them as background behaviours. 

\begin{figure}[thpb]
        \begin{minipage}{0.48\textwidth}
            \centering
            \includegraphics[height=0.6\linewidth]{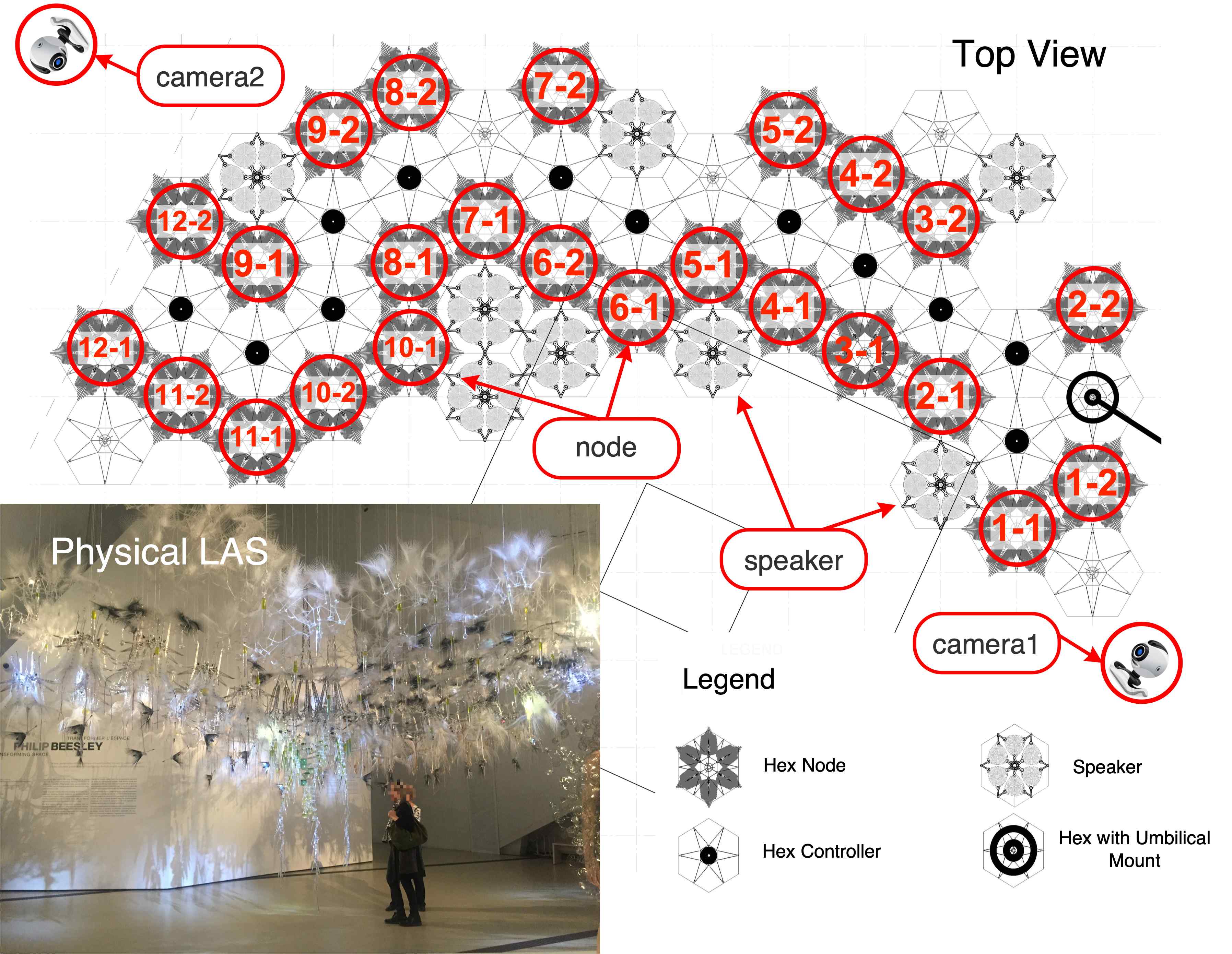}
            \caption{LAS: Aegis Canopy}
            \label{fig:Canopy}
        \end{minipage}
        \begin{minipage}{0.48\textwidth}
            \centering
            \includegraphics[height=0.6\linewidth]{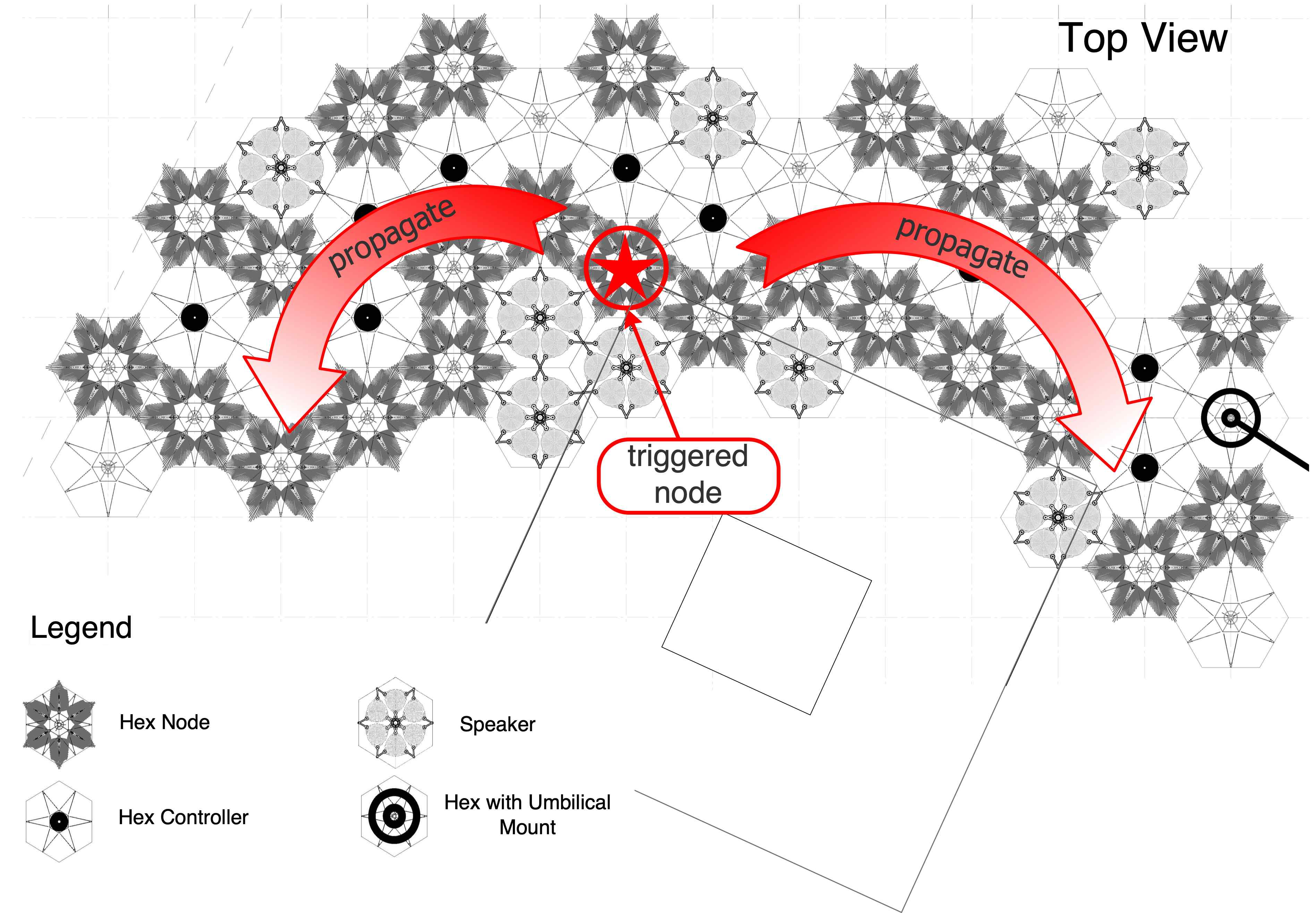}
            \caption{Pre-scripted Behaviour}
            \label{fig:Prescripted_Behaviour}
        \end{minipage}
    \end{figure}

The arrangement of the speakers and nodes is illustrated in Fig. \ref{fig:Canopy}, where the 24 nodes are highlighted by red circles. A photo of the physical LAS is shown in the bottom-left of Fig. \ref{fig:Canopy}. The 24 nodes are at varying height levels. Specifically, nodes at the left and right edges are slightly higher than those in the middle of the LAS. This spatial arrangement distinguishes three types of visitor engagement with the system. When visitors observe the LAS but are not underneath the LAS, no IR sensor is activated, i.e., visitors are observing the LAS but cannot be observed by the LAS sensors.  As shown in Fig. \ref{fig:Installation_Diagram_and_Interaction_Types}, when visitors walk or stand underneath the LAS, which we name \emph{Passive Interaction}, the IR sensors above them are activated, but the distance between the visitor and the system is still large, corresponding to a small reading of the IR sensor. Visitors engaging in \emph{Active Interaction} might also reach their hand upwards to interact with the LAS,  resulting in a higher activation value of the closest IR sensor.

Two web cameras (labeled Camera1 and Camera2 in Fig. \ref{fig:Installation_Diagram_and_Interaction_Types} and Fig. \ref{fig:Canopy}) are used to record video during our experiment and to calibrate sensory data. These two web cameras are mounted on the wall in the front-right and back-left corners of the LAS space.

\subsection{Pre-scripted Behaviour}
\label{subsec:Prescripted_Behaviour}

\emph{Pre-scripted behaviour (PB)} is the interactive behaviour manually designed by the architects, and it is also the baseline used for comparison with adaptive behaviours we will describe in Section \ref{sec:Proposed_Approaches}. Within the PB mode, the system can be in two types of states:  active  and  background, in which behaviours are mainly controlled by 17 parameters (shown in Table \ref{table:Related_Parameters_in_Prescripted_Behaviour}) specified by the architects. The values in the Default and Range columns are used in the PB and PLA modes, respectively.

The active state is entered if any of the IR sensors is triggered. In this state, the node corresponding to the triggered sensor will first activate its {\em local reflex behaviour}. In the local reflex behaviour, the Moth, the LED and six SMAs attached to the same node as the triggered IR sensor will be activated. When a Moth is activated, it will ramp up the vibration to its maximum $I_{max}$ over time $T_{ru}^m$, hold there for a period of time $T_{ho}^m$, then ramp down over ($T_{rd}^m$). After a waiting period ($T_{gap}^m$) following the sensor trigger, the LED on the same node is activated. It ramps up over time period ($T_{ru}^l$) to its maximum brightness ($I_{max}$), holds for a period of time  ($T_{ho}^l$) and then gradually dims over ($T_{rd}^l$). At the same time, the SMAs are activated one after another separated by ($T_{gap}^{sma}$). A step voltage is applied to contract the SMA, after which a cooling-down time is started during which this SMA will not be activated again. The activation profile of the SMA wires is fixed in order to protect them from overheating, so these are not included in the parameterization shown in Table \ref{table:Related_Parameters_in_Prescripted_Behaviour}. After the local reflex behaviour is triggered, the IR-detected event will be propagated from the triggered node to neighbouring nodes after a delay ($T_{gap}^n$), until the edge nodes of the LAS are reached (shown in Fig. \ref{fig:Prescripted_Behaviour}), causing a cascade of local reflex behaviours at each node.

If no IR sensor triggering happens for a random time within $\left [ T_{bg}^{min}, T_{bg}^{max}\right ]$, the system goes into the background state. In this state, the LEDs and moths will randomly activate their local reflex behaviours with probability $P$ every amount of time $T_w$. The SMAs are also activated independently with the same probability $P$ every $T_{sma}$. 

In either state, a sweep of LEDs in either direction along the longer axis of the installation happens at random time intervals within $\left [ T_{sw}^{min},T_{sw}^{max} \right ]$. During the sweep, each LED activates local reflex behaviour and propagates in the direction of the sweep.

    \begin{table}[h]
        \caption{Pre-scripted Behaviour Parameters}
        \label{table:Related_Parameters_in_Prescripted_Behaviour}
        \begin{center}
        \begin{tabular}{c | c | c | c}
        \hline\hline
        \textbf{\makecell{Parameters}}  & \textbf{Meaning} & \textbf{Default} & \textbf{Range} \\
        \hline
        \makecell{$T_{ru}^m$, $T_{ru}^l$} & \makecell{ramp up time: the time it takes for the Moths or \\LEDs to increase up to their maximum value} & 1.5 &$\left[0,5\right]$ \\ 
        \hline
        \makecell{$T_{ho}^m$, $T_{ho}^l$} & \makecell{hold time: the time that Moths and \\LEDs are held at their maximum value} &1 &$\left[0,5\right]$ \\ 
        \hline
        \makecell{$T_{rd}^m$, $T_{rd}^l$ } & \makecell{ramp down time: the time it takes for Moths and \\LEDs to fade down to 0} &2.5 &$\left[0,5\right]$ \\ 
        \hline
        \makecell{$I_{max}$} & \makecell{maximum percentage of duty cycle per PWM period} & 78 &$\left[0,100\right]$ \\ 
        \hline
        $T_{gap}^m$          & \makecell{the time gap between the Moth starting to ramp up \\and the LED starting to ramp up} &1.5 &$\left[0,5\right]$ \\ 
        \hline
        $T_{gap}^{sma}$      & \makecell{the time gap between activation of each SMA arm\\on the nodes} &0.3 &$\left[0,5\right]$ \\ 
        \hline
        $T_{gap}^{n}$      & \makecell{the time gap between activation of each node during \\neighbour behaviour} & 1.8 &$\left[0,5\right]$ \\ 
        \hline
        \makecell{$T_{bg}^{min}$} & \makecell{minimum time to wait before activating\\background behaviour} &45 &$\left[15,60\right]$ \\ 
        \hline
        \makecell{$T_{bg}^{max}$} & \makecell{maximum time to wait before activating\\background behaviour} &90 &$\left[60,100\right]$ \\ 
        \hline
        \makecell{$T_{w}$} & \makecell{time to wait before trying to pick a moth or LED} &5 &$\left[0,10\right]$ \\ 
        \hline
        \makecell{$P$} & \makecell{probability of successfully choosing an actuator during \\background behaviour} &0.4 &$\left[0,1\right]$ \\ 
        \hline
        $T_{sma}$              & \makecell{time between choosing SMAs to actuate} &0.7 &$\left[1,5\right]$ \\ \hline
        \makecell{$T_{sw}^{min}$} & \makecell{minimum time to wait before performing sweep} &120 &$\left[5,200\right]$ \\ \hline
        \makecell{$T_{sw}^{max}$} & \makecell{maximum time to wait before performing sweep} &240 &$\left[200,400\right]$ \\ \hline
        \multicolumn{2}{l}{\tiny The unit of all time parameters is seconds, except $I_{max}$ is percentage} and $P$ is probability. \\
        \end{tabular}
        \end{center}
    \end{table}

\section{PROPOSED APPROACH}
\label{sec:Proposed_Approaches}

In this section, we describe how the \emph{Parameterized Learning Agent (PLA)} is designed to automatically generate interactive actions. 

\begin{figure}[!htb]
        \centering
        \begin{minipage}{.48\textwidth}
            \centering
            \includegraphics[width=.8\linewidth]{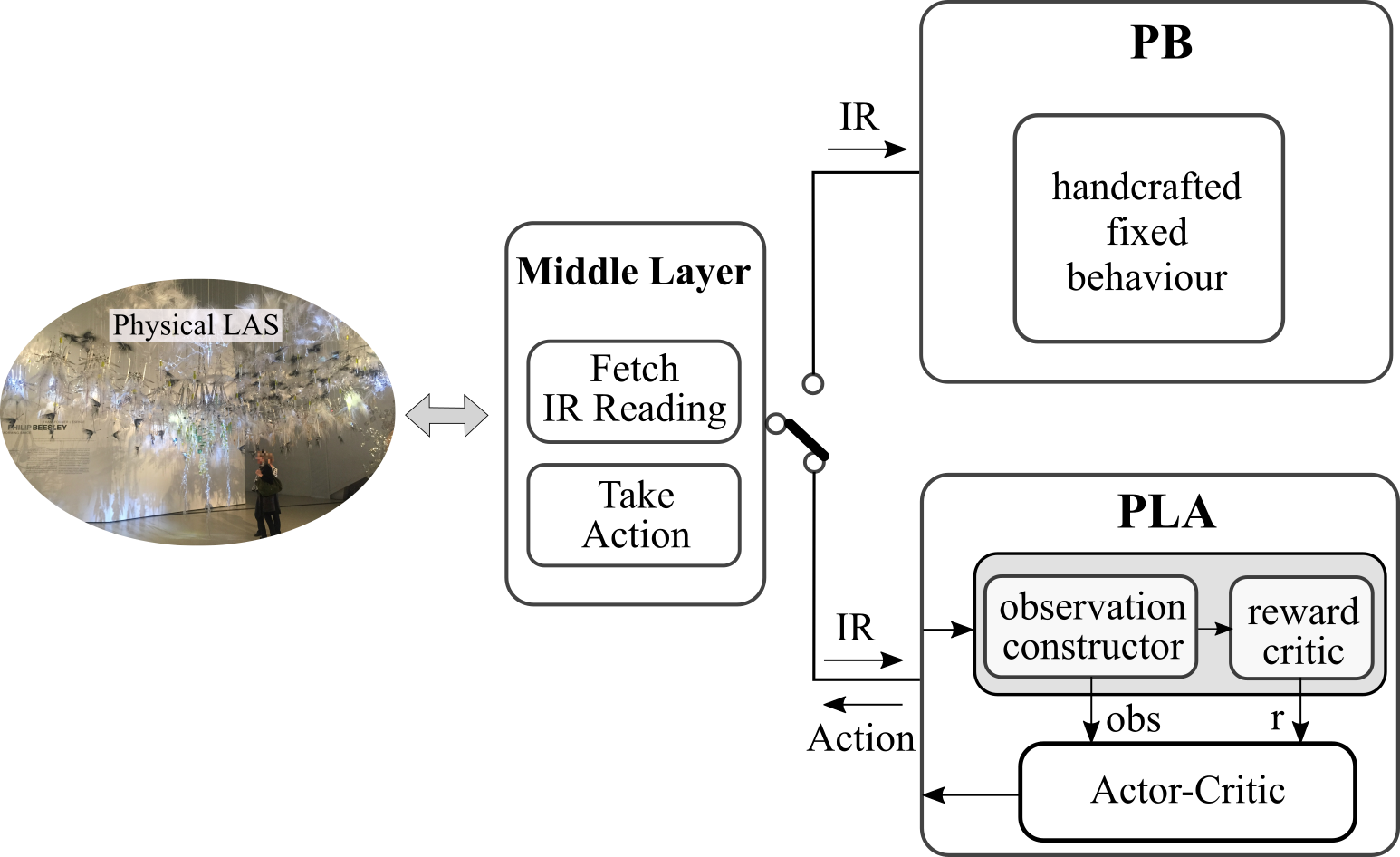}
            \caption{Interface Between LAS and PLA}
            \label{fig:Interface_Between_LAS_and_PLA}
        \end{minipage}%
        \hspace{2pt}
        \begin{minipage}{0.48\textwidth}
            \centering
            \begin{subfigure}[t]{.46\linewidth}
              \centering
              \includegraphics[width=1\linewidth]{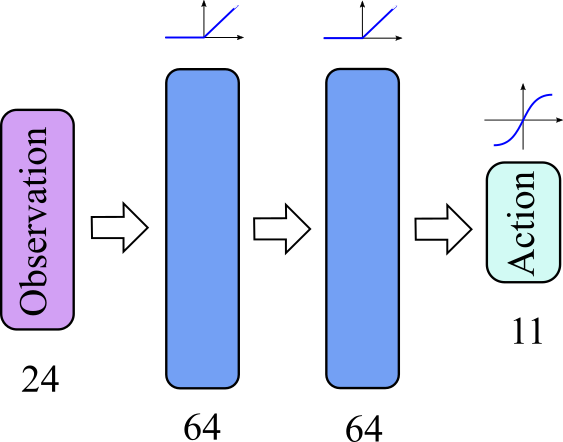}
              \caption{Actor}
              \label{fig:actor_PLA}
            \end{subfigure}%
            \hspace{8pt}%
            \begin{subfigure}[t]{.46\linewidth}
              \centering
              \includegraphics[width=1\linewidth]{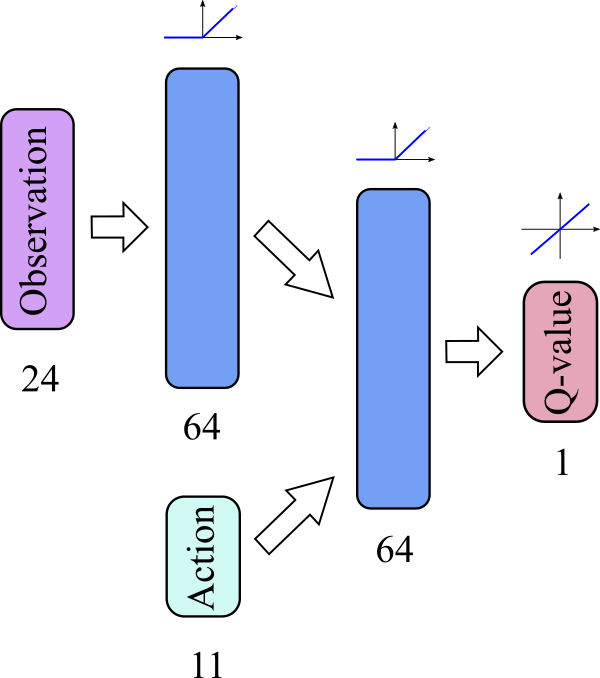}
              \caption{Critic}
              \label{fig:critic_PLA}
            \end{subfigure}
            \caption{Actor-Critic of PLA, where layer size and activation function are indicated under and above each layer respectively.}
            \label{fig:Actor_Critic_of_PLA}
        \end{minipage}
    \end{figure}

Fig. \ref{fig:Interface_Between_LAS_and_PLA} illustrates the software architecture connecting the physical hardware to the learning agents. The Middle Layer coordinates the sensor reading and action execution among the physical LAS and agents.  Only one agent type controls the LAS at any given time. PB only takes and stores IR readings without issuing actions because its action policy is fixed, while PLA is formulated as a standard RL framework where the Observation Constructor constructs the observation, the Reward Critic generates a reward that estimates the engagement of visitors, and Actor-Critic is the learned policy and Q-value function.
    
\subsection{Parameterized Learning Agent: Learning on Top of Pre-scripted Behaviour}
\label{subsec:Learning_on_Top_of_Prescripted_Behaviour}

PLA is designed to learn on top of PB, i.e., parameterized action space. The motivation for this approach is to bootstrap learning by exploiting the designer's knowledge of engaging behaviour, where we hypothesize the designer already has a good idea about what types of actions might be engaging to visitors and this can form a helpful starting point for the learner.

\subsubsection{Observation and Action Space Construction}

For PLA, we select 11 parameters from Table \ref{table:Related_Parameters_in_Prescripted_Behaviour} as the action space, i.e., the dimension of the action vector is 11. This is because some parameters do not take effect until a subsequent trigger or until the current propagation finishes. This could lead to obtaining an observation which is based on both previous and updated parameters. To avoid this issue, we exclude $T_{bg}^{min}$, $T_{bg}^{max}$, $T_{w}$, $P$, $T_{sw}^{min}$ and $T_{sw}^{max}$ from the action space. In this way, we make sure every observation is only related to the latest action. To attenuate IR sensor noise, the observation for PLA is an average over 20 IR readings as defined in Eq. \ref{eq:PLA_observation}: 

    \begin{equation}
        \bm{obs}^{\left(t\right)} =\frac{1}{20}\cdot \sum_{i=0}^{19} {\bm{ir}}^{(t-i\cdot\Delta t)}
        \label{eq:PLA_observation}
    \end{equation}
where ${\bm{ir}}^{(t-i\cdot\Delta t)}$ is the value array of 24 IR sensor readings at time $\left(t-i\cdot\Delta t\right)$, $\sum$ is element-wise summation, and $\Delta t\approx 0.1s$ is the time to retrieve one set of 24 IR values from the physical LAS. Thus the dimension of the observation vector is 24 and each observation vector represents the average IR readings over 2 seconds. As described in Section \ref{subsec:Data_Preprocessing}, IR values are converted and scaled so that 0 corresponds to the maximum distance reading (i.e., there is nothing in front of the sensor), and 1 corresponds to the minimum distance reading (i.e., there is something very close to the sensor).

\subsubsection{Estimating and Using Engagement as a Reward for Learning}
\label{sebsec:Estimating_and_Using_Engagement_as_a_Reward_for_Learning}

A key feature of our approach is the formulation of the reward function: we wish to learn and reward behaviours which foster visitor engagement.  Specifically, the extrinsic reward is computed by summing over the IR observations, which can be regarded as a rough estimate of occupancy and engagement, because: 1) more activated IRs means more people are standing under the LAS, thus indicating higher occupancy; 2) closer distance between visitors and IR sensors implies more active interaction, e.g., looking very closely or raising hands, which are higher engagement behaviours. Therefore, higher occupancy and more active interaction will cause higher extrinsic reward. Formally, given a new observation at time $t+1$, $\bm{obs}^{(t+1)}=(obs^{(t+1)}_{1}, obs^{(t+1)}_{2}, \cdots , obs^{(t+1)}_{n})$\footnote{In this paper, we will use normal lowercase for scalar and bold lowercase for vector.}, the reward $r^{(t)} $ for taking action $\bm{a}^{(t)}$ while observing $\bm{obs}^{(t)}$ can be expressed as Eq. \ref{eq:reward_signal}:
    \begin{equation}
        r^{(t)}= \sum_{i=1}^{n}obs^{(t+1)}_{i},
        \label{eq:reward_signal}
    \end{equation}
where the value of $n$ is the number of dimensions of the observation vector, e.g. $n$ is 24 for PLA.

Low cost embedded IR sensors are employed to estimate engagement for three reasons: First, the varying lighting conditions in the exhibition area do not allow us to use the webcam footage for reliable pose or facial expression analysis. Because of occlusions and the uncontrolled nature of the space and the varying number of occupants, an array of inexpensive IR sensors that can be precisely located within the space is more effective than lower numbers of more sophisticated equipment. Finally, due to privacy concerns we could not guarantee that we would be permitted to use camera data.

\subsubsection{Implementation}

Given the observation and extrinsic reward, the optimal policy can be learned by an RL algorithm\footnote{However, note that the interaction frequency is limited by the physical constraints of the LAS.}. In this paper, learning is implemented using the Deep Deterministic Policy Gradient (DDPG) algorithm~\cite{lillicrap2015continuous}, a variant of Deterministic Policy Gradient \cite{silver2014deterministic}, where both the Actor and Critic are approximated with deep neural networks. We chose DDPG because: 1) the action space is continuous, and compared with benchmarks reported in the literature \cite{duan2016benchmarking} \cite{juliani2018unity} the dimension of the action space is relatively large, which means that even if we discretize the continuous action, the dimension of the discretized action space will be very large; 2) compared with other RL algorithms for continuous action space, DDPG is easier to implement.

PLA uses the DDPG agent from OpenAI's Baselines package~\cite{baselines}, following the pseudo-code shown in Alg. \ref{alg:DDPG_based_PLA} and the hyper-parameters specified in Table \ref{table:Hyper_parameters_of_DDPG_based_PLA}. The structure of the neural network is shown in Fig. \ref{fig:Actor_Critic_of_PLA}. All layers are dense layers, with layer-norm applied, where the activation function and neurons for each layer are indicated above and under each layer in Fig. \ref{fig:Actor_Critic_of_PLA}, respectively. At the start of the field study, the actor-critic networks of PLA were randomly initialized.  On all subsequent deployment days, the learned actor-critic from the previous day was loaded and training continued from the previous day's network parameters. The system does not have a terminal state, so we manually set the maximum length of an episode to 100 steps, which corresponds to about 200s, and after each episode the start observation is the observation encountered at the end of the last episode. The actor-critic is trained 20 times on a randomly sampled mini-batch every 10 interactions, corresponding to approximately 20s. At every step, the exploratory action is generated by the current actor perturbed with parameter noise, i.e. Parameter Space Noise \cite{plappert2017parameter}. For more detail, please refer to Alg. \ref{alg:DDPG_based_PLA}.

\begin{algorithm}[htb!]
    \SetAlgoLined
    \SetNoFillComment
    \DontPrintSemicolon
        \KwIn{$first\_day$}
        Initialize: $train\_interval$, $train\_times$, $episode\_length$, replay buffer $\mathcal{D}$, mini-batch size $N$\;
        \eIf{$first\_day$}{
            Initialize Critic $Q(obs, a|\theta^Q)$ and Actor $\mu(obs|\theta^{\mu})$ with random parameters $\theta^Q$, $\theta^{\mu}$\;
        }{
            Load pretrained Critic $Q(obs, a|\theta^Q)$ and Actor $\mu(obs|\theta^{\mu})$\;
        }
        Target networks $\theta^{Q-}\leftarrow \theta^Q$, $\theta^{\mu-}\leftarrow \theta^{\mu}$\;
        $start\_of\_day \gets True$\;
        \While{True}{
            \eIf{$start\_of\_day$}{
                Receive initial observation $obs^{(1)}$\;
                Initialize parameter noise $\sigma$\;
                $start\_of\_day \gets False$\; 
            }{
                $obs^{(1)}\gets obs^{(t+1)}$, $\sigma \leftarrow \sigma_{j+1}$\;
            }
            
            \For{$t\gets1$ \KwTo $episode\_length$}{
                Select action $a^{(t)}=\mu(obs^{(t)}|\theta^{\mu}+\mathcal{N}\left ( 0, \sigma \right ))$ according to the current policy and exploration noise.\;
                Execute action $a^{(t)}$,  observe reward $r^{(t)}$ and observe new observation $obs^{(t+1)}$\;
                $\mathcal{D} \gets \mathcal{D}\cup(obs^{(t)}, a^{(t)}, r^{(t)}, obs^{(t+1)})$ \;
                \If{$t\%train\_interval==0$ and $\left | \mathcal{D} \right | \geq N$}{
                    \For{$j\gets1$ \KwTo $train\_times$}{
                        Sample a random minibatch of $N$ trainsitions $(obs^{(i)}, a^{(i)}, r^{(i)}, obs^{(i+1)})$ from $\mathcal{D}$\;
                        \tcc{Update adaptive parameter noise scale}
                        
                        \hspace{1cm} $\sigma_{j+1} \leftarrow   
                            \left\{\begin{matrix}
                                \alpha \sigma_{j} &  if \; d\left ( \mu_{\theta^{^{\mu}}} , \mu_{\theta^{^{\mu}}+\mathcal{N}\left ( 0, \sigma_j \right )}  \right ) \leq \delta \\ 
                                \frac{1}{\alpha } \sigma_{j} & otherwise.
                            \end{matrix}\right.$ \;
                        where $d\left ( \mu_{\theta^{^{\mu}}} , \mu_{\theta^{^{\mu}}+\mathcal{N}\left ( 0, \sigma_j \right )}  \right )=\sqrt{\frac{1}{N}\sum_{i=1}^{N}\mathbb{E}\left [ \left ( \mu\left ( s|{\theta^{^{\mu}}} \right ) - \mu \left ( s|\theta^{^{\mu}}+\mathcal{N}\left ( 0, \sigma_j \right )\right ) \right )^{2} \right ]}.$\;
                        \tcc{Update actor-critic}
                        Set $y^{(i)} = r^{(i)} + \gamma Q'(obs^{(i+1)}, \mu(obs^{(i+1)}|\theta^{\mu-})|\theta^{Q-})$\;
                        Update critic by minimizing the loss: $L=\frac{1}{N}\sum_{i} \left ( y^{(i)}- Q\left ( s^{(i)}, a^{(i)}|\theta^{Q} \right )\right )^{2}$\;
                        Update actor using the sampled policy gradient:\;
                        \hspace{1cm} $\nabla_{\theta ^{\mu }}J\approx \frac{1}{N}\sum _{i}\nabla_{\mu\left ( obs^{(i)}|\theta ^{\mu } \right )}Q\left ( obs^{(i)},\mu\left ( obs^{(i)}|\theta ^{\mu } \right ) | \theta ^{Q}\right )\nabla_{\theta ^{\mu }}\mu\left ( obs^{(i)}|\theta ^{\mu } \right )$\;
                        Update the target networks: \;
                        \hspace{1cm} $\theta^{Q-} \leftarrow \tau \theta ^{Q}+\left ( 1-\tau  \right )\theta^{Q-}$\;
                        \hspace{1cm} $\theta^{\mu -} \leftarrow \tau \theta ^{\mu}+\left ( 1-\tau  \right )\theta^{\mu -}$\;
                    }
                }
                $obs^{(t)}\gets obs^{(t+1)}$, $\sigma_{1} \leftarrow \sigma_{j+1}$\;
            }
            
        } 
    \caption{DDPG-based PLA}
    \label{alg:DDPG_based_PLA}
\end{algorithm}

All hyper-parameters used in this paper were empirically chosen via experiments on a simplified simulator where the agent directly controls all actuators and maximizes the reward based on IR readings, and visitors are simulated to approach LEDs with the highest intensity. Our tests on the simplified simulator showed that DDPG worked well with the chosen hyper-parameters and was able to smoothly converge to the optimal policy in 100 episodes where the maximum length of each episode is 1000. Because the simulator is an extremely simplified model of the actual environment, we mainly used it to make sure our code is bug free rather than to pre-train a policy. 

\begin{table}[htb!]
    \centering
    \caption{Hyper-parameters of DDPG-based PLA}
    \label{table:Hyper_parameters_of_DDPG_based_PLA}
    \begin{tabular}{ r | c }
        \hline\hline
        Hyper-parameters     &  value \\\hline
        actor learning rate  &  $10^{-4}$ \\
        critic learning rate &  $10^{-3}$ \\
        discount rate $gamma$ & 0.99  \\
        batch size $N$        & 64    \\
        replay buffer size $\left | \mathcal{D} \right |$    & $10^6$\\
        $train\_interval$     &  10   \\
        $train\_times$        &  20   \\
        $episode\_length$     &  100  \\
        parameter noise $\alpha$ &  1.01 \\
        parameter noise $\delta$ &  0.1  \\
        initial parameter noise $\sigma$ &  0.1  \\\hline
    \end{tabular}
\end{table}

\subsection{Learning in Raw Action Space}
\label{subsec:Learning_in_Raw_Action_Space}

During the field study, we also tested two additional learning systems  designed to act in raw action space, i.e. directly control actuators, rather than acting in parameterized action space. The \emph{Single Agent Raw Act (SARA)}  directly controls the 192 raw actuators of the LAS using a single agent, while the \emph{Agent Community Raw Act (ACRA)} controls the raw actuators in a decentralized way, where a distributed multi-agent learning system replaces the single large learner. Unfortunately, we were not able to allocate enough time for examining SARA and ACRA in the field study, so their performance is not analysed in the results section.

\section{EXPERIMENTS}
\label{sec:EXPERIMENTS}

\subsection{Experimental Procedure}
\label{subsec:Experimental_Procedures}

Our experiment was conducted for two weeks from Sept. 14 to Oct. 03, 2018, at the ROM. We were permitted by the ROM to collect data from 1 p.m. to 4 p.m. every day on weekdays. In addition, we conducted in-person surveys on Sept. 18, 20, and 27. During the entire experiment period, visitors were free to visit and interact with the installation without any interference from researchers.  

\begin{figure}[htpb]
        \centering
        \begin{minipage}{0.49\textwidth}
            \centering
            \includegraphics[width=1\linewidth]{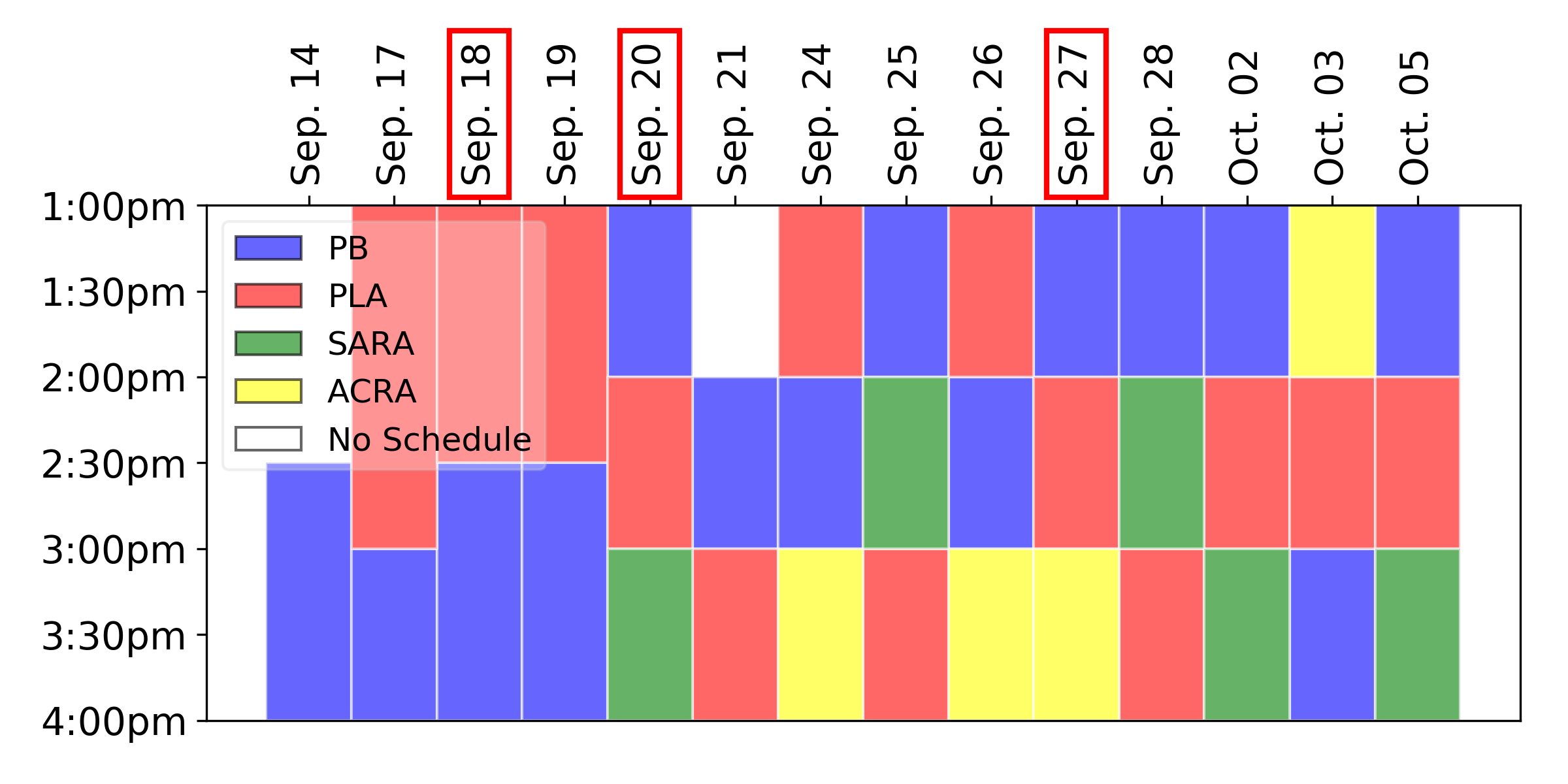}
            \begin{flushleft}
                {\tiny Video was not available on Sep. 14.}
            \end{flushleft}
            \caption{Experiment Schedule}
            \label{fig:Experiment_Schedule}
        \end{minipage}
        \begin{minipage}{0.5\textwidth}
            \centering
            \captionsetup{type=table} 
            \caption{Summary of Experiment Schedule and Data}
            \label{table:Summary_of_Experiment_Schedule_and_Collected_Data}
            \begin{tabular}{c | c | c | c}
            \hline\hline
            \textbf{Behaviour}  & \textbf{Days} & \textbf{Hours} & \makecell{\textbf{Survey }\\\textbf{Participants}} \\\hline
            PB     & 14  & 15.5 & 14 \\ \hline
            PLA    & 13  & 15   & 15 \\ \hline
            SARA   & 4  & 4 & 4 \\ \hline
            ACRA   & 4  & 4 & 3 \\ \hline
            \multicolumn{4}{l}{\tiny Video was not available on Sep. 14.}
            \end{tabular}
        \end{minipage}
    \end{figure}

For each day of the experiment, the following procedure was followed: 
    \begin{enumerate}
        \item Randomly schedule the different agent conditions into 1 or 1.5 hour time slots as shown in Fig. \ref{fig:Experiment_Schedule}. PB and PLA were scheduled on each day, while only one of SARA and ACRA were scheduled per day.
        \item Automatically run scheduled behaviour at each time slot, and save interaction data and learned models and videos at the end of each behaviour.
    \end{enumerate}

During days where no visitor surveys were collected, researchers were not present in the environment.  During the three survey days, researchers were present, but did not provide any additional instructions to visitors.  Researchers observed which visitors interacted with the LAS, passively or actively, within a specific behaviour period. When visitors were finished with their visit, researchers unobtrusively approached randomly selected visitors who had interacted with the system, and asked them if they were willing to participate in a survey.  If a visitor agreed to do the survey, they were guided to a table located around a corner and were provided with a tablet with a questionnaire (see Section \ref{subsec:Data_Collection}). The researchers also recorded which mode the visitor had interacted with. We only recruited visitors who had interacted with only one behaviour mode.

The overall experiment schedule is shown in Fig. \ref{fig:Experiment_Schedule}, where red, blue, green, yellow and white areas correspond to PB, PLA, SARA, ACRA and no schedule respectively. A summary of the experiment schedule and collected data is shown in Table \ref{table:Summary_of_Experiment_Schedule_and_Collected_Data}.

\subsection{Data Collection}
\label{subsec:Data_Collection}

The data collected comes in four types: sensor readings, learning agent logs, human survey data and video data from the two web-cams. 

Every raw sensor reading is logged. In addition to the raw sensor data, each agent also logs its own learning algorithm data collected during the course of learning.

For human survey data, 14, 15, 4, and 3 participants completed surveys in the PB, PLA, SARA and ACRA modes respectively, as summarized in Table \ref{table:Summary_of_Experiment_Schedule_and_Collected_Data}. The questionnaire used in our experiment is a standardized measurement tool for HRI: the Godspeed questionnaire \cite{bartneck2009measurement}. In addition to the 24 Godspeed questions, we asked participants about their interests and background, and their general feedback and comments. The Questionnaire consists of four types of questions: 
\begin{enumerate}
    \item Participants' interests and background (multiple-select multiple choice);
    \item Participants prior knowledge about interactive architecture and machine learning, including ``How familiar are you with interactive architecture?" and ``How familiar are you with machine learning algorithms?";
    \item 24 Godspeed questions namely Godspeed I: Anthropomorphism, Godspeed II: Animacy, Godspeed III: Likeability, Godspeed IV: Perceived Intelligence and Godspeed V: Perceived Safety \cite{bartneck2009measurement};
    \item Participants' general feedback, i.e., ``Any additional comments regarding your experience?" and ``Any overall feedback?".
\end{enumerate}

Video data is collected to calibrate sensory readings and validate occupancy estimates, which will be discussed in detail in Section \ref{subsec:Data_Analysis}. Video data is available for all the experiments except for Sep. 14. 

\subsection{Data Preprocessing}
\label{subsec:Data_Preprocessing}

The IR data is scaled so that all sensory observations for the learning agent are within $\left [ 0,1 \right ]$ and all actions that learning agent can take are within $\left [ -1,1 \right ]$. The IR readings are scaled into $\left [ 0,1 \right ]$ corresponding to the nearest object being at a detected distance of 80cm or more (no nearby humans detected), to the nearest object being 10cm (very close human detected). The action values for all raw actuators are scaled to $\left [ -1,1 \right ]$. For SMAs, values in the range $\left [ -1,0 \right )$ means off and values in the range $\left [ 0,1 \right ]$ means on, while for LED and Moths the continuous values from -1 to 1 are interpreted into 0 to 255, where 0 indicates off and 255 indicates the brightest light for LEDs, and the highest intensity vibration for the Moths. The 11 parameterized actions are scaled to $\left [ -1,1 \right ]$, where -1 corresponds to minimum and 1 corresponds to maximum, and their corresponding original values are shown in Table \ref{table:Related_Parameters_in_Prescripted_Behaviour}.

\subsection{Data Analysis}
\label{subsec:Data_Analysis}

\begin{figure}[thpb]
        \centering
        \begin{subfigure}[t]{.35\linewidth}
          \centering
          \includegraphics[width=0.95\linewidth]{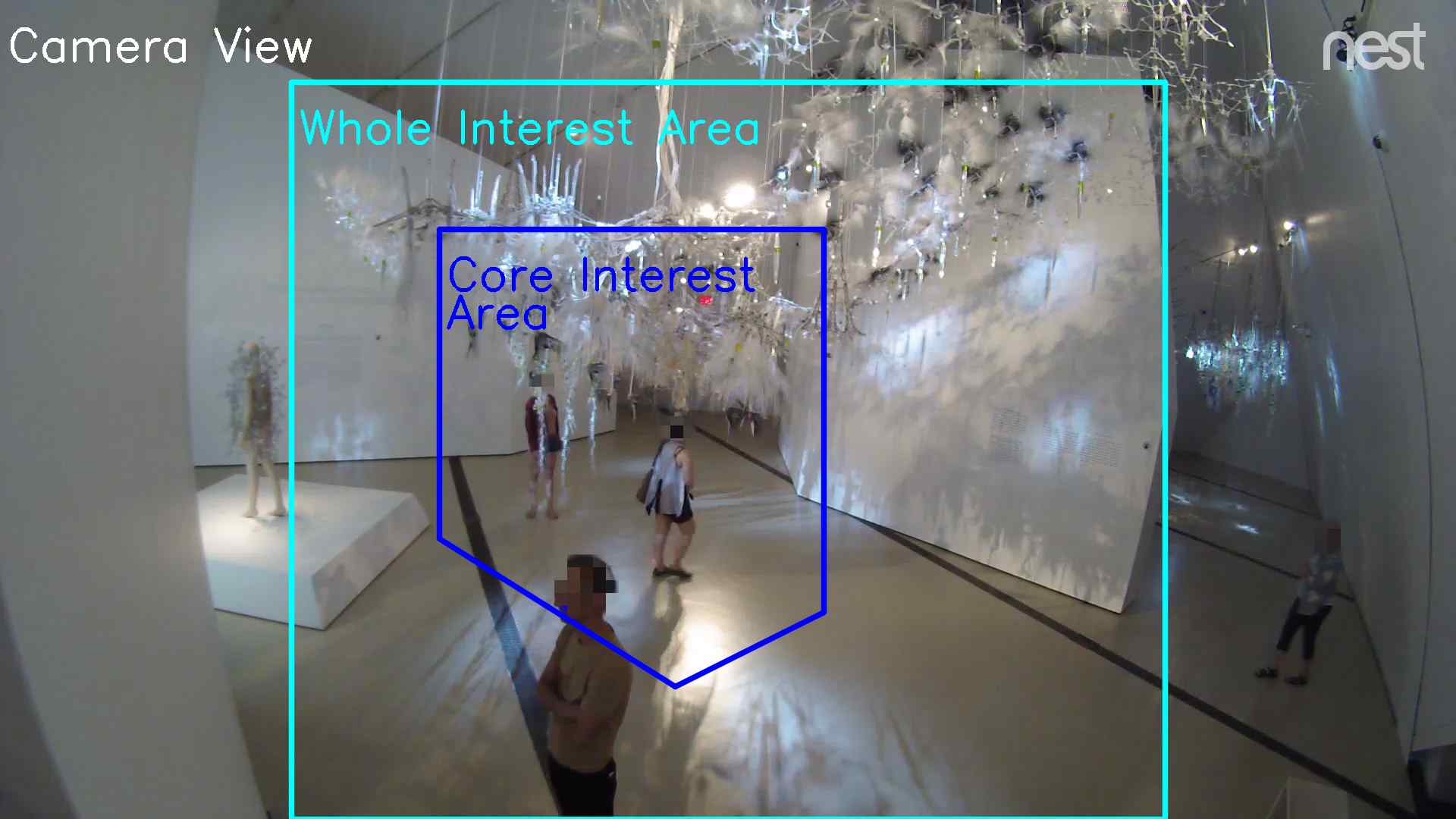}
          \caption{Camera1}
          \label{fig:interest_area_camera1}
        \end{subfigure}%
        \begin{subfigure}[t]{.35\linewidth}
          \centering
          \includegraphics[width=0.95\linewidth]{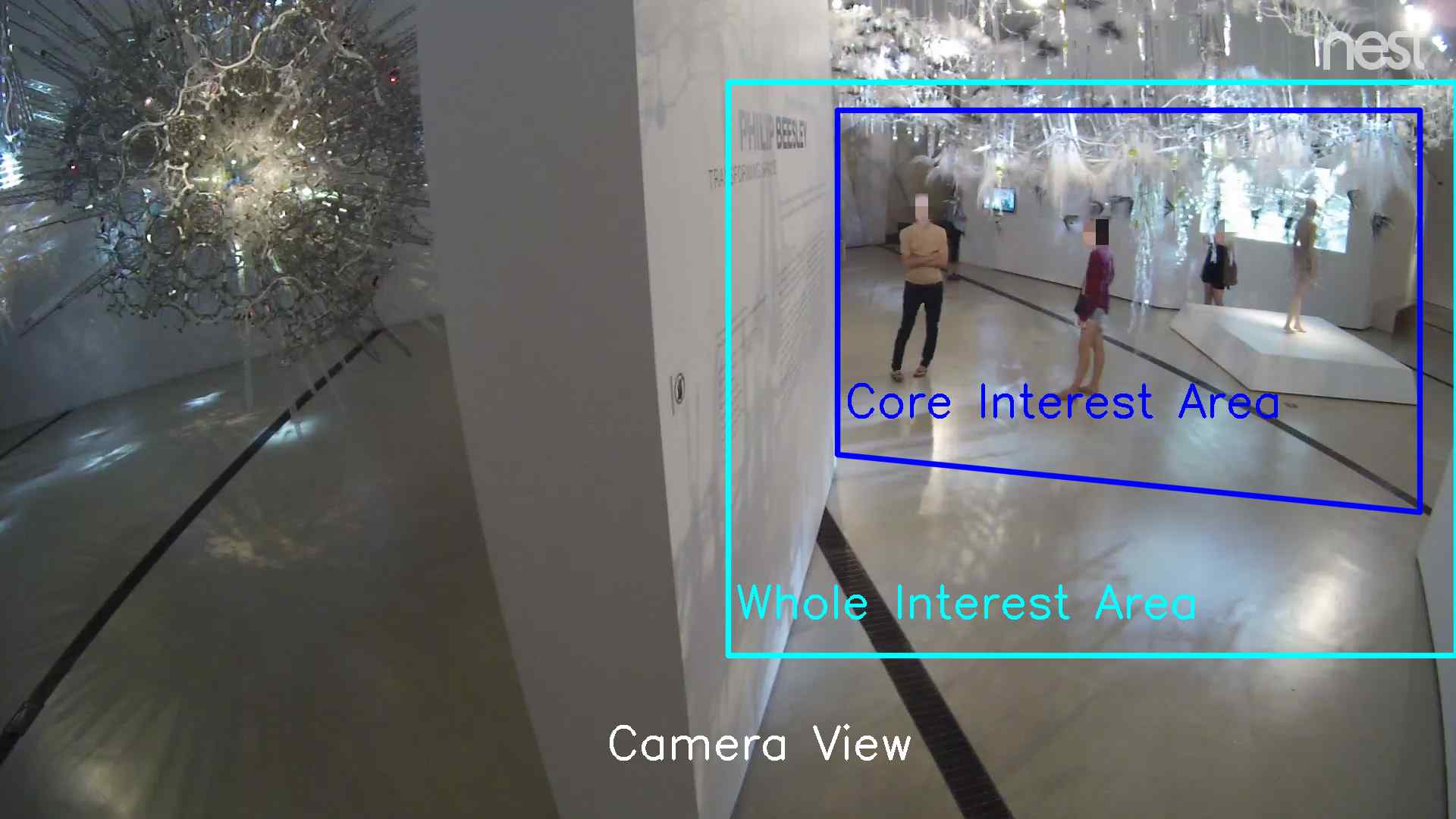}
          \caption{Camera2}
          \label{fig:interest_area_camera2}
        \end{subfigure}
        \begin{subfigure}[t]{.25\linewidth}
            \centering
            \includegraphics[width=0.9\linewidth]{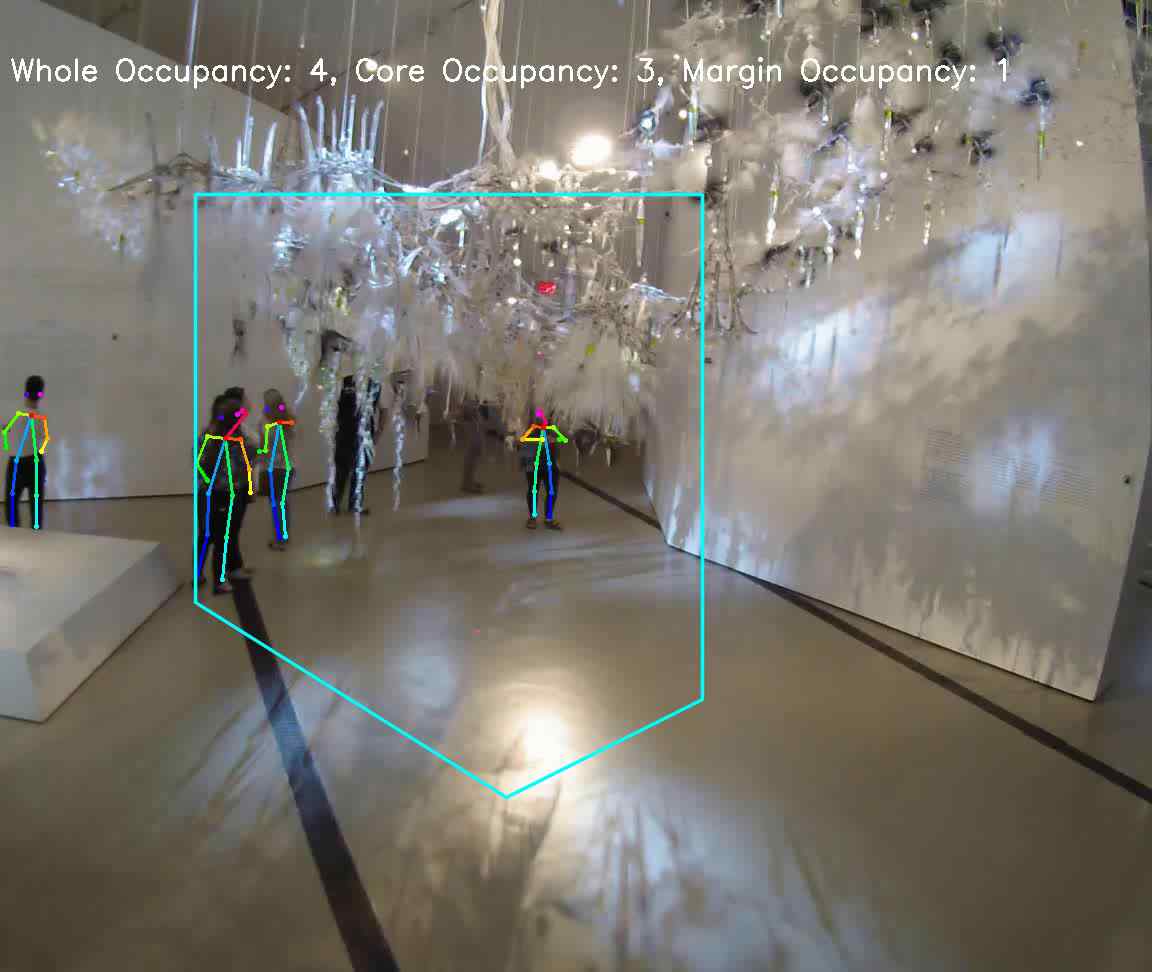}
            \caption{Sample of Estimated Occupancy from Camera1}
            \label{fig:Sample_of_Estimated_Occupancy_from_Camera1}   
        \end{subfigure}
        \caption{Interest Area Used to Estimate Occupancy}
        \label{fig:Interest_Area_Used_to_Estimate_Occupancy}
    \end{figure}

The camera view includes regions outside of the LAS itself. To only focus on areas directly related to the LAS, we define three parts of the whole camera view (as shown in Fig. \ref{fig:interest_area_camera1} and Fig. \ref{fig:interest_area_camera2} for Camera1 and Camera2 respectively). In Fig. \ref{fig:Interest_Area_Used_to_Estimate_Occupancy}, each camera view is divided into Camera View, Whole Interest Area and Core Interest Area.  For both IR Data Calibration (see Section \ref{subsubsec:IR_Data_Calibration}) and Occupancy Estimation (see Section \ref{subsubsec:Occupancy_Estimation}), we only consider the Whole Interest Area. Any visitors outside this interest area will be ignored for the purposes of occupancy estimation. The Core Interest Area approximates the space directly underneath the LAS. 

\subsubsection{IR Data Calibration}
\label{subsubsec:IR_Data_Calibration}

To enable comparison between different behaviour modes, the sensor data must be pre-processed to ensure consistency between conditions. Since visitors can physically interact with the system, it is possible that a visitor changes the direction of the IR sensor thus changing its field of view and subsequent readings. To calibrate the IR data, two calibration steps are taken: 1) IR sensors, whose value is relatively constant and effectively not responding to occupants (e.g., due to obstructions), are removed, 2) the baseline reading for each sensor is shifted to zero. Note that the calibration is only done for analysis, during the learning uncalibrated readings are used. To identify blocked IR and baseline shifts, we visually checked the videos recorded by the two web-cams, and selected a time period when there is no visitor within the whole interest area. Then, we find the IR data corresponding to the no-visitor time. Using the no-visitor time, we determine the thresholds for noise removal and blocked IR detection for the IR data. We use these thresholds to calibrate the raw IR data.

\subsubsection{Occupancy Estimation}
\label{subsubsec:Occupancy_Estimation}

We also use the camera data to generate a second estimate of occupancy. We estimate the number of people occupying the space during a one minute interval, using OpenPose\footnote{\url{https://github.com/CMU-Perceptual-Computing-Lab/openpose}} \cite{cao2016realtime} based on the videos recorded by one of the web-cams\footnote{Videos from Camera2 are highly affected by the changing light of the projector as shown in Fig. \ref{fig:interest_area_camera2}, so for occupancy estimation we only used videos from Camera1.}. When estimating occupancy, we only considered the Whole Interest Area. 

\subsubsection{Non-visitor Period Examination}

We also used the camera data to determine whether there are significant periods when no visitors are present. To identify the time periods with no visitors in Whole Interest Area, we manually labelled the time periods when no person is under either camera in the Whole Interest Area. If a person's body is partially visible in Whole Interest Area, we consider it as a person being in the area. The total amount of non-visitor time throughout the experiment is 1 hour, only 2.5\% of total experiment time. Therefore, we use the whole time period for analysis without removing any non-visitor intervals. 

\subsection{Quantitative Evaluation}
\label{subsec:Quantitative_Evaluation}

Two metrics, i.e. estimated engagement level and active interaction count, will be used to quantitatively evaluate the ability of PB and PLA to engage visitors. Although these two metrics both depend on IR readings, they emphasize different aspects of engagement. Specifically, estimated engagement level does not differentiate passive and active interaction (illustrated in Fig. \ref{fig:Installation_Diagram_and_Interaction_Types}), while active interaction count focuses on measuring active interaction.

We report the average estimated engagement and the average active interaction rather than accumulated reward commonly used in RL, because the test environment is non-episodic and non-stationary. In the natural setting of LAS, the number of visitors varies at different time periods and is highly irregular, which makes the evaluation of learned policy on a separate run unfeasible since the same environment will never be encountered twice. Similarly, comparing accumulated reward within a fixed-length episode is also unfair, because if we compare an episode from PLA during which there are no visitors with an episode from PB during which there are several visitors, we can not say PLA is worse than PB and vice versa, because no matter what PLA does there is no reward at all. Therefore, we regard the whole experiment as continuous learning and compare PLA and PB in terms of average estimated engagement and average active interaction.

\subsubsection{Estimated Engagement Level}
\label{subsubsec:Estimated_Engagement_Level}
We use raw IR readings recorded during each behaviour and Eq. \ref{eq:estimated_engagement} to calculate an estimated engagement for comparison among behaviours. Specifically,  given $M$ IR readings received within 1 minute (typically sampled at 10Hz) $\left \{ \bm{ir}^{\left ( 1 \right )},  \bm{ir}^{\left ( 2 \right )},..., \bm{ir}^{\left ( M \right )} \right \}$ where each IR reading $\bm{ir}^{\left ( i \right )}$ is a vector of 24 IR values, the estimated engagement level $e$ is defined by Eq. \ref{eq:estimated_engagement}:
    \begin{equation}
        e=\frac{1}{M}\frac{1}{24}\sum_{m=1}^{M}\sum_{i=1}^{24}ir_{i}^{\left ( m \right )}
        \label{eq:estimated_engagement}
    \end{equation}
where $ir_{i}^{\left ( m \right )}$ is the $i$th IR sensor in the $m$th IR reading. The estimated engagement is in the range [0,1], where the maximum 1 corresponds to a maximally engaging state, where all IR sensors are receiving maximum readings during the entire 1 minute window, while the minimum 0 corresponds to fully non-engaging state, where all IR sensors are receiving minimum readings for the duration of the one-minute window.

\subsubsection{Active Interaction Count Analysis}
\label{subsubsec:Active_Interaction_Count_Analysis}

In addition to the estimate of engagement, we separately estimate the level of active interaction. To capture active interactions, we count the number of IR readings having value $>=0.25$, which corresponds to a proximity of 35cm or less from an IR sensor, within 1 minute. Formally, given $M$ IR readings received within 1 minute (typically sampled at frequency $F=10Hz$) $\left \{ \bm{ir}^{\left ( 1 \right )},  \bm{ir}^{\left ( 2 \right )},..., \bm{ir}^{\left ( M \right )} \right \}$ where each IR reading $\bm{ir}^{\left ( i \right )}$ is a vector of 24 IR values, the number of active interactions $N_{active}$ is defined by Eq. \ref{eq:number_of_active_interaction}:

    \begin{equation}
        N_{active}=\frac{1}{F}\sum_{m=1}^{M}\sum_{i=1}^{24}\mathds{1}\left \{ ir_{i}^{\left ( m \right )} \geq 0.25  \right \}
        \label{eq:number_of_active_interaction}
    \end{equation}
where $ir_{i}^{\left ( m \right )}$ is the $i$th IR sensor in the $m$th IR reading, and $\mathds{1}\left \{ \cdot  \right \}$ is a indicator function. Therefore, $N_{active}$ is the total detected active interactions within 1 minute.

\section{RESULTS}
\label{sec:results}

In this section, we first quantitatively compare the performance of PB and PLA based on evaluation metrics introduced in Section \ref{subsec:Quantitative_Evaluation}. After that, we analyze the human survey data. 

\subsection{Quantitative Comparison Between PB and PLA}
\label{subsec:Quantitative_Comparison_Between_PB_And_PLA}

In this section, we quantitatively compare the performance of the two behaviour modes based on sensory data collected during the interaction between visitors and the LAS. We use two ways to quantitatively compare the two behaviours' performance: 1) comparing the estimated engagement level, as described in Section \ref{subsubsec:Estimated_Engagement_Level}, and 2) comparing the number of active interactions, as introduced in Section \ref{subsubsec:Active_Interaction_Count_Analysis}.

\begin{figure}[thpb]
        \centering
        \begin{subfigure}[t]{.5\linewidth}
            \centering
            \includegraphics[width=.95\linewidth]{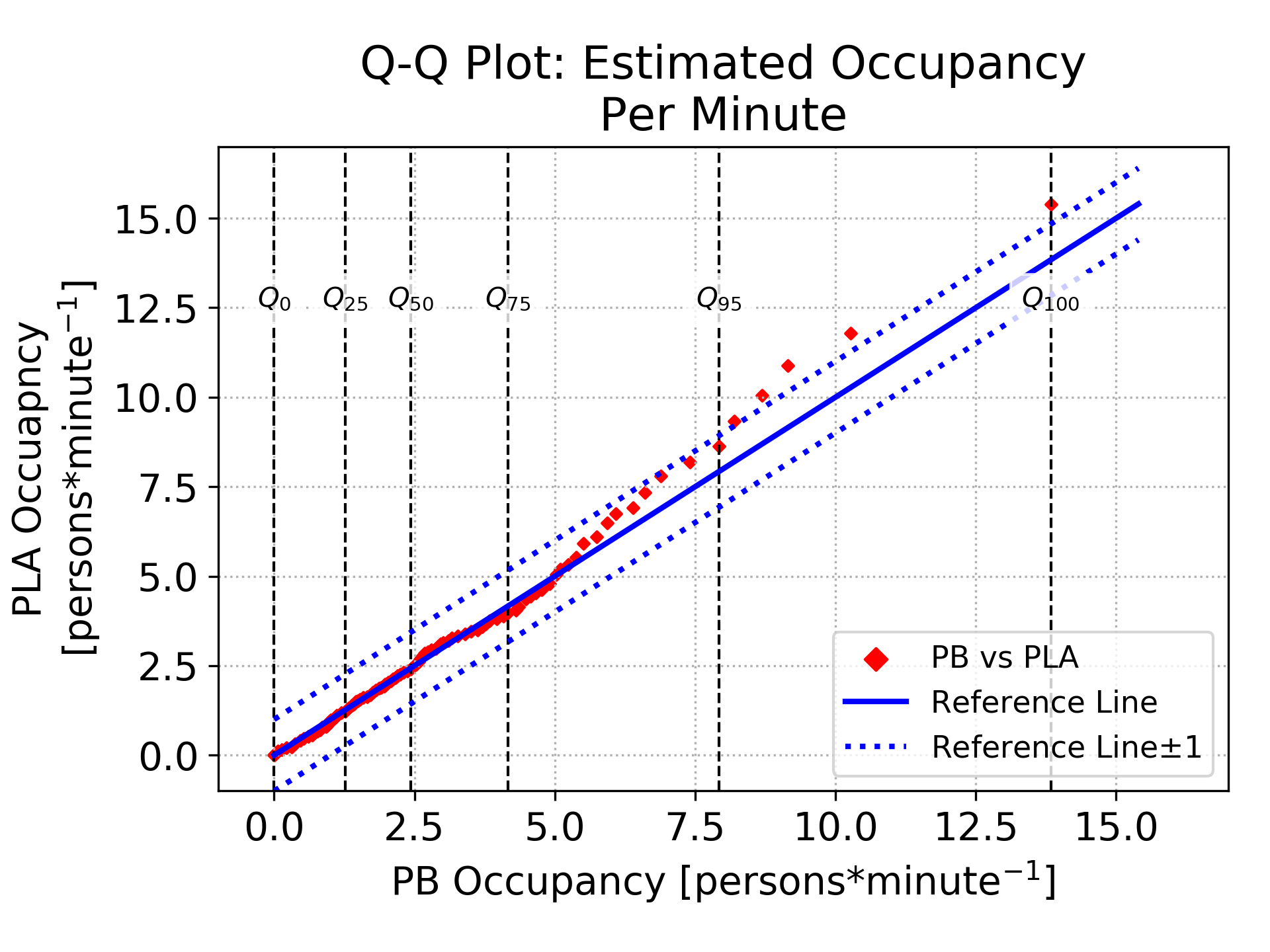}
            \caption{Comparison of Estimated Occupancy Distributions.}
            \label{fig:Comparison_of_Estimated_Occupancy_Distributions.}  
        \end{subfigure}
        \hspace{15pt}%
        \begin{subfigure}[t]{.38\linewidth}
            \centering
            \includegraphics[width=1\linewidth]{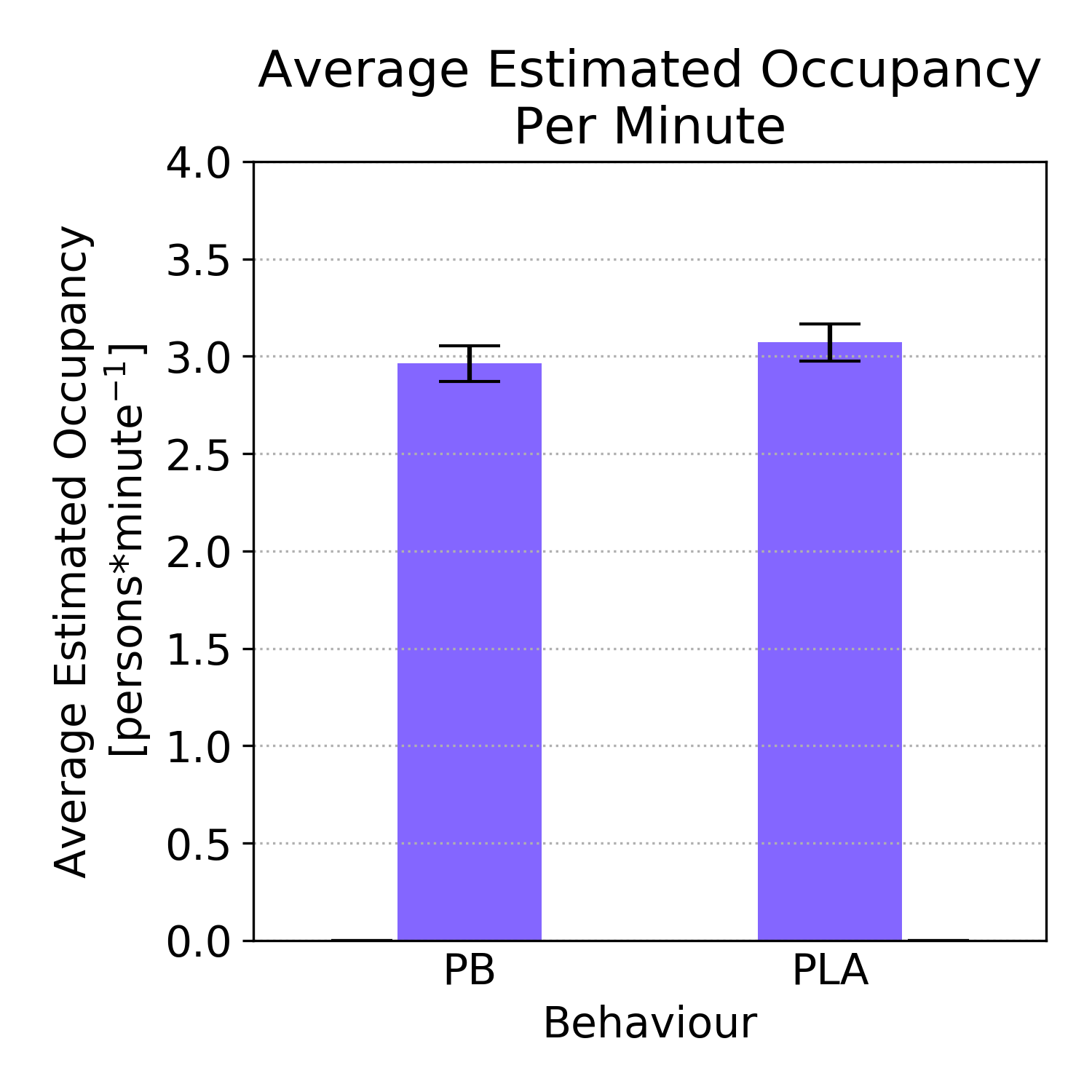}
            \caption{Average Estimated Occupancy}
            \label{fig:Average_of_Estimated_Occupancy}  
        \end{subfigure}
        \caption{Estimated Occupancy Comparison. (a) is a Q-Q (100-quantiles-100-quantiles) plot of estimated per-minute occupancy, using the method introduced in Section \ref{subsubsec:Occupancy_Estimation}, for PB and PLA, where the coordinate $(x, y)$ of the $q$-th point from bottom-left to up-right corresponds to the estimated occupancy of (PB, PLA) for the $q$-th percentile, i.e. $Q_{q}, q=0,1,...,100$, and the reference line indicates a perfect match of distribution between PB and PLA. For example, the point $(4.3,4)$ for PB vs PLA at the $Q_{75}$ means that 75\% of observations for PB and PLA are less than 4.3 and 4, respectively. (b) shows the average estimated per-minute occupancy and its standard error for PB and PLA}.
        \label{fig:Estimated_Occupancy_Comparison}
    \end{figure}

Our experiment is run in a natural setting, i.e., a publicly accessible museum, so it is possible that there are different occupancy levels in the space due to factors not related to the behaviour mode. To check whether there are different occupancy levels between conditions (which might be caused either by some behaviours being more attractive to visitors, or factors not related to system behaviours), we analyze the overall occupancy level for PB and PLA, as described in Section \ref{subsubsec:Occupancy_Estimation}. Fig. \ref{fig:Estimated_Occupancy_Comparison} shows a comparison of the estimated occupancy between PB and PLA, where (a) shows that, in only about 5\% of data, PLA has approximately 1 more visitor than PB, and (b) shows that the average occupancy between PB and PLA is very similar. A Mann-Whitney U test indicates that there is no significant difference between PB and PLA in terms of occupancy level, $U=239030.5$, $p=0.92$ (two-sided). 

\subsubsection{Estimated Engagement Level Comparison}
\label{subsubsec:Estimated_Engagement_Level_Comparison}

\begin{figure}[!htb]
        \centering
        \begin{minipage}[t]{.48\textwidth}
            \centering
            \includegraphics[width=1\linewidth]{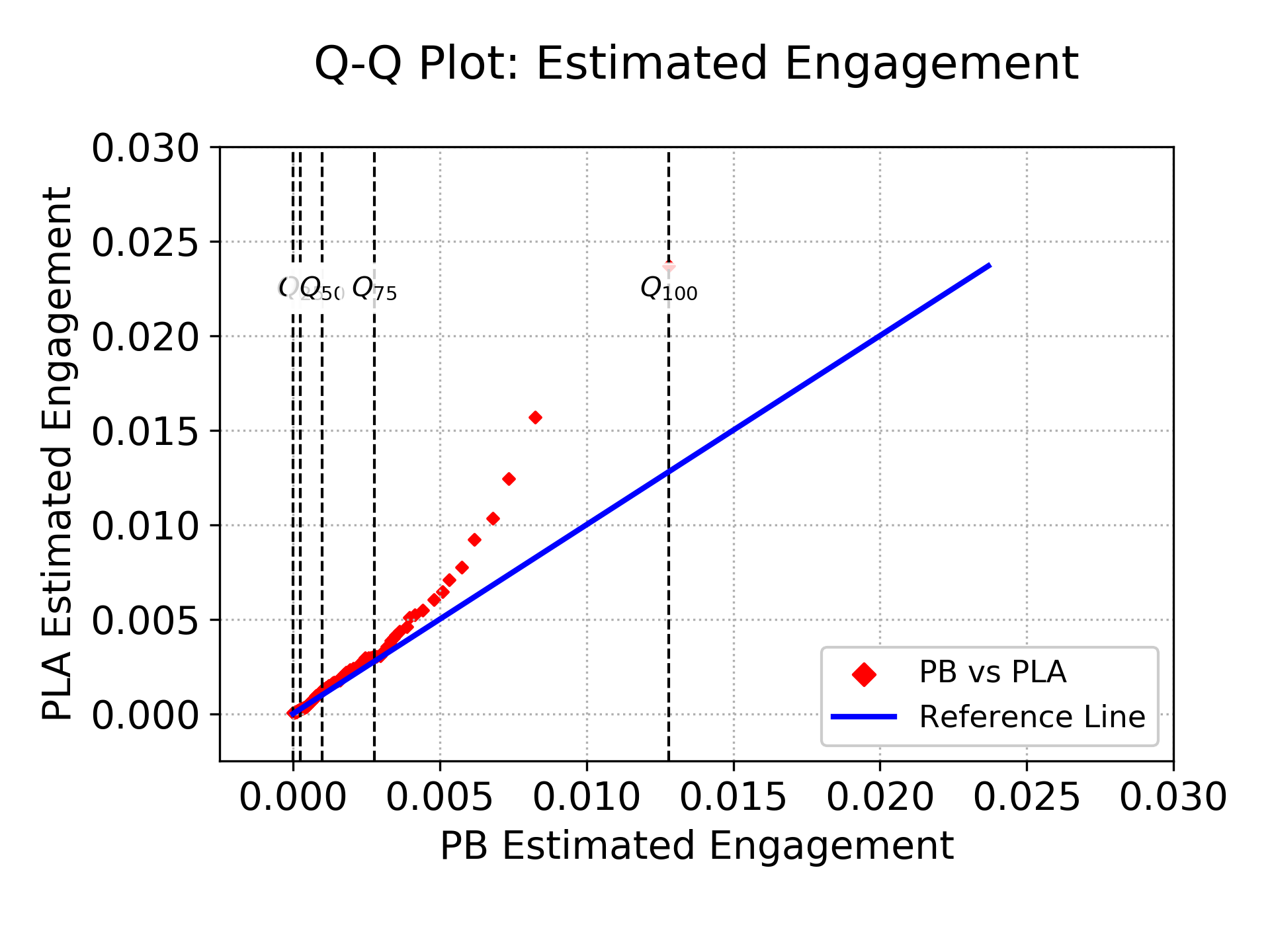}
            \caption{Estimated Engagement Distribution Comparison. The Q-Q (100-quantiles-100-quantiles) plot between PB and PLA  is based on average estimated engagement, where the coordinate $(x, y)$ of the $q$-th point from bottom-left to up-right corresponds to the estimated engagement level of (PB, PLA)  for the $q$-th percentile, i.e. $Q_{q}, q=0,1,...,100$, and the reference line indicates a perfect match of distributions between PB and PLA.}
            \label{fig:Estimated_Engagement_Distribution_Comparison}
        \end{minipage}%
        \hspace{8pt}
        \begin{minipage}[t]{0.48\textwidth}
            \centering
            \includegraphics[width=1\linewidth]{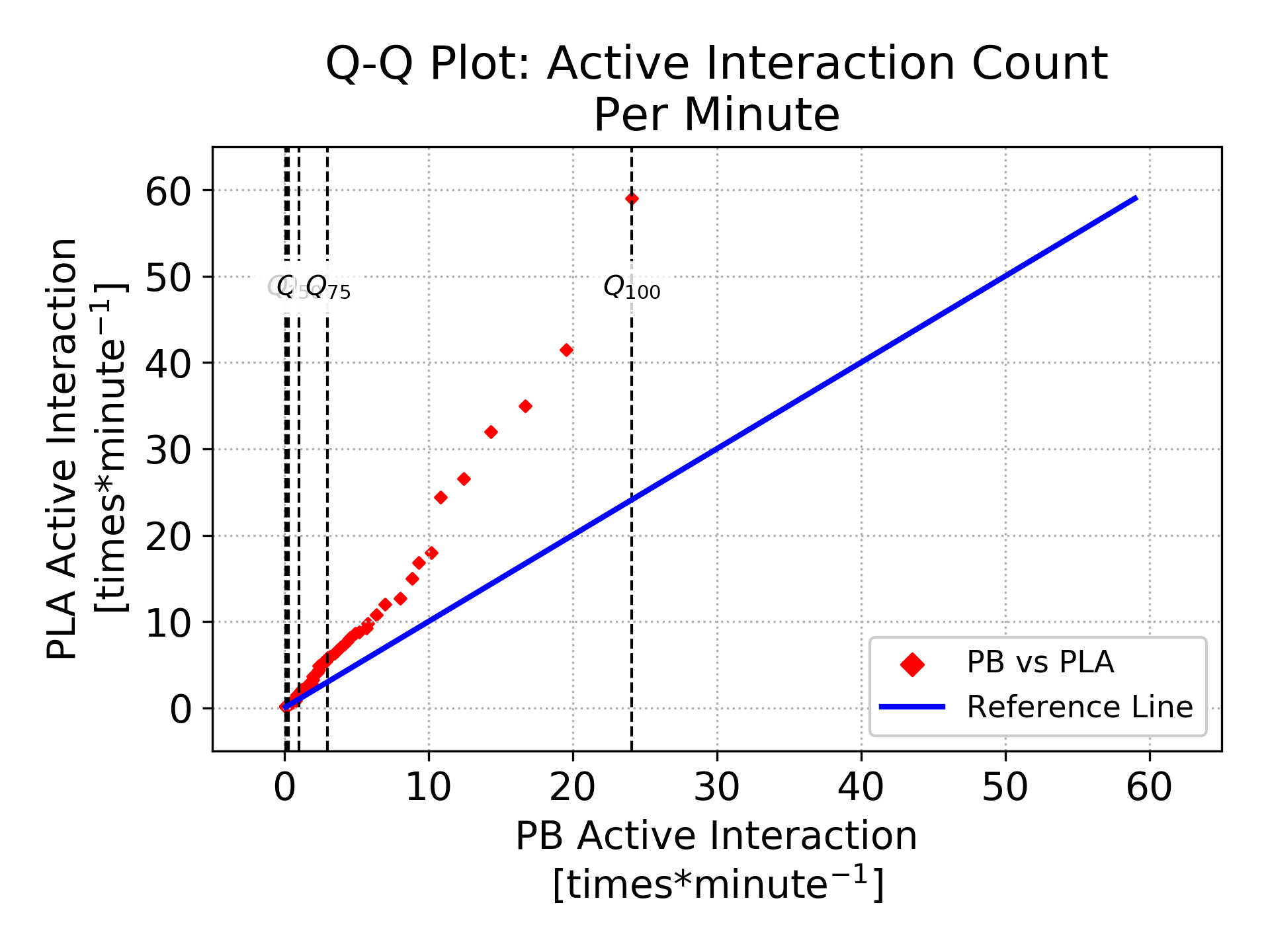}
            \caption{Active Interaction Count Comparison.  The Q-Q Plot is based on active interaction count per minute obtained using Eq. \ref{eq:number_of_active_interaction}, where every (x, y) corresponds to the active interaction count of PB in one percentile, and the line indicates a perfect match of distributions between PB and PLA.}
            \label{fig:Active_Interaction_Count_Distribution_Comparison}
        \end{minipage}
    \end{figure}

\begin{figure}[!htb]
        \centering
        \begin{minipage}[t]{.48\textwidth}
            \centering
            \includegraphics[width=.8\linewidth]{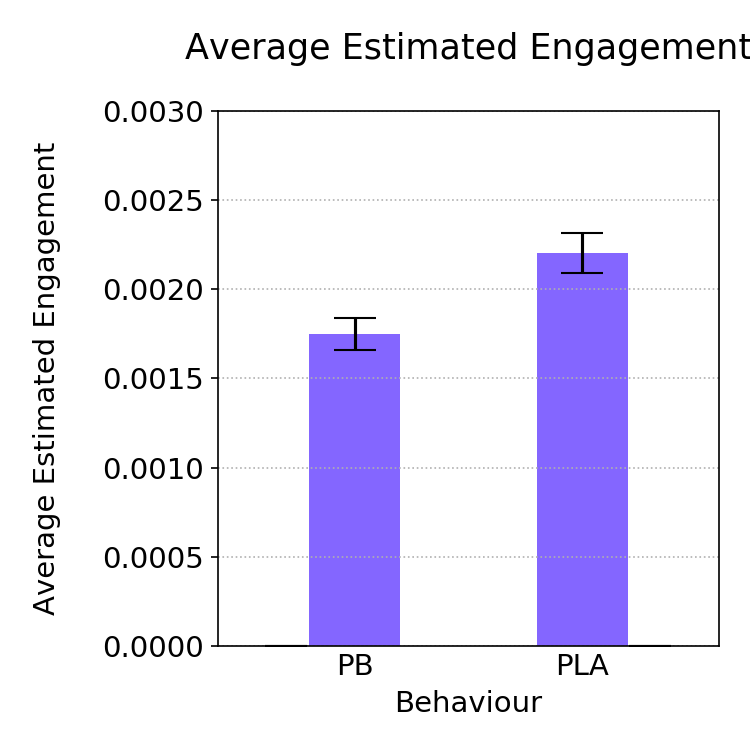}
            \caption{Average Estimated Engagement Comparison, where blue bars with standard errors show the average estimated engagement and its corresponding standard error.}
            \label{fig:Average_Estimated_Engagement_Comparison}
        \end{minipage}%
        \hspace{8pt}
        \begin{minipage}[t]{0.48\textwidth}
            \centering
            \includegraphics[width=.8\linewidth]{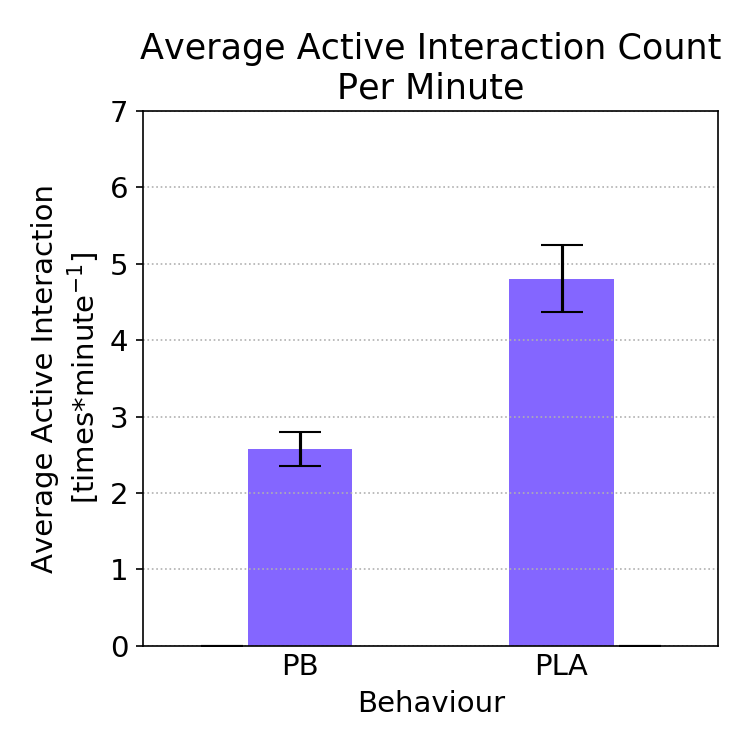}
            \caption{Average Active Interaction Count Comparison, where blue bars with standard errors show the average number of active interactions per minute for each behaviour (the counting is based on 1Hz interaction frequency).}
            \label{fig:Average_Active_Interaction_Count_Comparison}
        \end{minipage}
    \end{figure}

Fig. \ref{fig:Estimated_Engagement_Distribution_Comparison} compares the distributions of estimated engagement (see Section \ref{subsubsec:Estimated_Engagement_Level}) between PB and PLA. From Fig. \ref{fig:Estimated_Engagement_Distribution_Comparison}, we can observe that for the first 75\% of data there is no noticeable difference between PB and PLA, while for the last 25\% of data PLA has larger estimated engagement than PB. Fig. \ref{fig:Average_Estimated_Engagement_Comparison} shows the average estimated engagement, which shows PLA achieves higher average engagement than PB.
    
\subsubsection{Active Interaction Comparison}
\label{subsubsec:Active_Interaction_Comparison}

Fig. \ref{fig:Active_Interaction_Count_Distribution_Comparison} compares the active interaction count based on Eq. \ref{eq:number_of_active_interaction} between PB and PLA. From this figure, we can see that for about 50\% of observations, PLA achieves higher active interaction than PB. Fig. \ref{fig:Average_Active_Interaction_Count_Comparison} compares PB with PLA in terms of average raw active interaction count. As shown in the figure, PLA almost doubles the PB average active interaction count.

\subsubsection{Evolution of Average Estimated Engagement and Active Interaction}

\begin{figure}
    \begin{subfigure}{0.45\textwidth}
        \centering
        \includegraphics[width=1\linewidth]{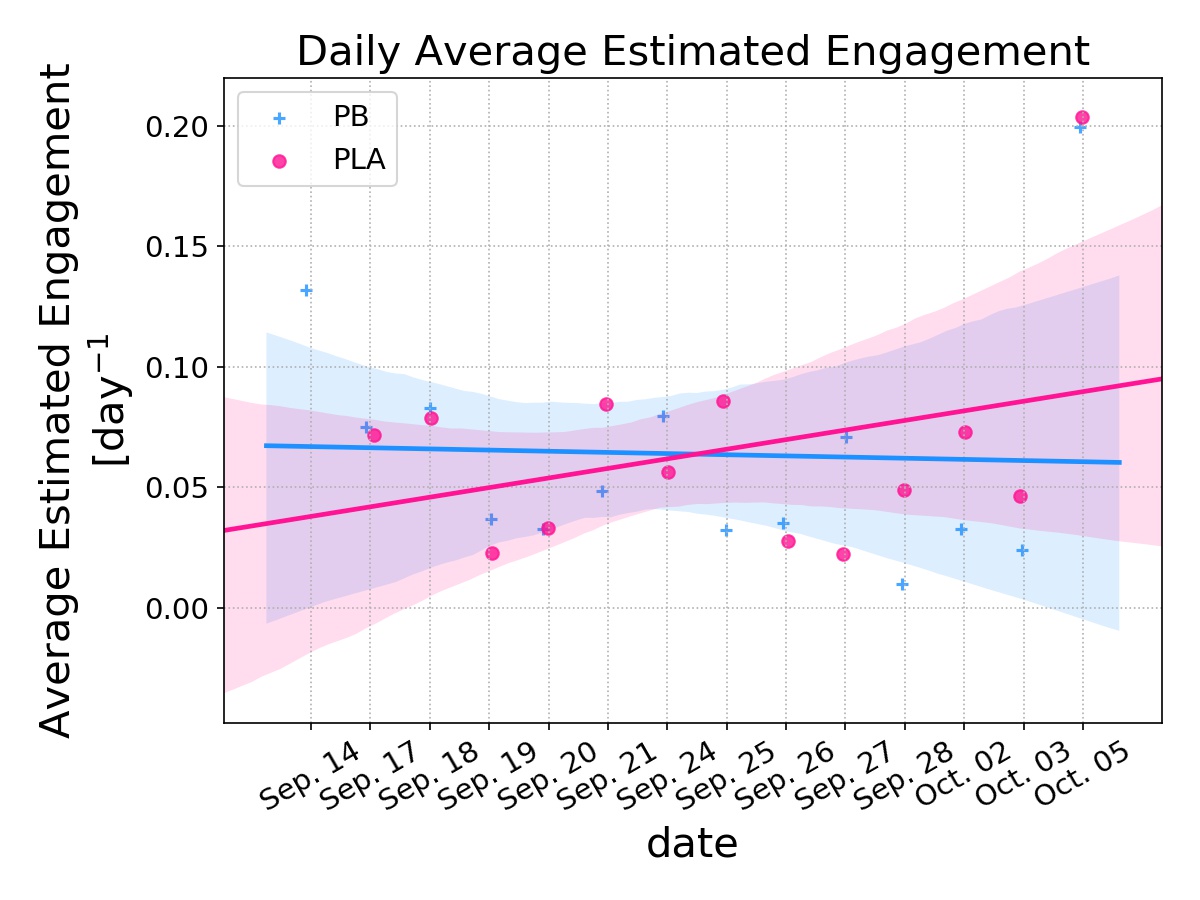}
        \caption{Estimated Engagement}
        \label{fig:Daily_Average_Estimated_Engagement_Comparison}
    \end{subfigure}
    \begin{subfigure}{0.45\textwidth}
        \centering
        \includegraphics[width=1\linewidth]{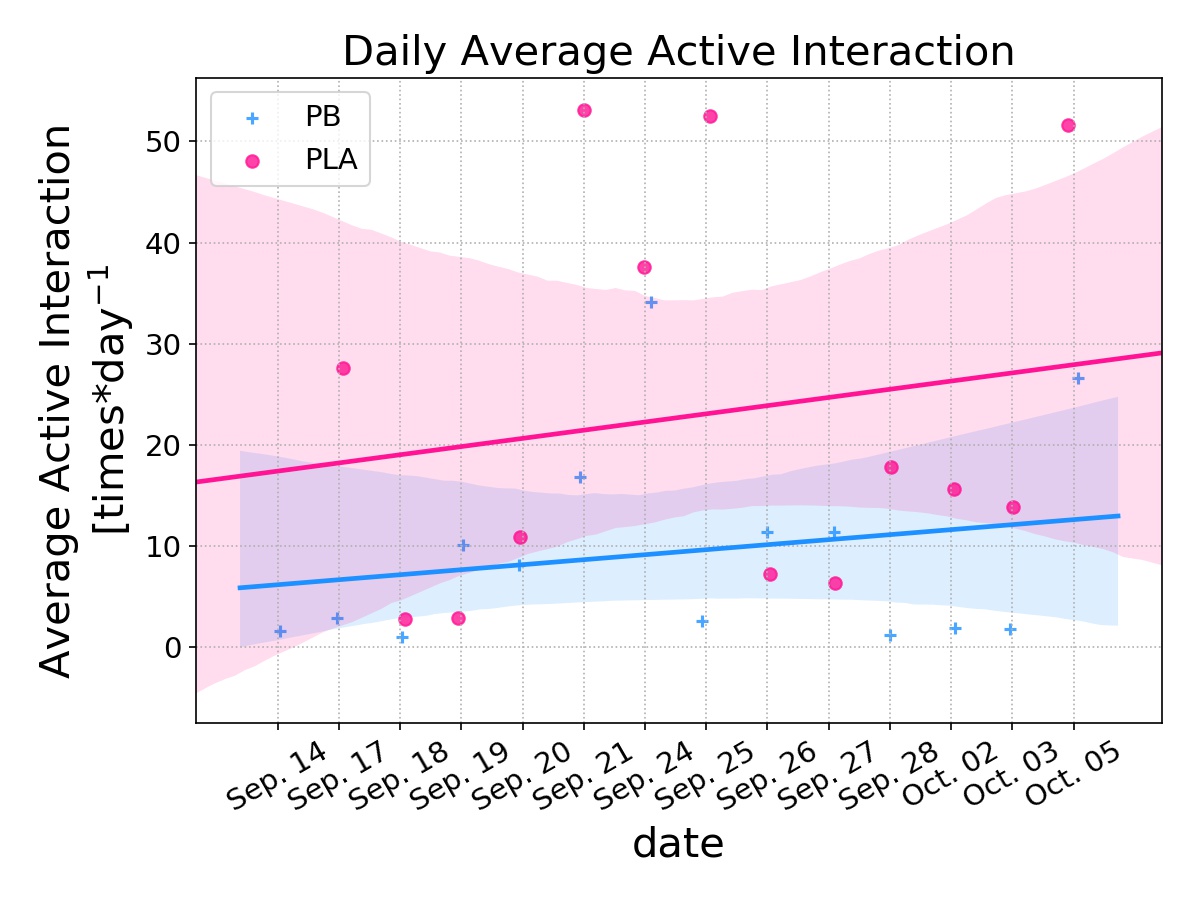}
        \caption{Active Interaction}
        \label{fig:Daily_Average_Active_Interaction_Comparison}
    \end{subfigure}
    \caption{Trajectory of Daily Average Metrics. (a) Daily Average Estimated Engagement and (b) Daily Average Active Interaction, where each data point is the corresponding average on each day, the lines are linear regressions of these data and the translucent bands around the regression line are the 95\% confidence intervals for the regression estimate.}
    \label{fig:Daily_Average_Metrics}
\end{figure}

To analyse how performance evolved over the 3 week experiment, we plot the daily average engagement and active interaction over the whole experiment. Fig. \ref{fig:Daily_Average_Metrics} shows daily average metrics of PB and PLA. In terms of daily estimated engagement, Fig. \ref{fig:Daily_Average_Estimated_Engagement_Comparison} shows that during the first two days, PB outperforms PLA, while after Sep. 25 PLA overtakes PB for the remainder of the experiment. A similar trend can be seen in terms of daily active interaction as shown in Fig. \ref{fig:Daily_Average_Active_Interaction_Comparison}.   PLA receives more active interaction than PB from the very beginning and keeps expanding the gap between PLA and PB. Even though it seems PLA is improving, due to the uncontrolled experimental setting, we cannot be certain whether this is caused by continuous adapting of PLA, or due to factors independent from the interactive action of the LAS.

    \begin{figure}[thpb]
        \centering
        \begin{subfigure}[c]{0.25\textwidth}
            \centering
            \includegraphics[width=1\linewidth]{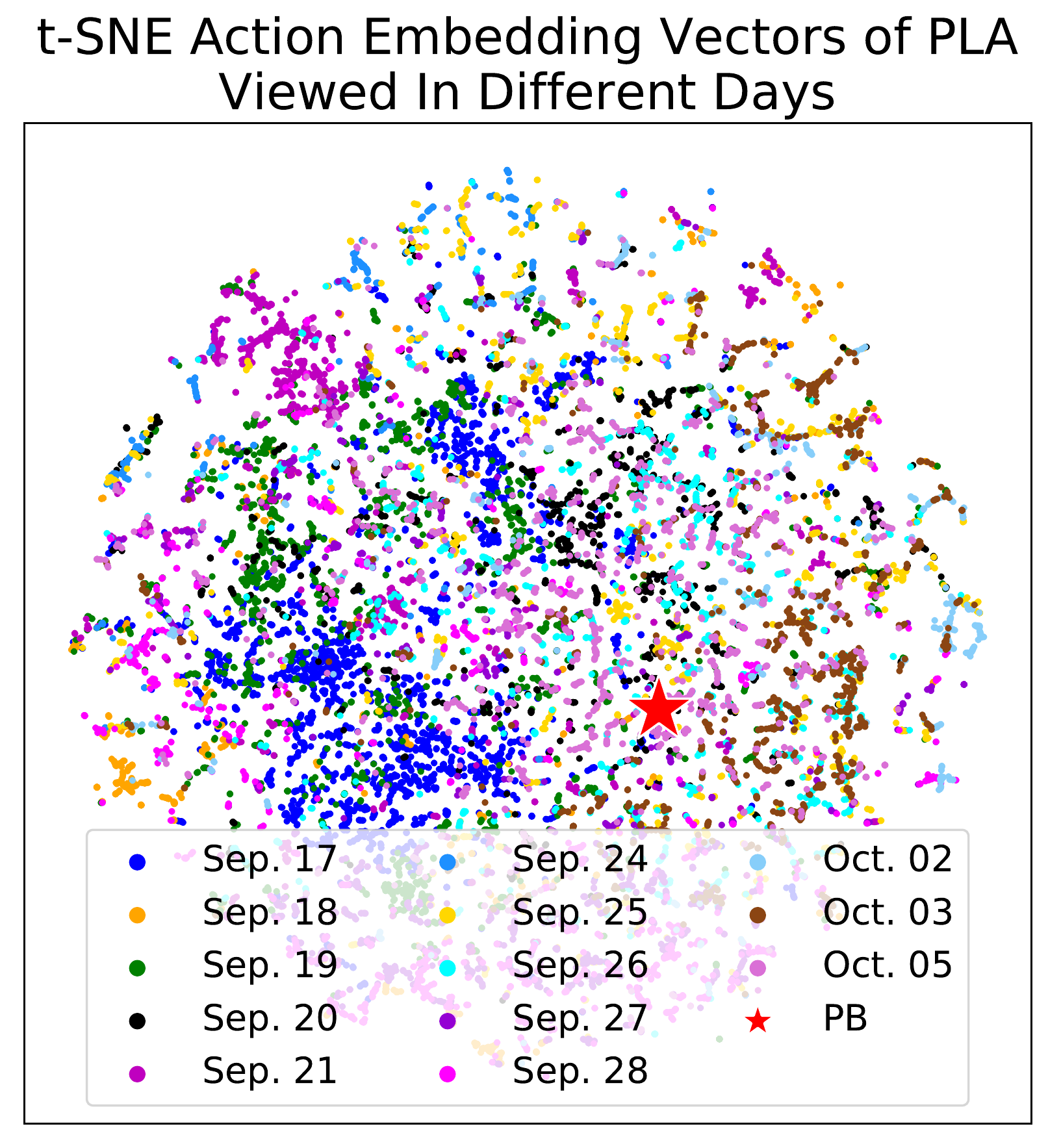}
            \caption{Viewed in Days}
            \label{fig:t_SNE_action_embedding_vectors_of_PLA_all_days_On_Different_Days}
            \hspace{15pt}%
            \begin{subfigure}[b]{1\linewidth}
                \includegraphics[width=1\linewidth]{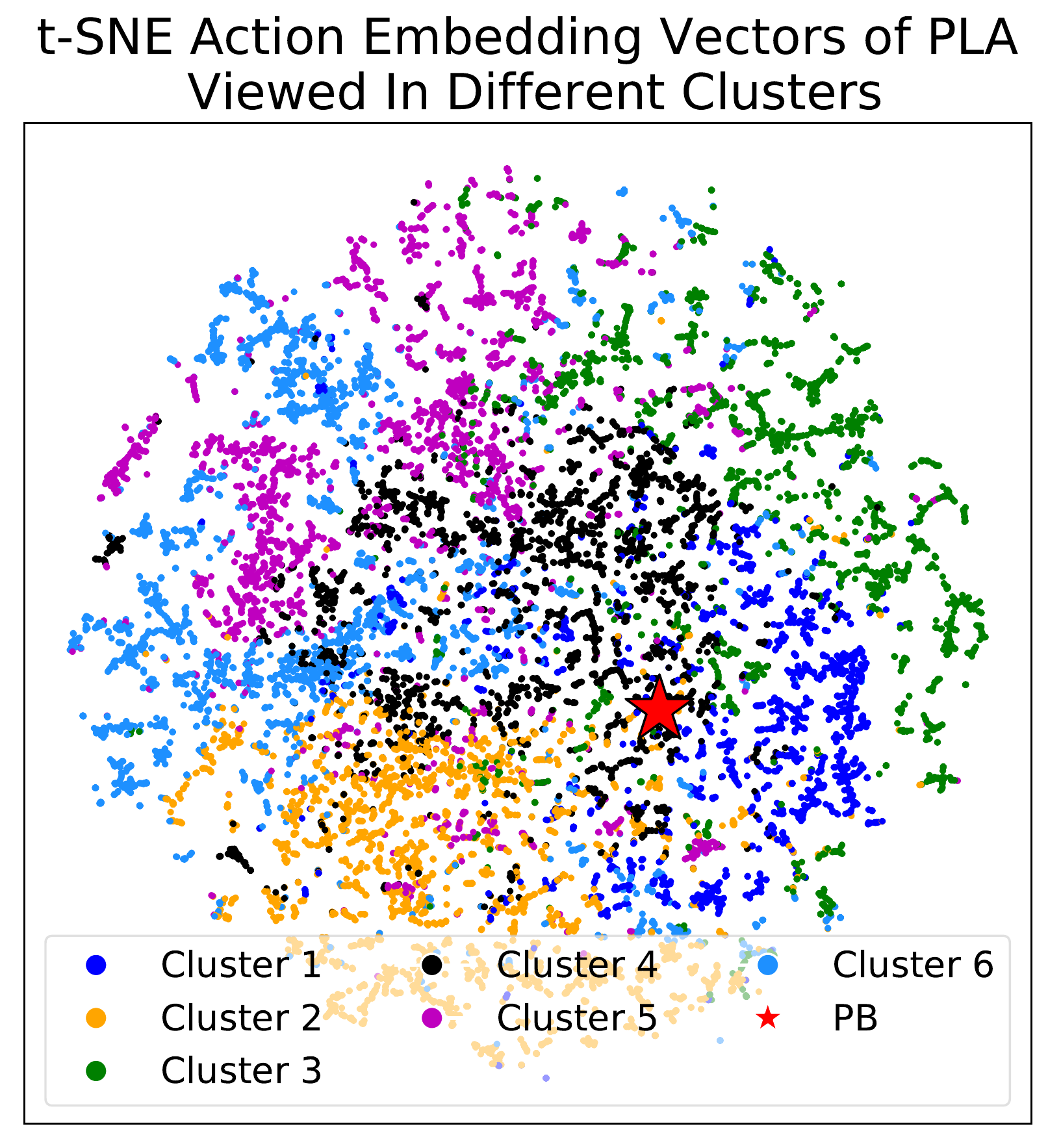}
                \caption{Viewed in Clusters}
                \label{fig:t_SNE_action_embedding_vectors_of_PLA_all_days_in_Different_Clusters}
            \end{subfigure}%
        \end{subfigure}
        \begin{subfigure}[c]{.6\linewidth}
            \centering
            \includegraphics[width=1\linewidth]{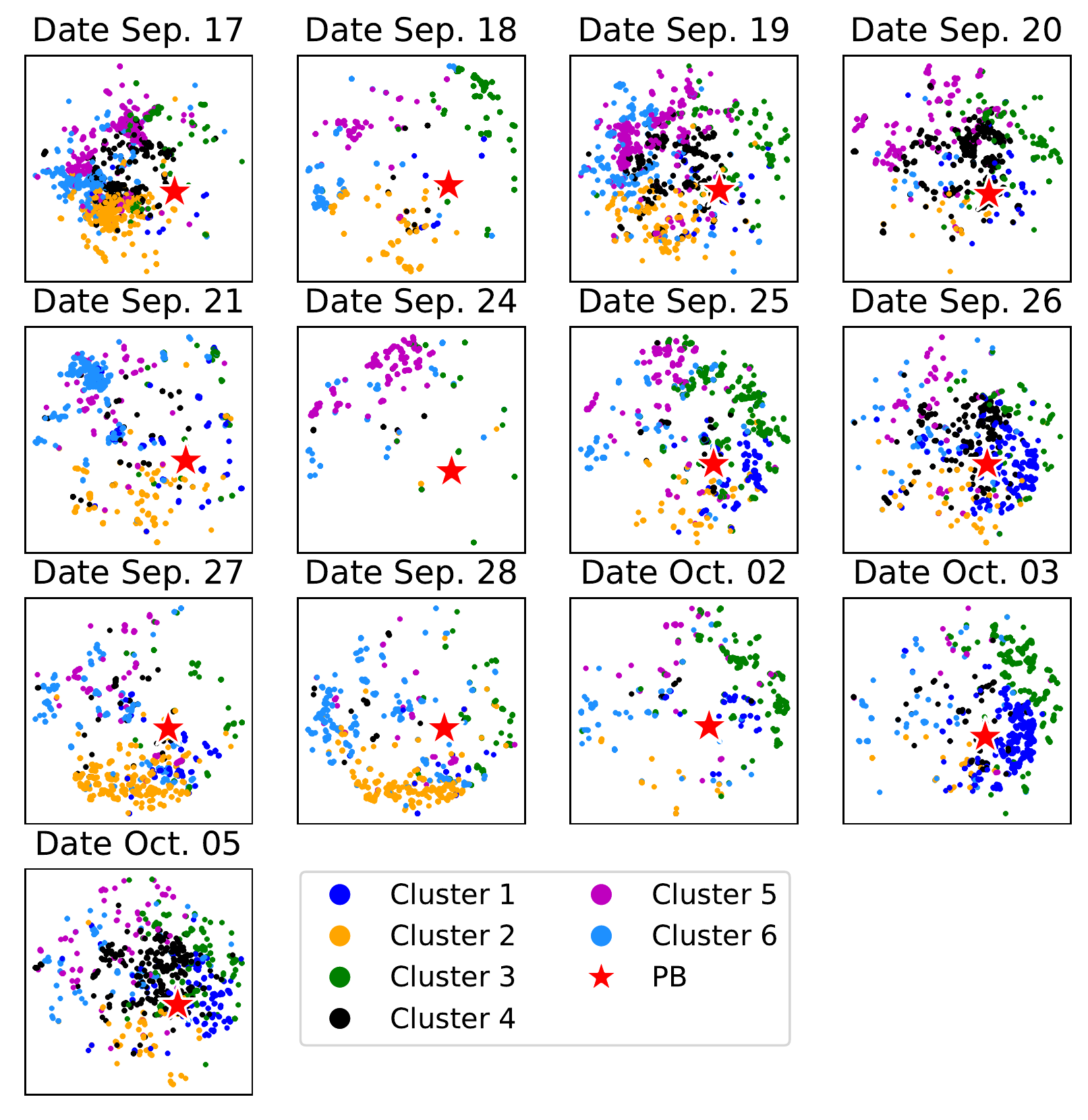}
            \caption{On Each Day in 6 Clusters}
            \label{fig:t_SNE_action_embedding_vectors_of_PLA_on_each_day}
        \end{subfigure}
        
        \begin{subfigure}[c]{.9\linewidth}
            \centering
            \includegraphics[width=1\linewidth]{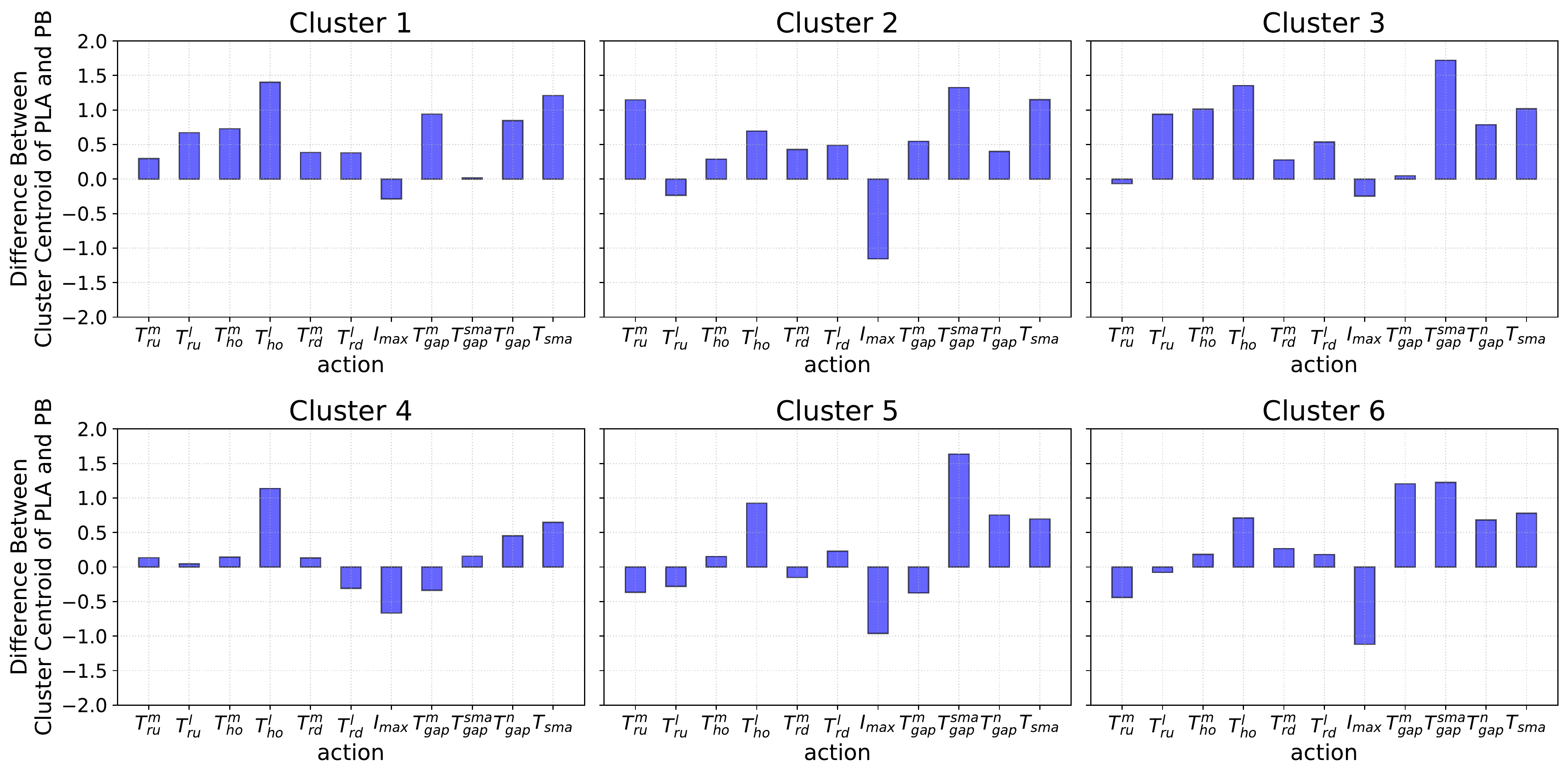}
            \caption{Difference between the Cluster Centroid of PLA and PB}
            \label{fig:Difference_between_the_Cluster_Centroid_of_PLA_and_PB}   
        \end{subfigure}
        \caption{Visualizing actions taken by PLA embedded in a two dimensional space, where each axis corresponds to one embedded dimension. (a) shows actions on all days with different colors for different days. (b) shows actions on all days with different colors for different clusters where K-Means was used to generate the 6 clusters. (c) shows actions taken by PLA on each day, with different colors for different clusters. Each panel in (d) shows the difference $a_{PLA}-a_{PB}$ between the cluster centroids of PLA $a_{PLA}$ and the default value of PB $a_{PB}$ for each dimension of the action space elaborated in Table \ref{table:Related_Parameters_in_Prescripted_Behaviour}, where a positive value means that the corresponding dimension of the cluster centroid of PLA is greater than the default value of PB, while a negative value means the contrary.}
        \label{fig:t_SNE_action_embedding_vectors_of_PLA}
    \end{figure}
    
\subsection{Analysis of Actions Automatically Generated by PLA}
\label{subsec:Analysis_of_Actions_Automatically_Generated_by_PLA}

In this section, we analyze actions automatically generated by PLA. Since the dimensionality of the action space is 11 and each dimension is continuous, visualisation and analysis of the action trajectory is difficult.  For ease of visualisation, we first cluster actions into 6 clusters using K-Means \cite{arthur2007k}, then use t-SNE~\cite{maaten2008visualizing} to embed actions generated by PLA into a 2 dimensional space, where actions which are close in high dimensional space are modeled by nearby points and dissimilar actions are modeled by distant points with high probability.  The resulting visualisation allows us to compare actions over all days or between specific days.  Note that the clustering is done in the 11-dimensional action space of PLA rather than the 2-dimensional embedding space, and the centroid of each cluster is an 11-dimensional data point which can be thought of as the multi-dimensional average of the points in the cluster. Fig. \ref{fig:t_SNE_action_embedding_vectors_of_PLA_all_days_On_Different_Days} and \ref{fig:t_SNE_action_embedding_vectors_of_PLA_all_days_in_Different_Clusters} show embedded actions generated by PLA viewed in different colors for different days and different clusters, respectively, where PB is also embedded for comparison and is indicated by the red star. Fig. \ref{fig:t_SNE_action_embedding_vectors_of_PLA_on_each_day} shows the actions taken by PLA separately by day, with the different clusters labeled. Fig. \ref{fig:Difference_between_the_Cluster_Centroid_of_PLA_and_PB} depicts, for each PLA action cluster, the difference between each of the 11 dimensions of the cluster centroid and the corresponding default value of PB. 

From Fig. \ref{fig:Difference_between_the_Cluster_Centroid_of_PLA_and_PB}, we can roughly observe that: (1) most action dimensions of the centroids of clusters 1, 2 and 3 are greater than the default values of PB; (2) cluster 4 is very similar to PB on many action dimensions; and (3) the centroid of cluster 5 has more action dimensions with values lower than the default value of PB. Taking cluster 1 of PLA as a concrete example, most dimensions of the centroid are larger than the default value of PB, indicating actions in cluster 1 are slower and smoother than PB since Moths and LEDs take more time to ramp up ($T^{m}_{ru}$, $T^{l}_{ru}$), hold on ($T^{m}_{ho}$, $T^{l}_{ho}$) and ramp down ($T^{m}_{rd}$, $T^{l}_{rd}$), and the time gaps between activating Moths and LEDs and neighbour nodes are longer ($T^{m}_{gap}$, $T^{n}_{gap}$, $T_{sma}$). Cluster 1 seems to be preferred on Sep. 26th, Oct. 3rd and Oct. 5th as shwon in Fig. \ref{fig:t_SNE_action_embedding_vectors_of_PLA_on_each_day}, as many actions in cluster 1 are taken. Cluster 4 shares more similarity with PB, as seen from the small differences in most dimensions shown in the 4th panel in Fig. \ref{fig:Difference_between_the_Cluster_Centroid_of_PLA_and_PB}. Actions in cluster 4 are taken densely on Sep. 17th, 19th, 20th, 26th and Oct. 5th, as shown in Fig. \ref{fig:Difference_between_the_Cluster_Centroid_of_PLA_and_PB}. As an interesting and slightly contrary example to cluster 1, cluster 5 of PLA shows abrupt action where Moths and LEDs take less time than PB to ramp up ($T^{m}_{ru}$, $T^{l}_{ru}$), which seems to be preferred on Sep. 24th. Note that here we arbitrarily set the total cluster number to 6 for a relatively clear visualization.  As the clusters still have considerable within-cluster variance, the analysis of centroids only provides an approximate analysis of the diversity of actions taken by PLA. 


Combining Fig. \ref{fig:t_SNE_action_embedding_vectors_of_PLA_all_days_On_Different_Days}, \ref{fig:t_SNE_action_embedding_vectors_of_PLA_on_each_day} and \ref{fig:Difference_between_the_Cluster_Centroid_of_PLA_and_PB}, we do not see specific types of action which are dominant. Nevertheless, it seems that PLA continuously adapts, because actions generated by PLA on each day show slightly different coverage as shown in Fig. \ref{fig:t_SNE_action_embedding_vectors_of_PLA_on_each_day} and demonstrated in previous paragraph. Overall, we can observe that PLA covered a wide range of actions on each day and no obvious dominant actions were reached by the end of the experiment.

\subsection{Human Survey Results}
\label{subsec:Human_Survey_Results}

In this section, we analyze the visitor responses to the survey for PB and PLA. We first examine if there are any differences in the population characteristics between the participants who engaged with the system in PB or PLA behaviour modes. Then, we compare the PB and PLA responses for each Godspeed category. Finally, we compare PB and PLA for each question in Godspeed Likeability category individually.

We first analyze whether there are population differences between conditions. In Section \ref{subsubsec:Occupancy_Estimation}, we confirmed that there were no significant differences between PB and PLA in terms of estimated occupancy. To test for differences in participant background and interest, we performed a $\chi^2$-test on participants' background and interests based on the first two questions in our questionnaire (see Section \ref{subsec:Data_Collection}), and found no statistically significant differences between the two groups.

Cronbach's $\alpha$-test was conducted on each category of Godspeed for both PB and PLA to examine the reliability of participants' responses, results are shown in Table \ref{table:Cronbach_alpha_on_Godspeed_for_PB_and_PLA}. Although $\alpha$ on Anthropomorphism and Perceived Safety is low, $\alpha$ on others is in the acceptable range, especially for Likeability $\alpha\geq0.85$.

    \begin{table}[htb!]
        \caption{Cronbach's $\alpha$ on Godspeed  for PB and PLA}
        \label{table:Cronbach_alpha_on_Godspeed_for_PB_and_PLA}
        \begin{tabular}{ c | c | c | c | c | c }
        \hline\hline
         &  \small \textbf{Anthropomorphism}   & \small \textbf{Animacy}  & \small \textbf{Likeability}   & \small \textbf{\makecell{Perceived\\Intelligence}}  & \small \textbf{\makecell{Perceived\\Safety}} \\\hline
        PB  & 0.74 & 0.77 & 0.85 &  0.89 & 0.52 \\\hline
        PLA & 0.64 & 0.80 & 0.93 & 0.85 & 0.27  \\\hline
        \multicolumn{6}{c}{\tiny \makecell{A commonly accepted rule \cite{devellis2016scale}: $0.9\leq\alpha$: Excellent; $0.8\leq\alpha<0.9$: Good; $0.7\leq\alpha<0.8$: Acceptable;\\ $0.6\leq\alpha<0.7$: Questionable; $0.5\leq\alpha<0.6$: Poor; $\alpha<0.5$: Unacceptable.}} \\
        \end{tabular}
    \end{table}

    \begin{figure}[!hb]
        \begin{minipage}{0.58\textwidth}
            \centering
            \includegraphics[width=.9\linewidth]{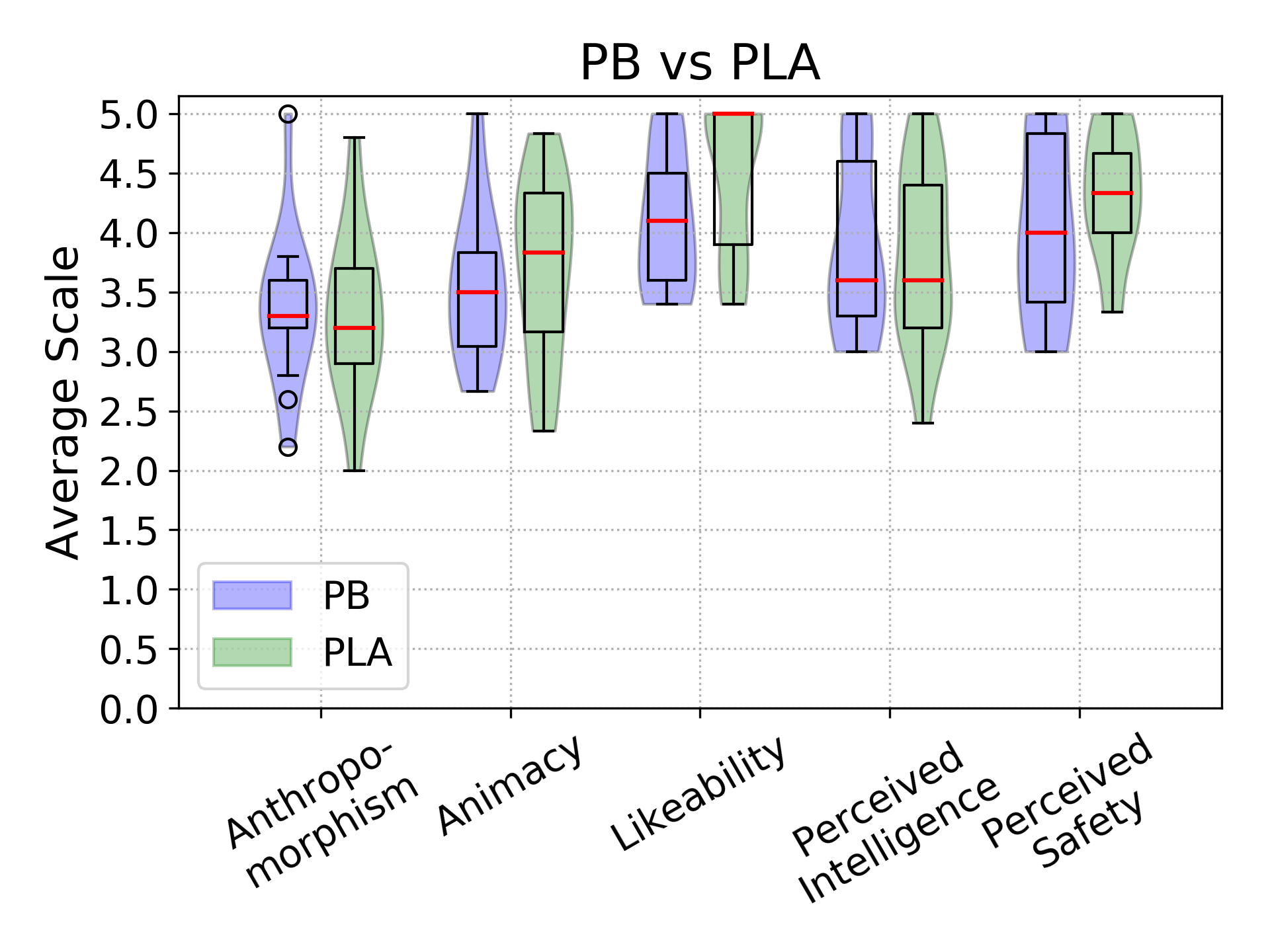}
            \caption{Boxplot and Violinplot of Average Scale of each Godspeed Category over Participants within PB or PLA.}
            \label{fig:Boxplot_and_Violinplot_of_Average_Scale_of_each_Godspeed_Category_over_Participants_within_PB_or_PLA}
        \end{minipage}
        \hspace{10pt}
    \end{figure}

\begin{figure}[!h]
        \centering
        \begin{subfigure}[c]{0.32\textwidth}
            \includegraphics[width=1\linewidth]{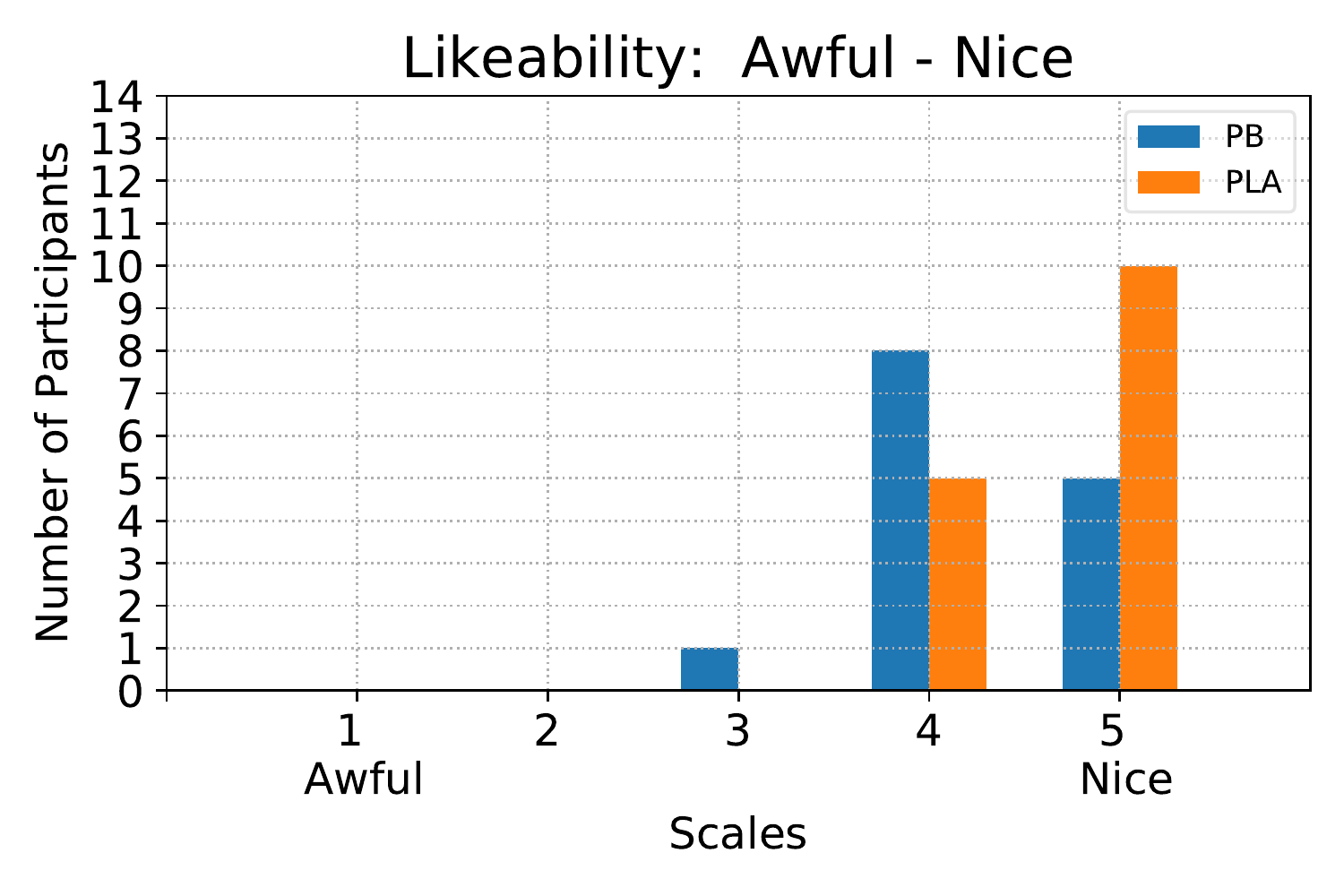}
        \end{subfigure}
        \begin{subfigure}[c]{0.32\textwidth}
            \includegraphics[width=1\linewidth]{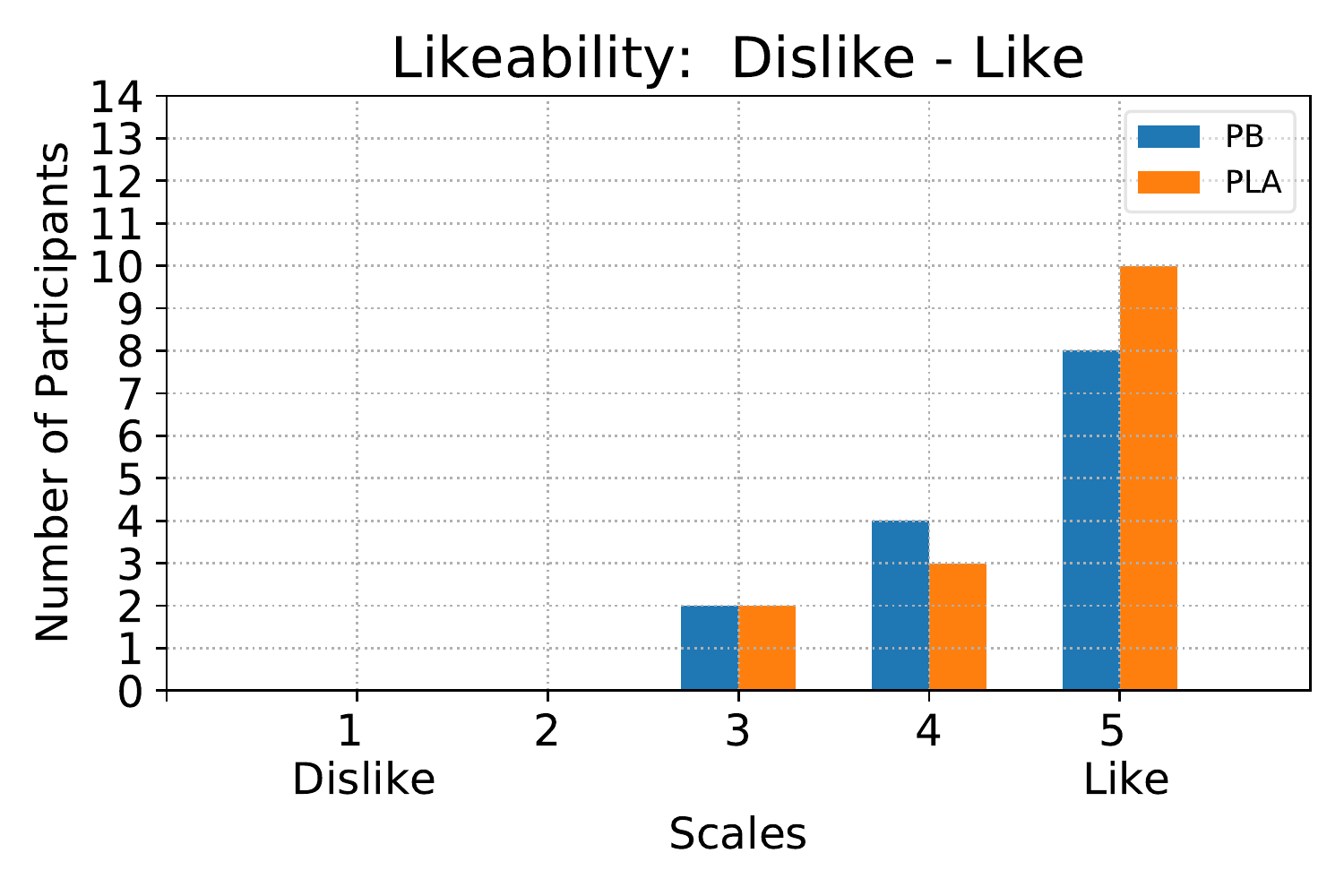}
        \end{subfigure}
        \begin{subfigure}[c]{0.32\textwidth}
            \includegraphics[width=1\linewidth]{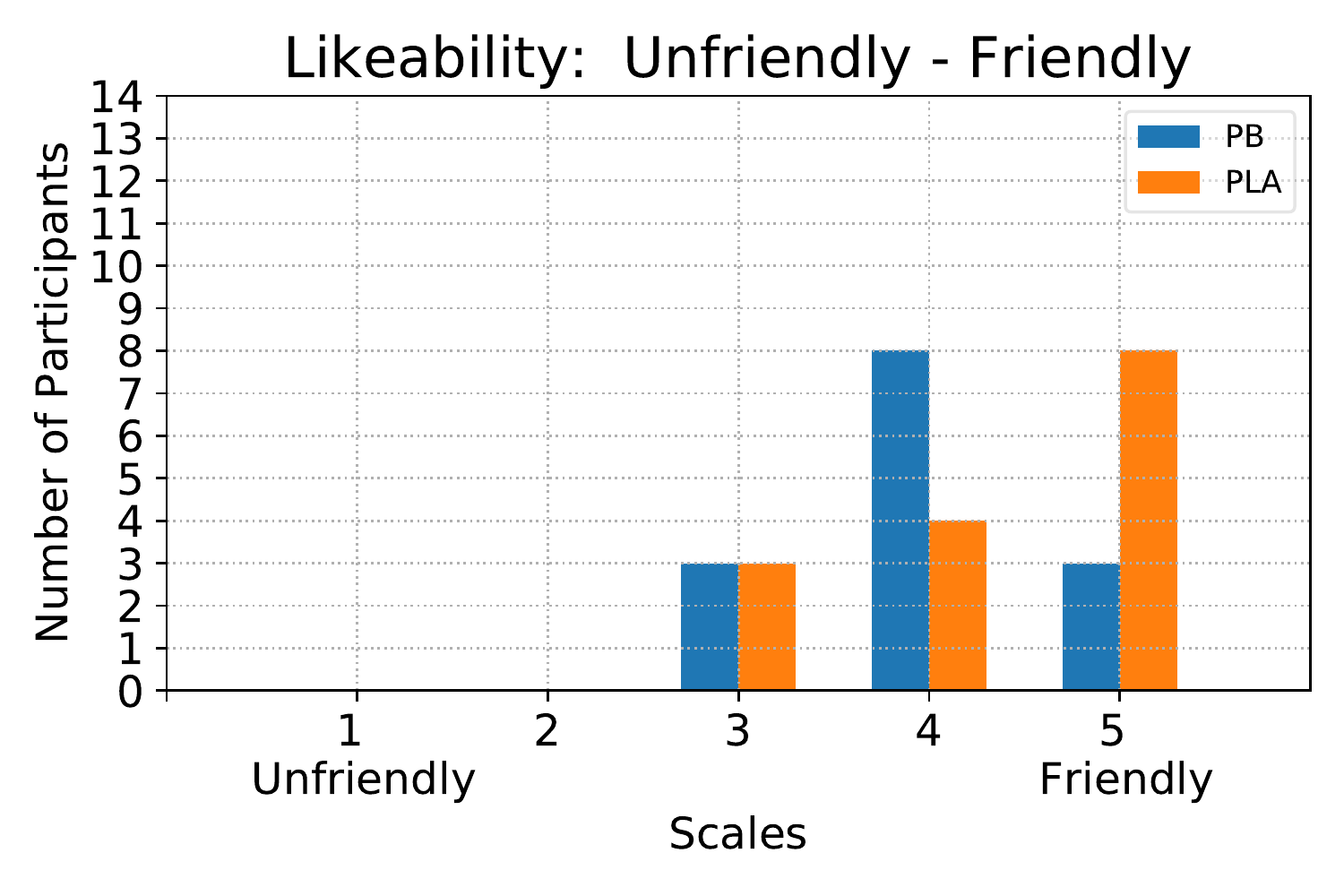}
        \end{subfigure}
        \begin{subfigure}[c]{0.32\textwidth}
            \includegraphics[width=1\linewidth]{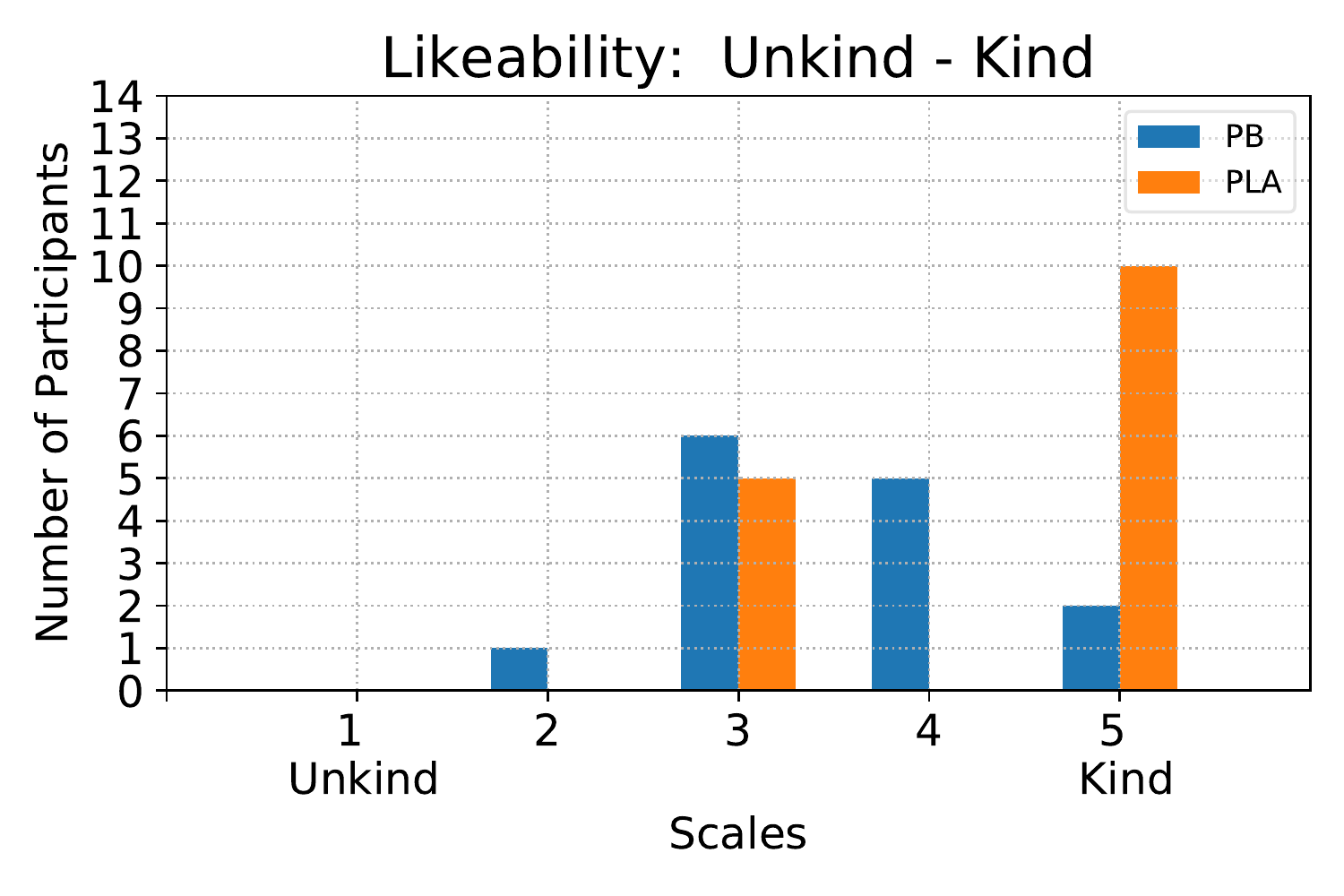}
        \end{subfigure}
        \begin{subfigure}[c]{0.32\textwidth}
            \includegraphics[width=1\linewidth]{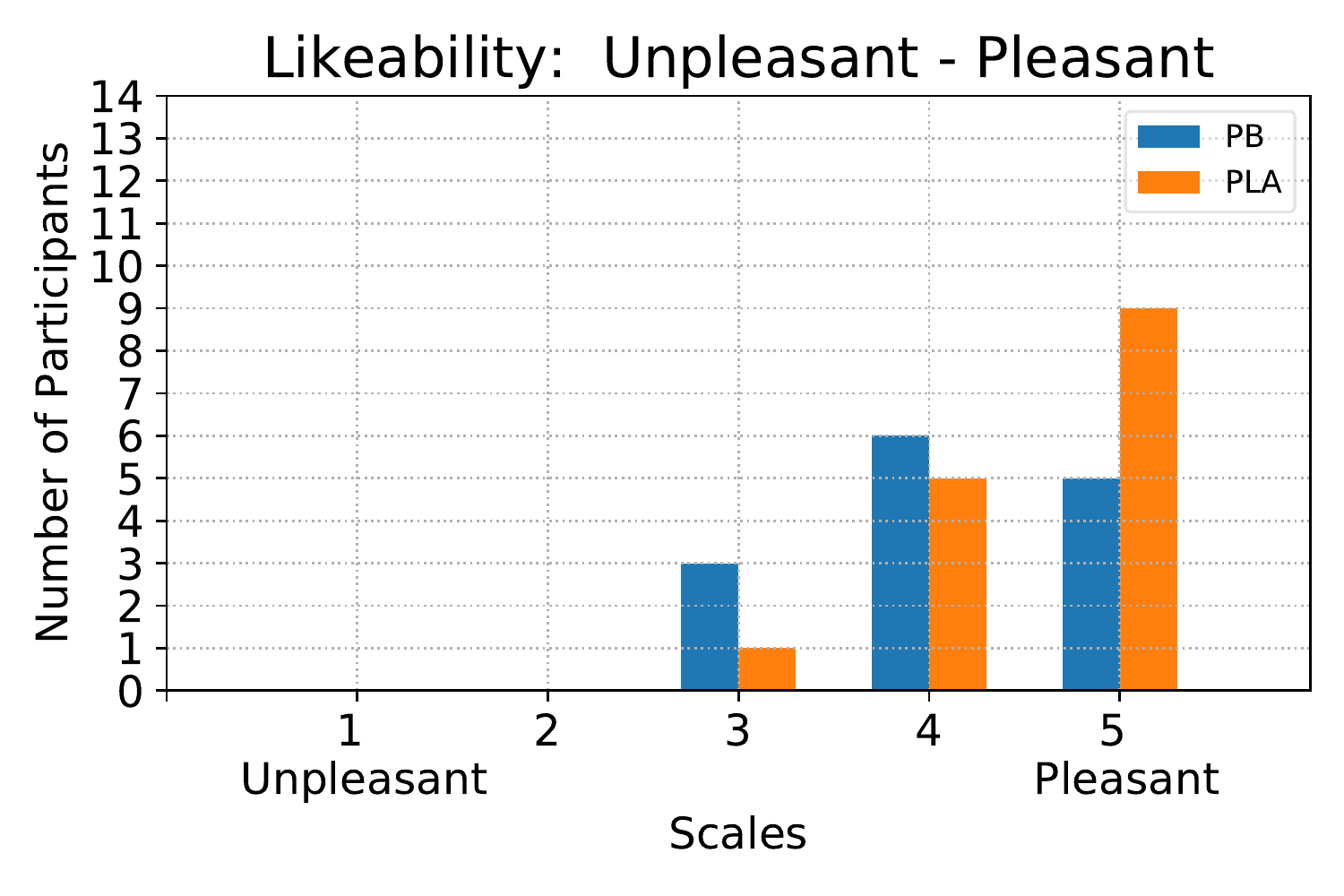}
        \end{subfigure}
        \caption{Histograms of Questions in Godspeed Likeability. }
        \label{fig:Histogram_of_Godspeed_Likeability}
    \end{figure}

Fig. \ref{fig:Boxplot_and_Violinplot_of_Average_Scale_of_each_Godspeed_Category_over_Participants_within_PB_or_PLA} shows the Box-plot and Violin-plot of the calculated average scale over each Godspeed category for PB and PLA. Within the five Godspeed categories, only \emph{Likeability} has a relatively large gap between the medians of PB and PLA. In addition, Likeability has a relatively small variance, whereas other categories have large variance. A $t$-test finds a significant difference in the mean Likeability score between PB (\textit{M=4.09}, \textit{SD=0.56}) and PLA (\textit{M=4.48}, \textit{SD=0.64}; \textit{t(25)=-1.69}, \textit{p=0.05}), whereas for other categories there is no significant difference between PB and PLA. A normality test was conducted for the Likeability category for participants from PB and PLA, respectively. Shapiro-Wilk Test \cite{shapiro1965analysis} indicates PB ($p=0.12$) is normally distributed, while PLA ($p=0.0008$) is not. The histograms of the responses for Godspeed Likeability are shown in Fig. \ref{fig:Histogram_of_Godspeed_Likeability}. From this figure, we can see that on all questions PLA has more participants selecting the value of 5 compared to PB.
    
In summary, PLA is rated higher than PB by the participants in terms of Likeability, while there are no significant differences between PB and PLA in the other Godspeed categories.

\section{DISCUSSION}

In this paper, we investigated how an interactive system can learn to engage with visitors in a natural setting, where no constraints are imposed on visitors and group interaction is accommodated. Relying on the standard RL framework and a novel measure of engagement as reward, an adaptive behaviour \emph{PLA} was compared to a pre-scripted behaviour \emph{PB}. Our results show that PLA outperforms PB in terms of estimated engagement level, active interaction count, and human survey data. We hypothesize that the PLA configuration outperforms PB because it benefits from both human expert input, such as parameterized action space and manual reward function, and learning. During the design process, architects design PB based on their expertise.  We exploit this human expertise to effectively restrict the parameterized action space of PLA into a region where we know good solutions can be found, and at the same time limit the dimension of the PLA action space. Therefore, PB and PLA both incorporate human intelligence, but compared with PB, PLA is endowed with adaptability by applying RL to its parameterized action space.

Our approach uses low-cost IR sensors for engagement estimation.  Compared to other social HRI work with rich sensing such as cameras and microphones, we were still able to estimate engagement with limited sensing and generate engaging behaviours accordingly. This might be helpful to other large scale interactive systems where having sophisticated measurement may be unfeasible. In summary, this work provides two useful generalizable guidelines for designing engaging behaviour for long-term interaction: (1) in a group interaction setting, group engagement from low-cost ambient sensors can be used either standalone, or as a complement of individual engagement measures, and (2) such a measure of engagement can be used as a reward signal to generate customised and evolving behaviour.

Creating engaging behaviours for LAS requires learning algorithms that continue to adapt rather than optimizing to a single best policy. First, the Markov Decision Process and Stationary Environment Dynamics assumptions are broken because of the complicated interaction environment. In addition, interaction time cannot be assumed to have a constant length, and the interaction speed is bound to physical interaction and cannot be sped up. In addition, since LAS is an architectural scale interactive system, perception of the environment is more complex and may need to include both proprioception and exteroception. Therefore, although we exploit a RL framework in our work, the role it plays is different from that of standard testbeds such as OpenAI Gym~\cite{brockman2016openai} or Arcade Learning Environment (ALE)~\cite{bellemare2013arcade}.  In this work, RL is used to introduce adaptability, but there is no guarantee that the learning leads to optimal policy. This is illustrated by the observation that compared with PB, PLA shows very flexible action patterns, and some are very different from PB. Specifically, one observed behaviour generated by PLA is LEDs turned on and propagated quickly from one node to another back and forth multiple times accompanying activated SMAs and Moths (see \url{https://youtu.be/2tICanYEpoo}) for the video comparing PB and PLA, which makes the LAS look like a thunderstorm. This novel behaviour illustrates that the sculpture has taken the primitives composed by the designers and evolved engaging and interesting behaviour from those.

\begin{figure}[thpb]
        \centering
        \begin{subfigure}[t]{.24\linewidth}
            \centering
            \includegraphics[width=1\linewidth]{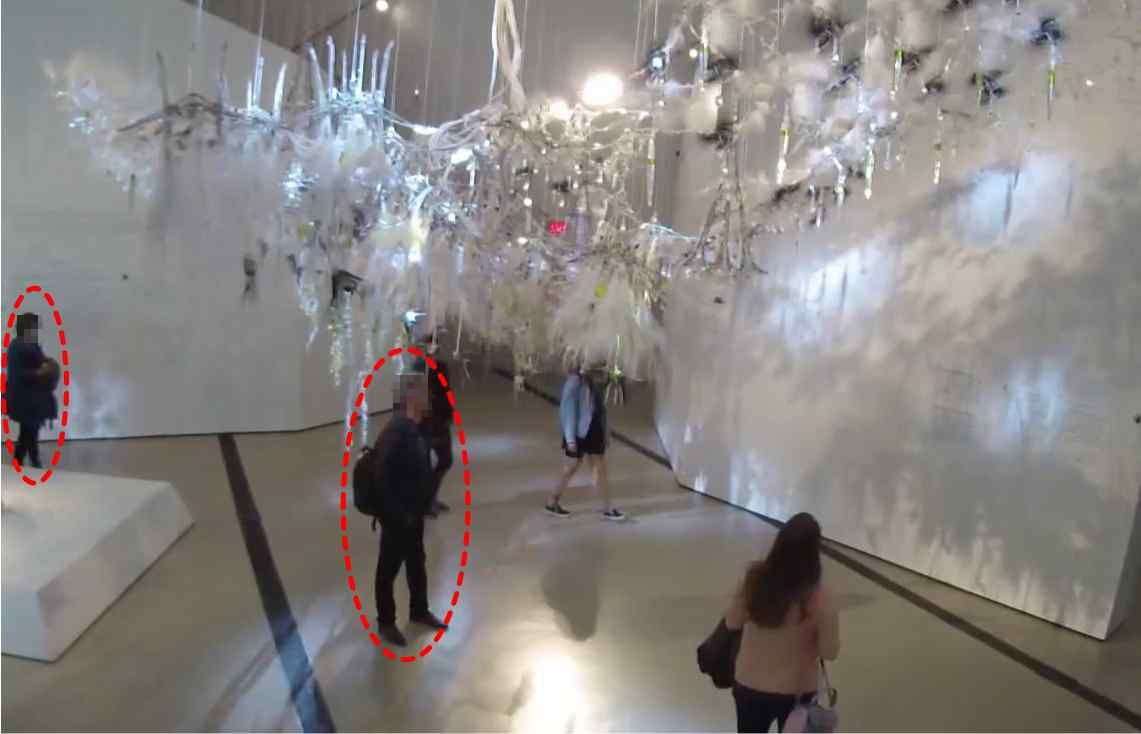}
            \includegraphics[width=1\linewidth]{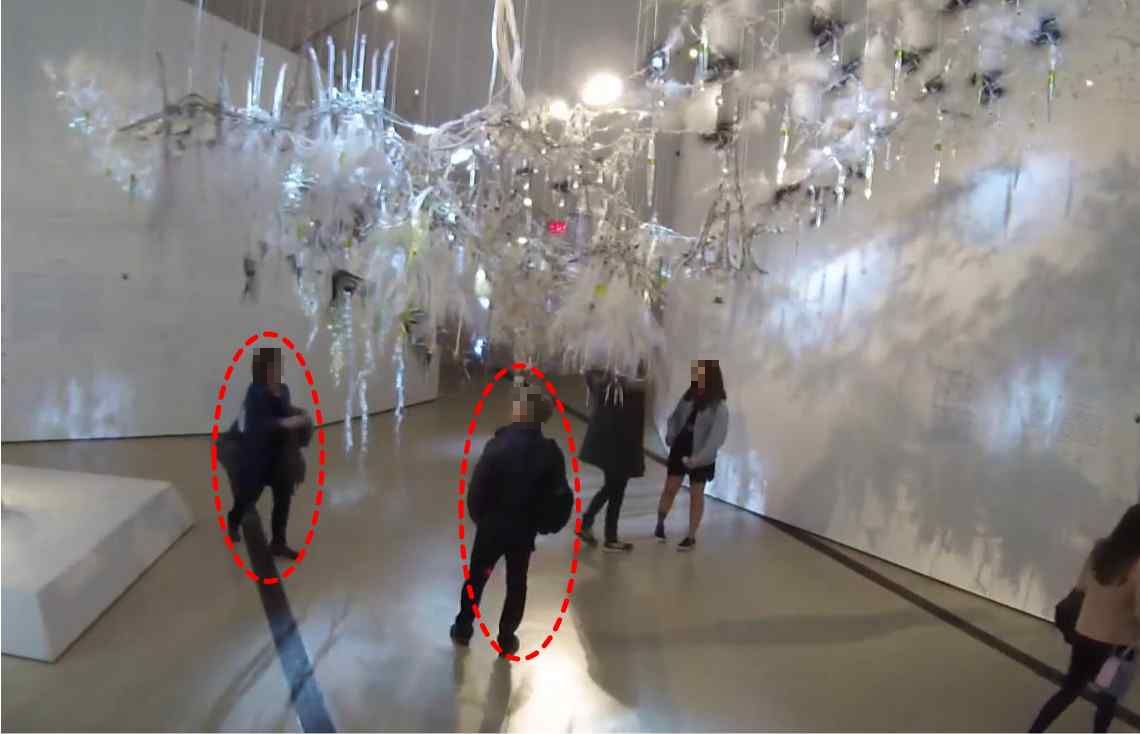}
            \includegraphics[width=1\linewidth]{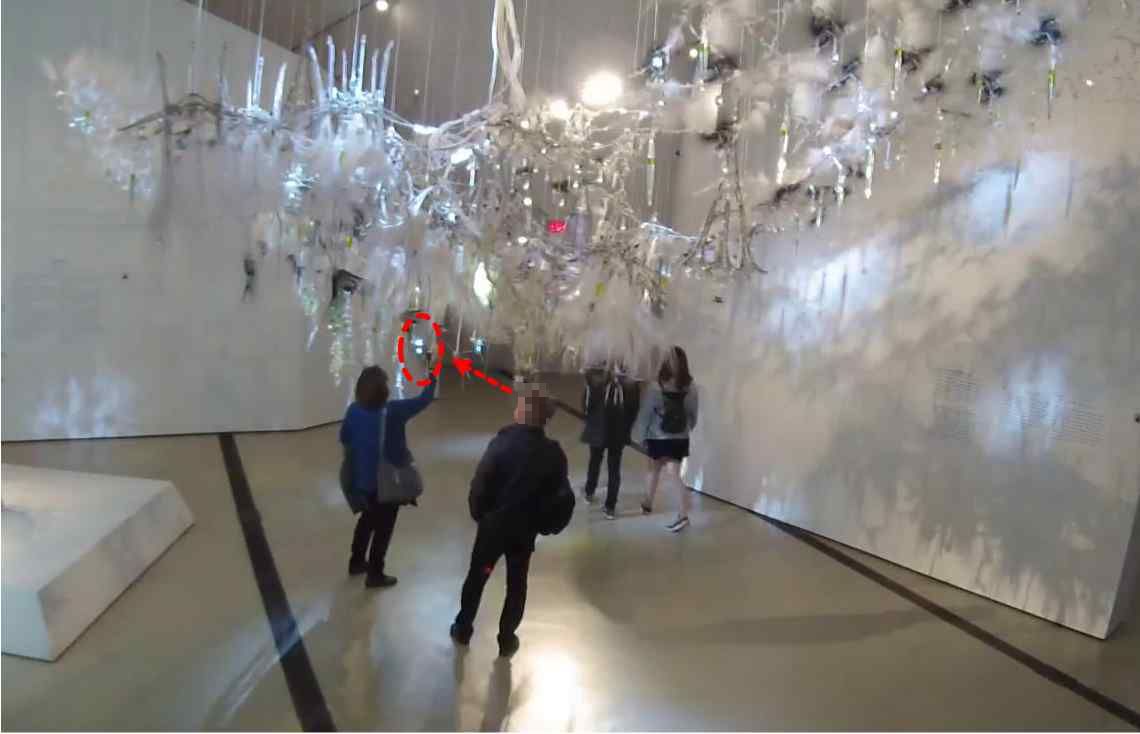}
            \includegraphics[width=1\linewidth]{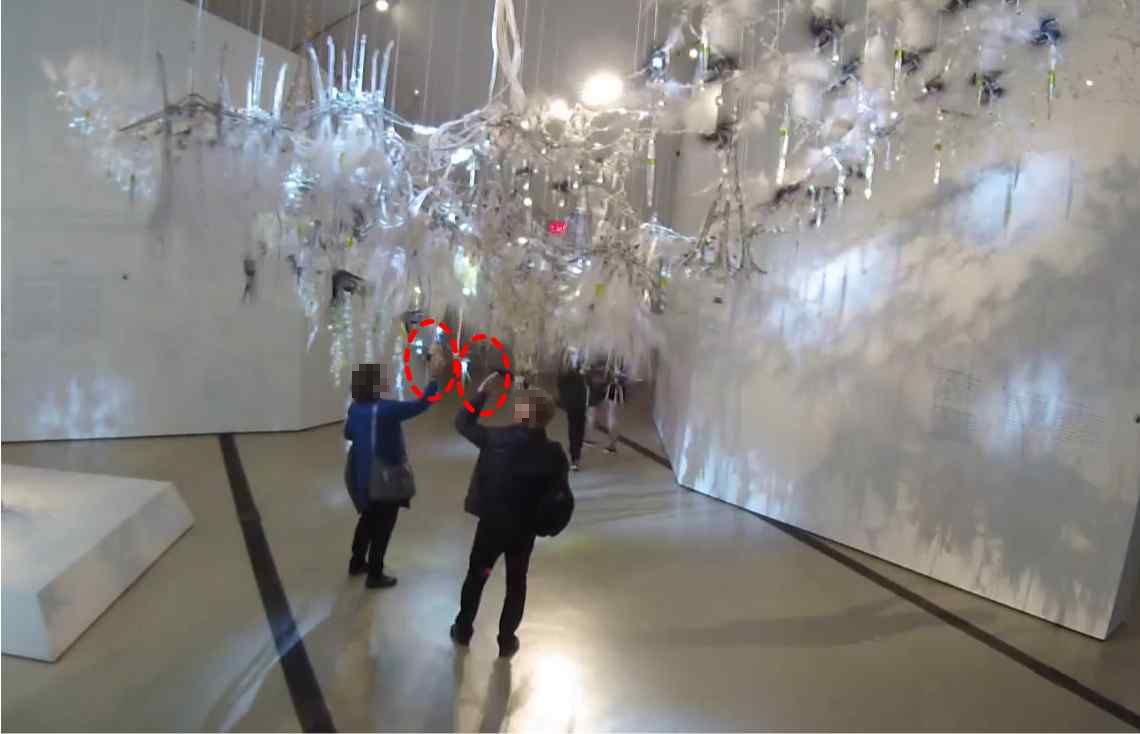}
            \caption{Learn how to interact from other visitor}
            \label{fig:Scenario_Learn_how_to_interact_from_other_visitor}
        \end{subfigure}%
        \hspace{3pt}%
        \begin{subfigure}[t]{.24\linewidth}
            \centering
            \includegraphics[height=.645\linewidth]{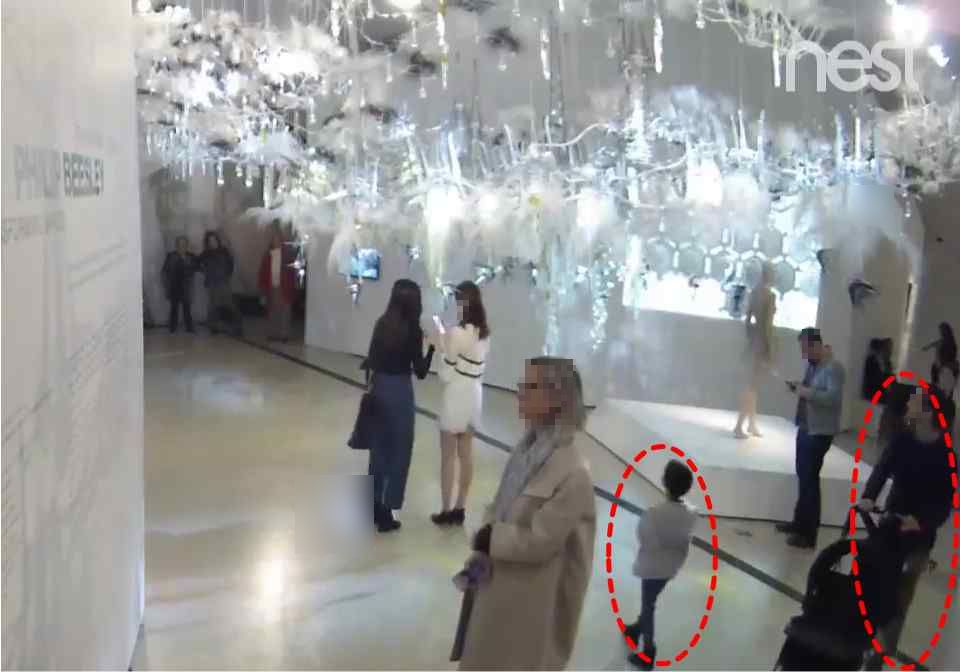}
            \includegraphics[height=.645\linewidth]{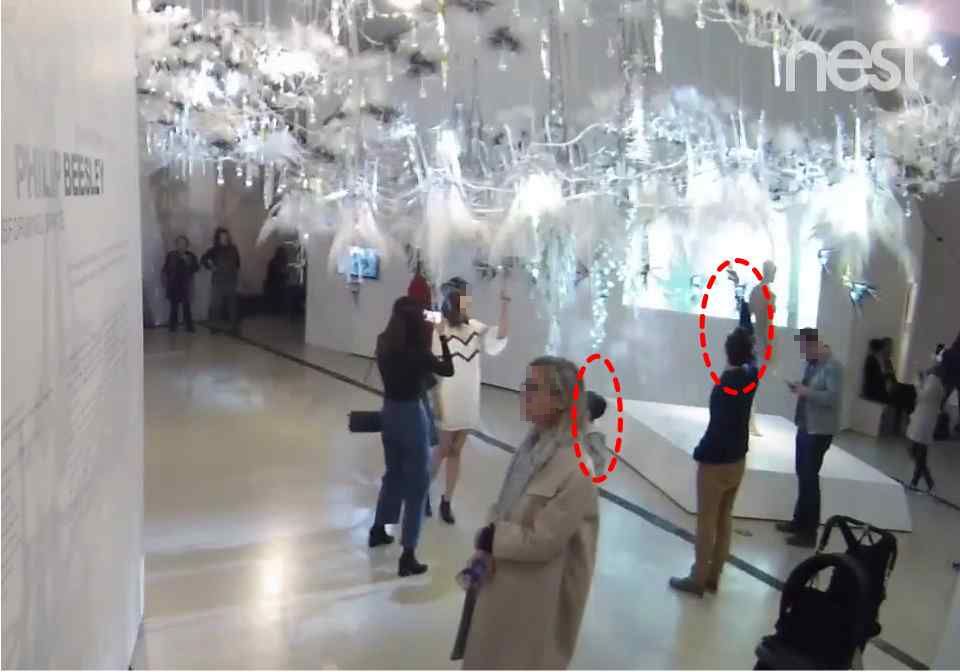}
            \includegraphics[height=.645\linewidth]{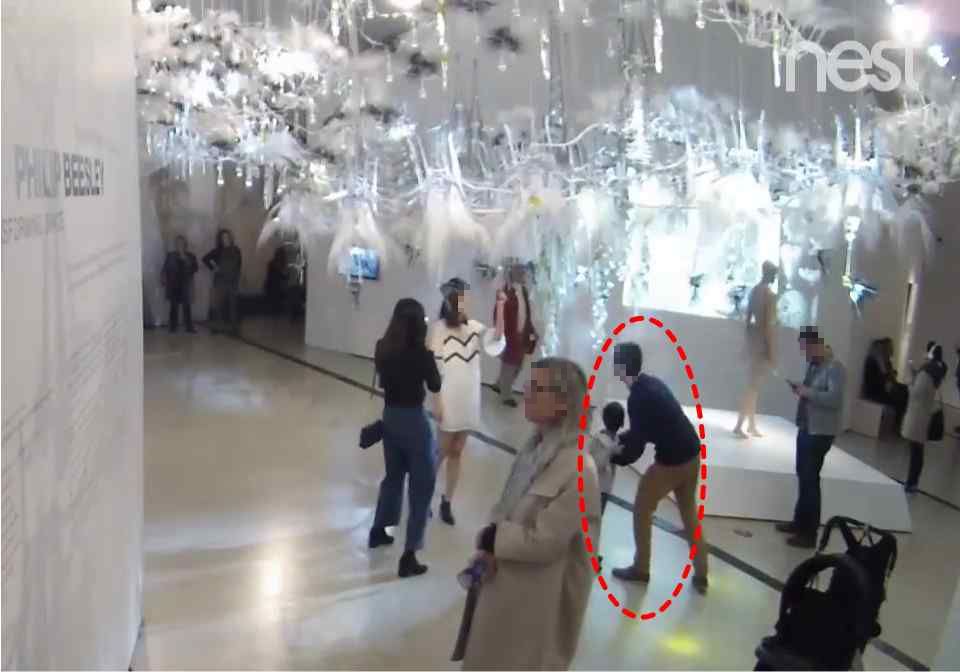}
            \includegraphics[height=.645\linewidth]{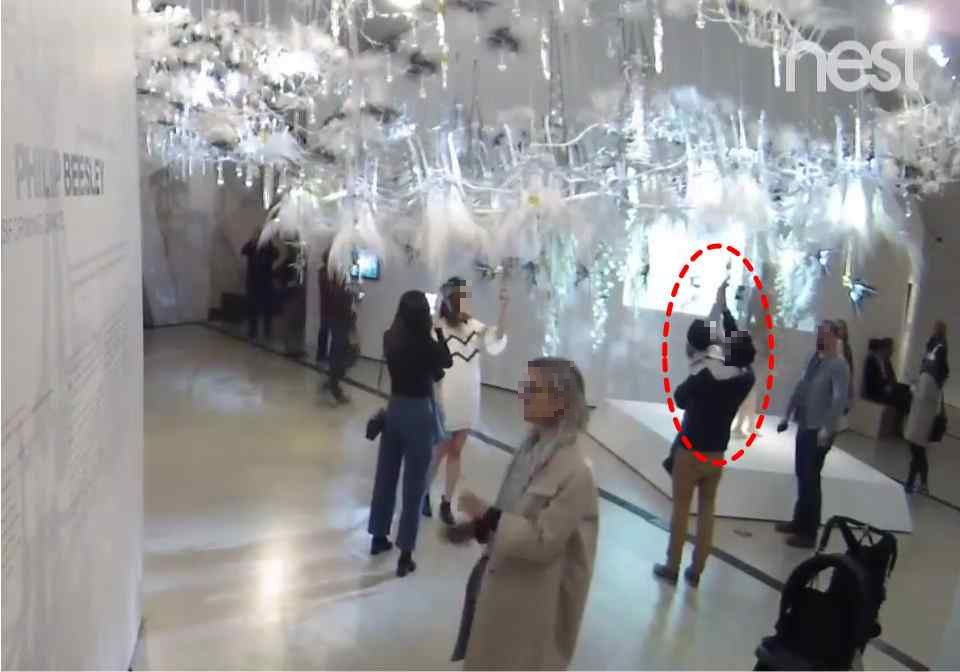}
            \caption{Parent lifts child}
            \label{fig:Scenario_Parent_lifts_child}
        \end{subfigure}
        \hspace{3pt}%
        \begin{subfigure}[t]{.24\linewidth}
            \centering
            \includegraphics[width=1\linewidth]{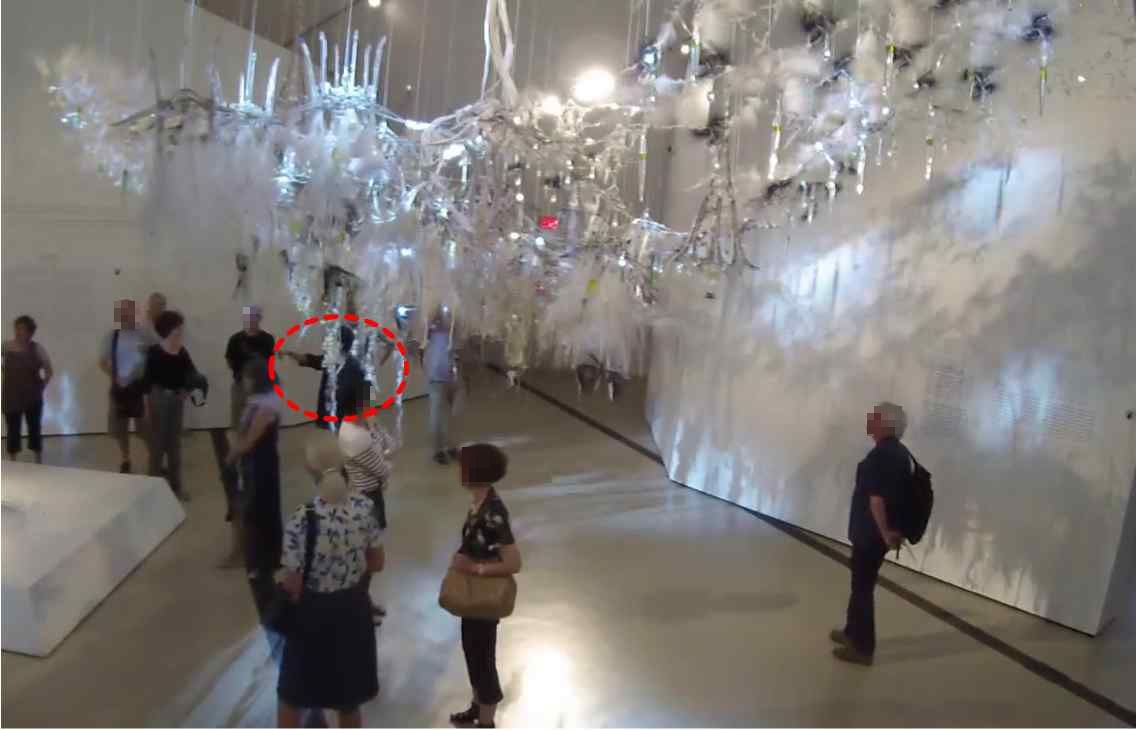}
            \includegraphics[width=1\linewidth]{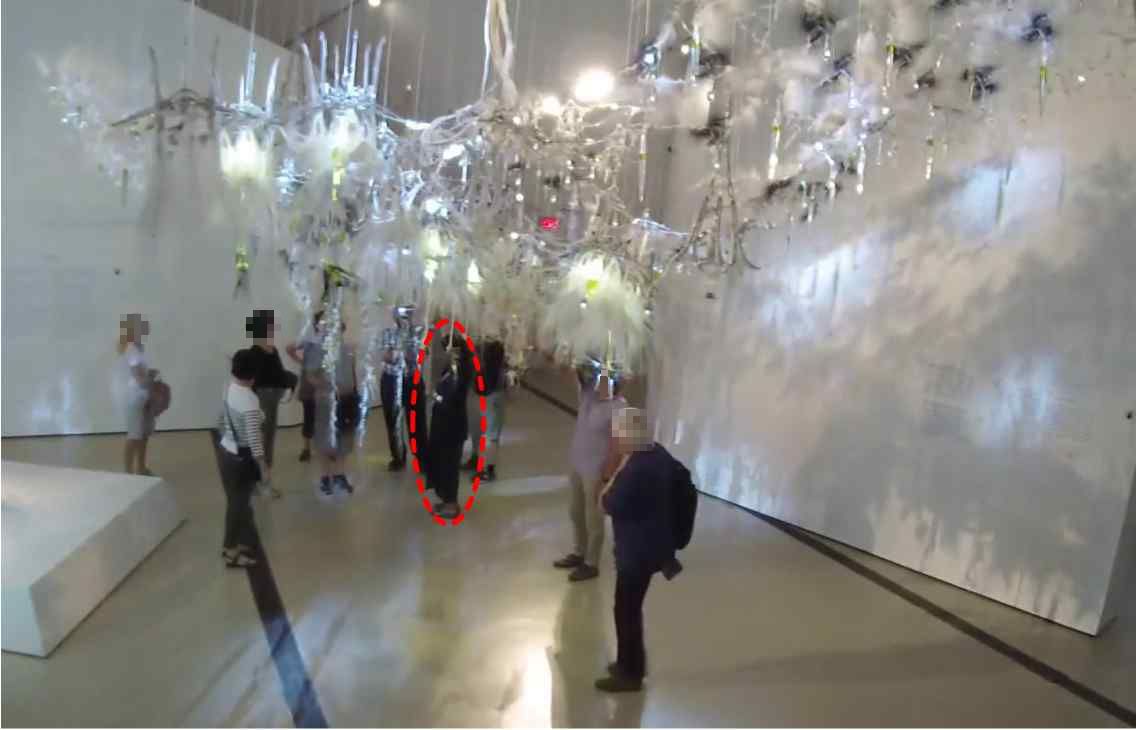}
            \includegraphics[width=1\linewidth]{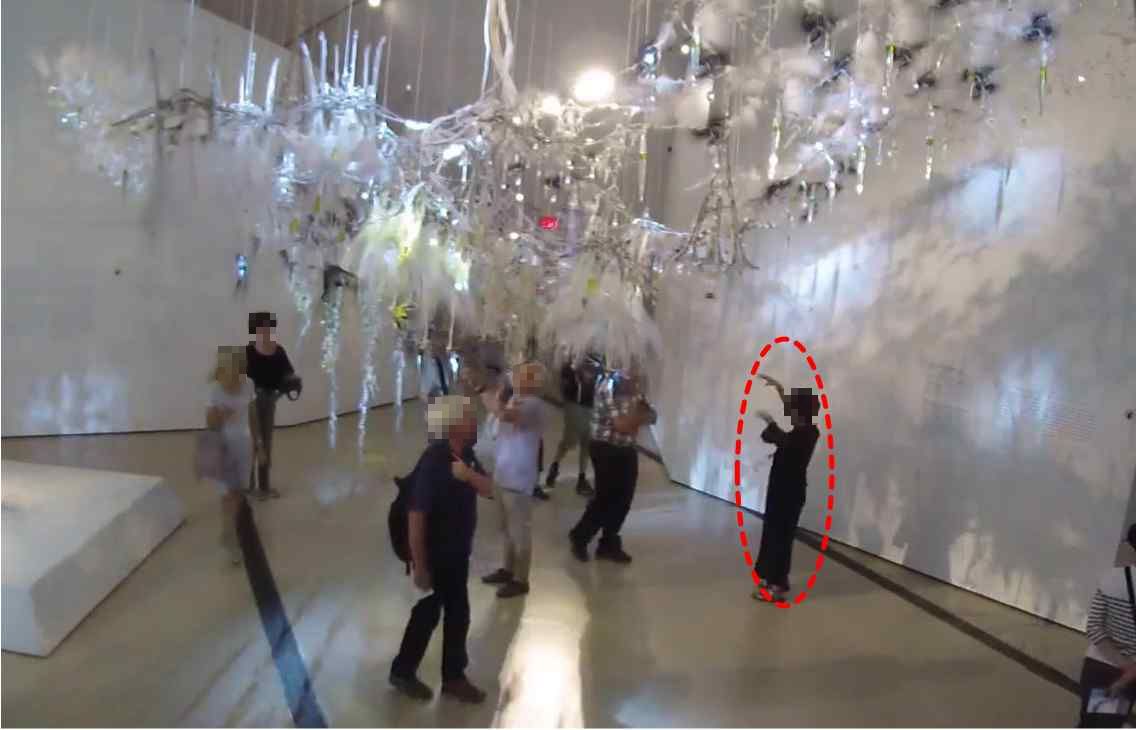}
            \includegraphics[width=1\linewidth]{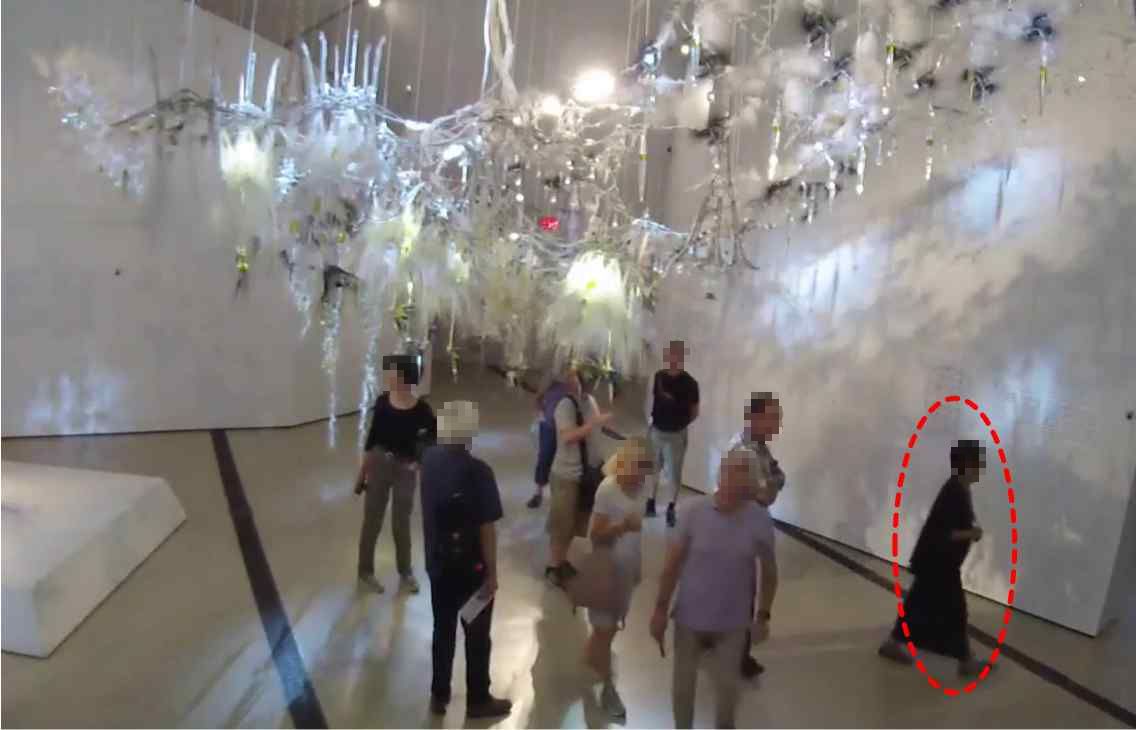}
            \caption{Group visit lead by a guide}
            \label{fig:Scenario_Group_visit_led_by_a_guide}
        \end{subfigure}
        \hspace{3pt}%
        \begin{subfigure}[t]{.24\linewidth}
            \centering
            \includegraphics[height=.645\linewidth]{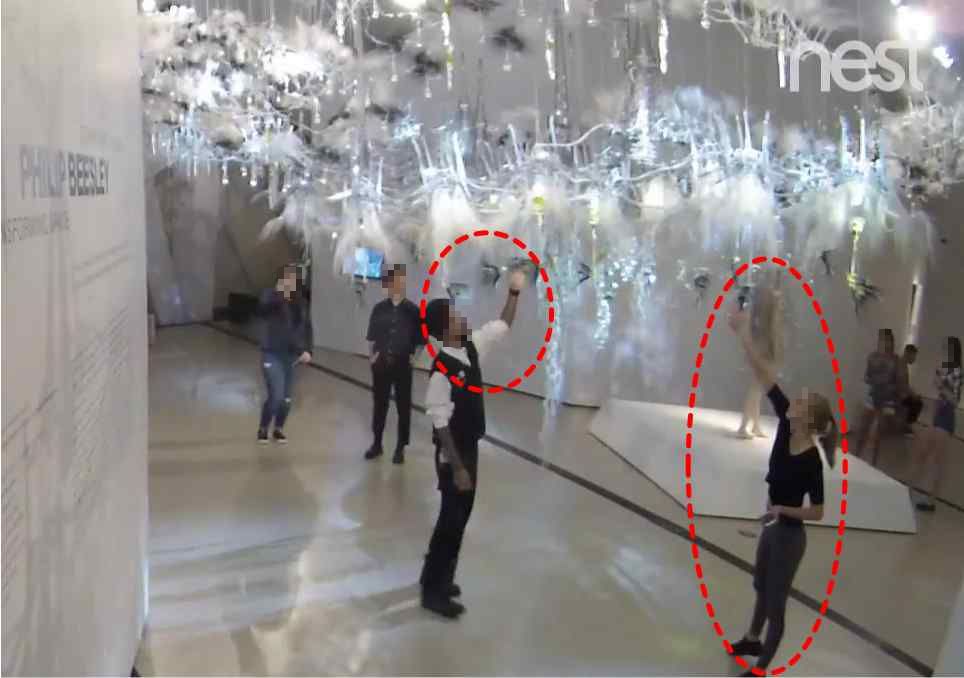}
            \includegraphics[height=.645\linewidth]{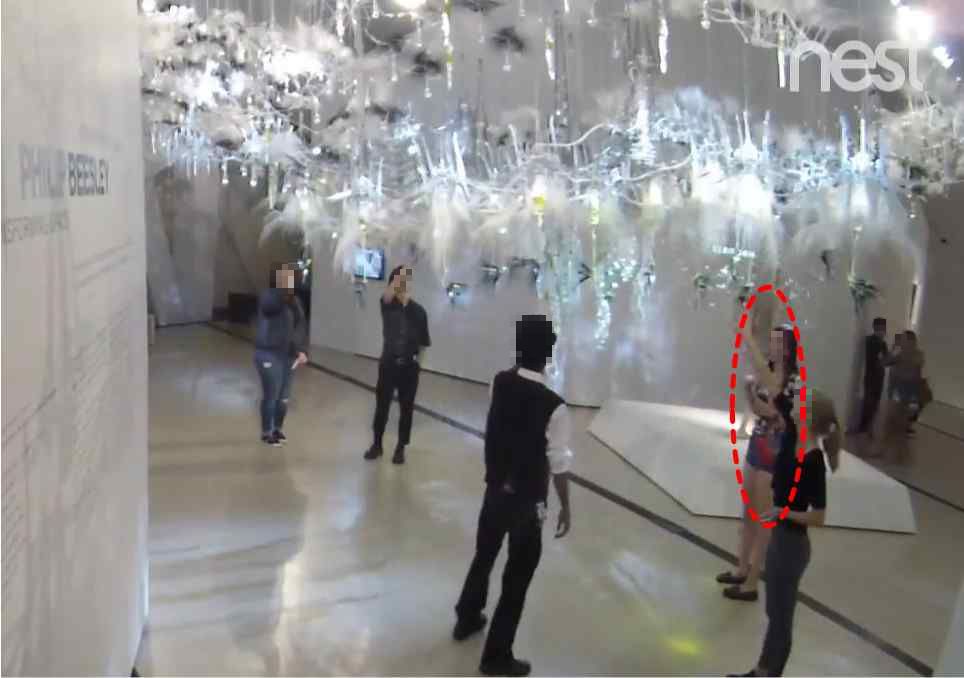}
            \includegraphics[height=.645\linewidth]{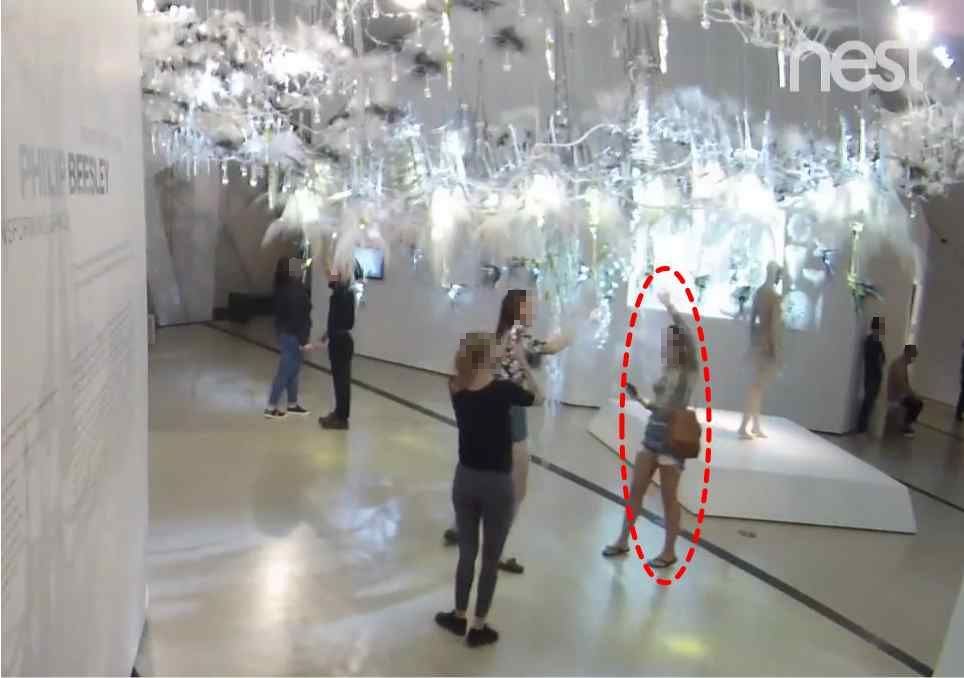}
            \includegraphics[height=.645\linewidth]{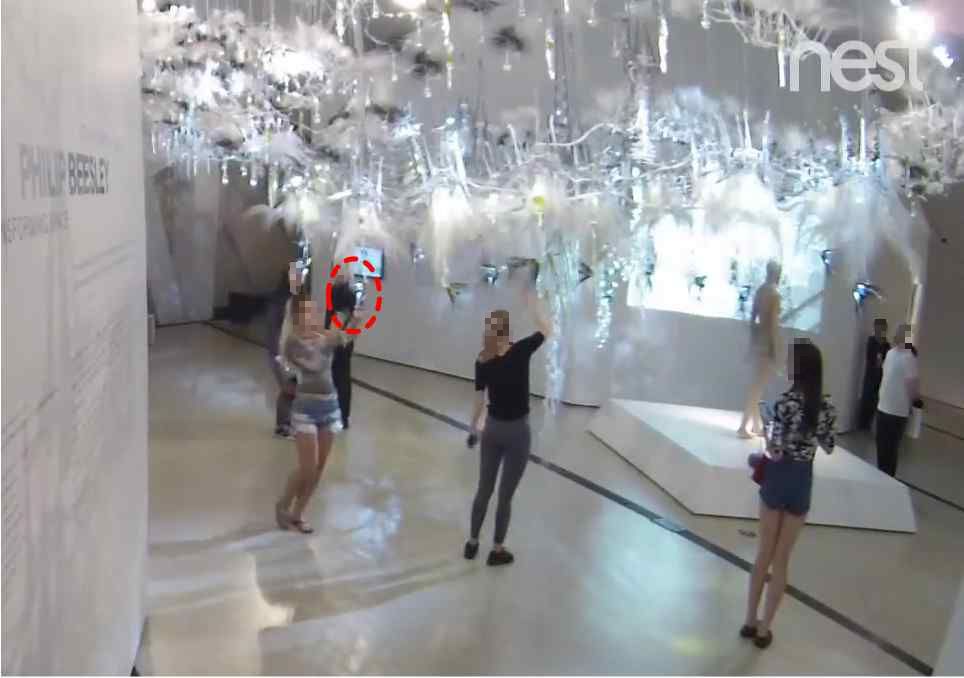}
            \caption{Visitors taught by security guard but misunderstood touch IR sensors}
            \label{fig:Scenario_Visitors_taught_by_security_guard_but_misunderstood_touch_IR_sensors}
        \end{subfigure}
        \caption{Sample Interesting Scenario. (a) shows how visitors can affect each other. Before the woman in blue arrives, the man in black had spent a while just looking around and did not know how to interact with the LAS. Then coincidentally he saw how the woman raised her hand and how the LAS responded. After that he copied her action to interact with the LAS. (b) shows a parent helping his child experience the interaction after he explored how to interact. Since the child could not reach the IR sensor, the parent lifted his child. (c) shows a group visit where a group of visitors is led by a guide. (d) shows a more complex scenario about misunderstanding shared information. In this case, three girls were taking photos without interacting. Then a security guard taught one of the three girls how to interact. At the same time, the second girl saw and joined them. After that the third girl learned this from her friends. However, there was a misunderstanding of the security guard's instruction, so they directly touched the IR sensors rather than just waving hand closely.}
        \label{fig:Sample_Interesting_Scenario}
    \end{figure}

The group setting presents a challenging environment for learning. During the entire experiment, we found some scenarios that highlight the complexity of using RL in LAS, such as interactions between visitors and the possibility that the LAS could be physically changed by touching as shown in Fig. \ref{fig:Sample_Interesting_Scenario}. In addition, there are many other examples of complicated environment dynamics that present challenges to a learning algorithm. Examples include visitors who use alternate interaction strategies, change their interaction strategy over time or because they are influenced by other visitors, as well as visitors who raise their hands for reasons other than interaction. These observations illustrate the non-stationarity of the environment, and the influence of human-human interaction in group scenarios during HRI.  They also emphasize the importance of developing and testing these algorithms ``in the wild.''

Selecting RL algorithms and hyper-parameters with real experiments is challenging, since we do not have a "validation set" that allows us to do multiple runs and we cannot accelerate learning. Due to the non-stationary nature of the investigated environment in our case, this is also impossible to do with a simulator, because unlike some fields in robotics where a realistic simulator can be devised, in our case the diversity of visitors' interaction styles, social influence of human-human interaction and variations in visitor numbers during different time periods, etc. make it impossible to devise such a good simulator. Therefore, in this work we only used a simplified simulator to make an initial hyperparameter selection and confirm our code is bug free rather than to pre-train a policy or fine-tune hyper-parameters. Finally, although DDPG was selected as the learning algorithm as it is easy to implement and works on continuous action space, the best choice of RL algorithm requires further investigation.  DDPG may not be sample efficient and can be susceptible to overestimation and sensitive to hyper-parameters \cite{henderson2018deep}. More advanced continuous state space algorithms, such as Proximal Policy Optimization (PPO) \cite{schulman2017proximal}, Soft Actor-Critic (SAC) \cite{haarnoja2018soft}, and Twin Delayed DDPG (TD3) \cite{fujimoto2018addressing}, should be investigated in the future. 

\section{CONCLUSIONS AND FUTURE WORK}

In this paper, we developed and evaluated algorithms for generating interactive behaviours in group environments. Specifically, we provide a way to estimate engagement during group interaction based on multiple IR sensors, where both individual engagement, passive and active interaction, and group engagement, i.e. occupancy, are taken into account. PB and PLA were examined to evaluate how the use of human knowledge influences interaction. By analyzing interaction and human survey data, we found that learned interactive behaviours, i.e. PLA, result in higher engagement and perceived likeability than pre-scripted behaviour, i.e., PB.

Even though PLA received higher average engagement and perceived likeability than PB, we cannot be certain about the cause of this difference. Therefore, a baseline with random policy can be tested to see if there is a  difference between this baseline and PLA to confirm that the learning agent is indeed learning from and adapting to its interaction experience. Other advanced continuous control DRL algorithms such as PPO, SAC and TD3 are also worth investigating. Upcoming installations of even larger LAS are planned and this decentralization will become necessary as the number of actuated elements increases beyond one thousand in a single installation. In addition, hierarchical RL with PB bootstrapping could be a promising extension, where we could design a pool of PBs and various levels of reward functions, and see how complicated action patterns could emerge. We also plan to introduce intrinsic motivation and a learning algorithm driven both intrinsically and extrinsically for LAS. To tackle the low pace of interaction in LAS and high sample requirement of RL, we also plan to investigate how to transfer learned models from simulation to physical LAS. 

\begin{acks}

The authors would like to thank Philip Beesley Architect Inc. and the Royal Ontario Museum for allowing us to conduct research with the exhibit, Salvador Breed  and 4DSOUND for developing the sound system, and Joslin Goh and Pavel Shuldiner from the Statistical Consulting and Collaborative Research Unit (SCCR) at the University of Waterloo for insightful discussions on statistical analysis.

This work was supported by the Living Architecture Systems Group, funded by the Social Sciences and Humanities Research Council of Canada.

\end{acks}

\bibliographystyle{ACM-Reference-Format}
\bibliography{bibliography}

\end{document}